\documentclass[final]{jfm}

\usepackage[dvipsnames,xcdraw]{xcolor}
\usepackage{graphicx}
\usepackage{newtxtext}
\usepackage{newtxmath}
\usepackage{natbib}
\usepackage[hidelinks]{hyperref}
\usepackage[noabbrev,nameinlink]{cleveref}
\usepackage{bm}
\usepackage{amsfonts,amsmath} 
\usepackage{textcomp}
\usepackage{verbatim}
\usepackage{physics}
\usepackage{subcaption}
\usepackage[normalem]{ulem}
\usepackage{mathtools,tikz,caption}
\usepackage{afterpage}
\usepackage{pgfplots}\pgfplotsset{compat=1.17}
\usepackage{dcolumn}
\usepackage{float}
\usepackage{hhline}
\newcolumntype{P}[1]{>{\centering\arraybackslash}p{#1}}
\newcommand{\RNum}[1]{\uppercase\expandafter{\romannumeral #1\relax}}
\crefformat{section}{\S#2#1#3} 
\crefformat{subsection}{\S#2#1#3}
\crefformat{subsubsection}{\S#2#1#3}

\newcommand{\RomanNumeralCaps}[1]
\linenumbers

\title{Machine-learned  control-oriented flow estimation for multi-actuator multi-sensor systems exemplified for the fluidic pinball}

\author{Songqi Li\aff{1},
  Wenpeng Li\aff{1}
 \and Bernd R. Noack\aff{1} \corresp{\email{bernd.noack@hit.edu.cn}}}

\affiliation{\aff{1}School of Mechanical Engineering and Automation, Harbin Institute of Technology, 518055 Shenzhen, PR China.}

\begin{document}
\maketitle

\begin{abstract}
We propose the first machine-learned control-oriented flow estimation
for multiple-input multiple-output plants.
\textcolor{black}{
Starting point 
is constant actuation
with  open-loop actuation commands
leading to a database with simultaneously 
recorded actuation commands, sensor signals and flow fields.}
A key enabler is an estimator input vector 
comprising sensor signals and actuation commands.
The mapping from the sensor signals and actuation commands to the flow fields
is realized in an analytically simple, data-centric and general nonlinear approach.
The analytically simple estimator generalizes 
Linear Stochastic Estimation (LSE) for actuation commands.
The data-centric approach yields flow fields from estimator inputs
by interpolating from the database
--- similar to \citet{Loiseau2018jfm} for unforced flow.
The interpolation is performed with $k$ Nearest Neighbors ($k$NN).
The general global nonlinear mapping from inputs to flow fields
is obtained from a Deep Neural Network (DNN) via an iterative training approach.
The estimator comparison is performed 
for the fluidic pinball plant, which is a multiple-input, multiple-output wake control benchmark \citep{Deng2020jfm} featuring rich dynamics under steady controls.
We conclude that the machine learning methods clearly outperform the linear model.
The performance of $k$NN and DNN estimators are comparable for periodic dynamics.
Yet,  DNN performs consistently better when the flow is chaotic.
Moreover, a thorough comparison regarding the complexity, computational cost, and prediction accuracy is presented 
to demonstrate the relative merits of each estimator.
\textcolor{black}{The proposed method can be generalized for closed-loop flow control plants.} 
\end{abstract}



\section{Introduction}
\label{sec:intro}

Taming turbulence in the flow represents significant economic and environmental benefit in most industrial applications \citep{Brunton2015}.
\textcolor{black}{Turbulence control plays a significant role in drag reduction (for example, see \citealt{Choi1994,Lee1997,Chamorro2013}).}
For the example of transport vehicles, aerodynamic drag accounts for about 15\% of the round-trip fuel consumption for a coal train \citep{Stodolsky2003}, and for cars overcoming aerodynamic drag takes up most of the power consumption at highway speeds \citep{Sudin2014}.
A drag reduction of around 25\% in a 3D bluff body reported by \citet{Pfeiffer2012} indicates turbulence control a remarkable energy saving strategy which is potentially applicable to all relative industries.

To tackle the turbulence control problems, myriads of control strategies have been proposed \citep{Joslin2009} and these strategies can be principally classified into passive and active categories.
For passive control, small modifications of the flow configuration are implemented to improve the surrounding flow, such as the installation of vortex generators to reduce drag on a car \citep{Mukut2019}, and the application of chevrons nozzle to enhance the jet mixing \citep{Zaman2011}.
Active control requires energy input to the flow domain from the actuators. 
Different types of active actuators such as synthetic jets \citep{Amitay2001}, plasma actuators \citep{Wang2013}, and moving surfaces \citep{Modi1997} have been widely applied in the flow control experiments with significant improvement of flow performance \citep{Cattafesta2011}. 
For either control strategy, the control devices can be described in a parameterized manner, such as the shape parameters of the vortex generator or the duty cycles of the plasma actuators.
These parameterized flow control devices will alter the dominant turbulent structures in the original flow, and generates distinct flow patterns according to the selection of the control parameters.

In the typical flow control setup, point-wise sensors, such as hotwire probes, pressure transducers, and shear stress sensors, are sparsely placed in the flow or on the boundary.
The deployment of these sensors makes it possible to recover the flow state from sparsely sampled sensor signals which comprises partial information of the flow state.
The full state estimation from sparse sensors is of great importance in terms of flow control \textcolor{black}{\citep{Taylor2004}}.
The accurate estimation of the full or partial flow state has the potential to notably improve the control outcome via a carefully designed controller, 
leading to remarkable drag reduction, mixing enhancement, noise attenuation, etc. \citep{Brunton2015}.
\textcolor{black}{The machine learning architecture which
couples recurrent neural network (RNN) and Model Predictive Control (MPC) 
in \cite{Bieker2020} is also proven to accurately predict and control the flow state from limited sensor data.
For oscillatory dynamics, 
 flow state estimation techniques based on reduced-order models (ROM) have also been widely applied in the active flow control plants (see, for example,  \citealt{Tadmor2004} and \citealt{Barbagallo2009}).
However, 
a sensor-based dynamic observer looses 
its accuracy with increasing state-space dimension even for a simple unforced oscillatory wake transient \citep{Gerhard2003aiaa}. 
Every extra degree of freedom may act as potential noise amplifier.
Dynamic ROM, let alone estimation, 
is also a challenge for wall-bounded actuators 
with corresponding actuation commands 
as free input as exemplified by  \citealt{Weller2009physd} for oscillatory dynamics.}
\textcolor{black}{These challenges are even more pronounced for sensor-based estimation using ROM for  complex dynamics with multiple independent actuators.}

As the most commonly used technique, the Linear Stochastic Estimation (LSE, \citealt{Adrian1994}) as well as its variants \citep{Tinney2006,Lasagna2013}, establish a linear model between sensor inputs and the flow state output \citep{Clark2014}.
This method has been widely applied to either distill flow structures \citep{Bonnet1994} or construct models for accurate flow estimation \citep{Stalnov2007}.
The well-trained linear model can serve as state estimation function in feedback flow control experiments (for example, see \citealt{Samimy2007}).
\textcolor{black}{
The Quadratic Stochastic Estimation (QSE) includes a
nonlinear second order term.
These terms have been shown to be important
for some wake estimation problems \citep{Bourgeois2013jfm}.
Yet, QSE calibration is computationally expensive  \citep{Murray2002}.
Challenges of QSE have already been reported in \cite{Tung1980}. 
In contrast, neural network regression
has become a mature methodology.}

The fast development of machine learning methods \citep{Brunton2020} makes it possible to estimate the flow state via a spectrum of non-linear approaches.
A state estimation method based on the autoencoder and the recurrent neural network is proposed by \textcolor{black}{\citet{Kumar2022}} in which the flow field can be accurately predicted from few sensors.
The comparison of neural network, LSE, and gappy-POD is reported by \citet{Nair2020} for the prediction of the laminar wake behind a fat plate from sparse sensor measurements.
A series of CNN-based estimation architectures are proposed and evaluated by \citep{Guemes2019,Guemes2021,Guastoni2021} to estimate the turbulent flow from wall measurements and the results yield promising accuracy.
The accurate estimation of the flow state from time-resolved sensor measurements and non-time-resolved velocity snapshots using long-short-term-memory (LSTM) is reported in \cite{Deng2019}.
Nevertheless, neither of these recent advancements in machine learning is naturally designed for multiple-input, multiple-output flow control.
In the case of model-based feedback flow control, it is necessary to establish an estimation method which can accurately infer the flow state from sparse measurements under adjustable control commands.

For the present study, we propose the first machine learning, sensor-based flow estimation which is capable to infer the flow state from sparse measurements for multiple-input, multiple-output plants.
The $k$ Nearest Neighbors ($k$NN) regression and Deep Neural Network (DNN) serve as two non-linear approaches, which represents a data-centric and a global non-linear approach, respectively.
In addition, LSE is employed as a linear reference point to highlight the significance of non-linearity in the estimation problem.
These methods are examined in the fluidic pinball plant, a cluster of three equal parallel cylinders centered on an equilaterial triangle pointing upstream in a uniform flow \cite{Bansal2017,Chen2020}.
This configuration is becoming an increasingly popular benchmark for 
 multiple-input multiple-output (MIMO) flow control.  
The flow may be monitored using the forces on the cylinders or the velocity signals at selected locations.
Actuation is performed with 
the rotation speed of each cylinders as control input. 
These actuations give rise to rich dynamics in the flow and consequently it becomes challenging to obtain accurate flow estimation under a wide range of steady control commands.
Meanwhile, a large spectrum of control algorithms have been applied to this plant \citep{Peitz2020,Raibaudo2020,Raibaudo2021,Blanchard2021,Ghraieb2021,Maceda2021,Li2022,Castellanos2022}.
The fluidic pinball wake has been investigated in detail by \citet{Noack2016,Ishar2019,Deng2020jfm,CortinaFernndez2021,Deng2021,Deng2021epl,Deng2022jfm,Chen2022,Menier2022} among many others.

This manuscript is organized as follows. The problem setup of the control-oriented flow estimation problem and the estimation procedure \textcolor{black}{are detailed in} in \Cref{ssec:problem}, followed by the introduction of $k$NN (\Cref{ssec:kNN}), DNN (\Cref{ssec:dnn}), LSE (\Cref{ssec:lse}).
Section \ref{sec:plant} introduces the fluidic pinball plant as well as the numerical database that will be used in this study.
The comparison of three estimation methods exemplified for the fluidic pinball plant is presented in \Cref{sec:results}. 
Sections \ref{ssec:case1} to \ref{ssec:case3} show the estimation results for three representative cases which corresponds to the specific flow control strategies.
The overall performance of each estimation method is concluded in \Cref{ssec:evaluation}, followed by the conclusions and outlook in \Cref{sec:conclusions}.

\section{Non-Linear, Control-Oriented Flow Estimation}
\label{sec:estimation}

In this section we introduce control-oriented, sensor-based flow estimation which is capable to evaluate the flow state under various steady control commands.
The setup of estimation problem, as well as the general estimation procedure, are described in \Cref{ssec:problem}, and the data pre-processing technique is briefly introduced in \Cref{ssec:pre-pod}.
To tackle the estimation problem, we propose a spectrum of data-driven estimators in this work. 
Two machine learning-based approaches, i.e., $k$ Nearest Neighbors ($k$NN) regression and Deep Neural Network (DNN), will be introduced in \cref{ssec:kNN} and \ref{ssec:dnn}, respectively. 
In addition, the traditional Linear Stochastic Estimation (LSE) method, which serves as a reference point in the present work, will be explained in \Cref{ssec:lse}.

\subsection{Problem setup and estimation procedure}\label{ssec:problem}

For stationary flow under a range of steady control commands, our objective is to estimate the instantaneous two-dimensional velocity field $\pmb{u_b}(\pmb{x})$ from the known information, which includes the instantaneous sensor signal $\pmb{s}$ and the control command $\pmb{b}$.
Here the velocity vector is defined as $\pmb{u_b}=(u,v)$, and the spatial domain $\pmb{x}=(x,y)$ is bounded in $\mathbb{R}^2$.
The control command $\pmb{b}=[b_1,...,b_M]^\text{T}$ is composed of all parameterized actuation instructions from a total of $M$ actuators in the flow. 
Each actuation instruction is bounded in $\mathbb{R}$ such that the control command $\pmb{b}$ is located in a finite high-dimensional space $\mathcal{D} \in \mathbb{R}^M$.
For the fluidic pinball plant that will be examined in this study, $\pmb{b}$ represents the rotation speeds of cylinders which are placed in the flow.
For other relevant flow control applications, $\pmb{b}$ may also represent duty cycles of the mini-jet/plasma actuators, shape parameters of the passive control devices, etc.
Local sensor signal $\pmb{s}=[s_1,...,s_L]^\text{T}$ contains partial information of the flow field.
The sensors are placed at sparse locations $\{\pmb{x}_{\pmb{s}}^{(i)}\}^{L}_{i=1}$, 
and the measurement is equivalent to a masking function $\mathcal{G}$ which preserves the flow state at sensor locations.
The challenge of this study is to establish a mapping function $\mathcal{F}$ from local sensor signal $\pmb{s}$ and the actuation command $\pmb{b}$, to the estimation of the corresponding flow state $\hat{\pmb{u}}_{\pmb{b}}(\pmb{x})$ under control.

\begin{figure}
    \centering
    \includegraphics[width=.8\linewidth]{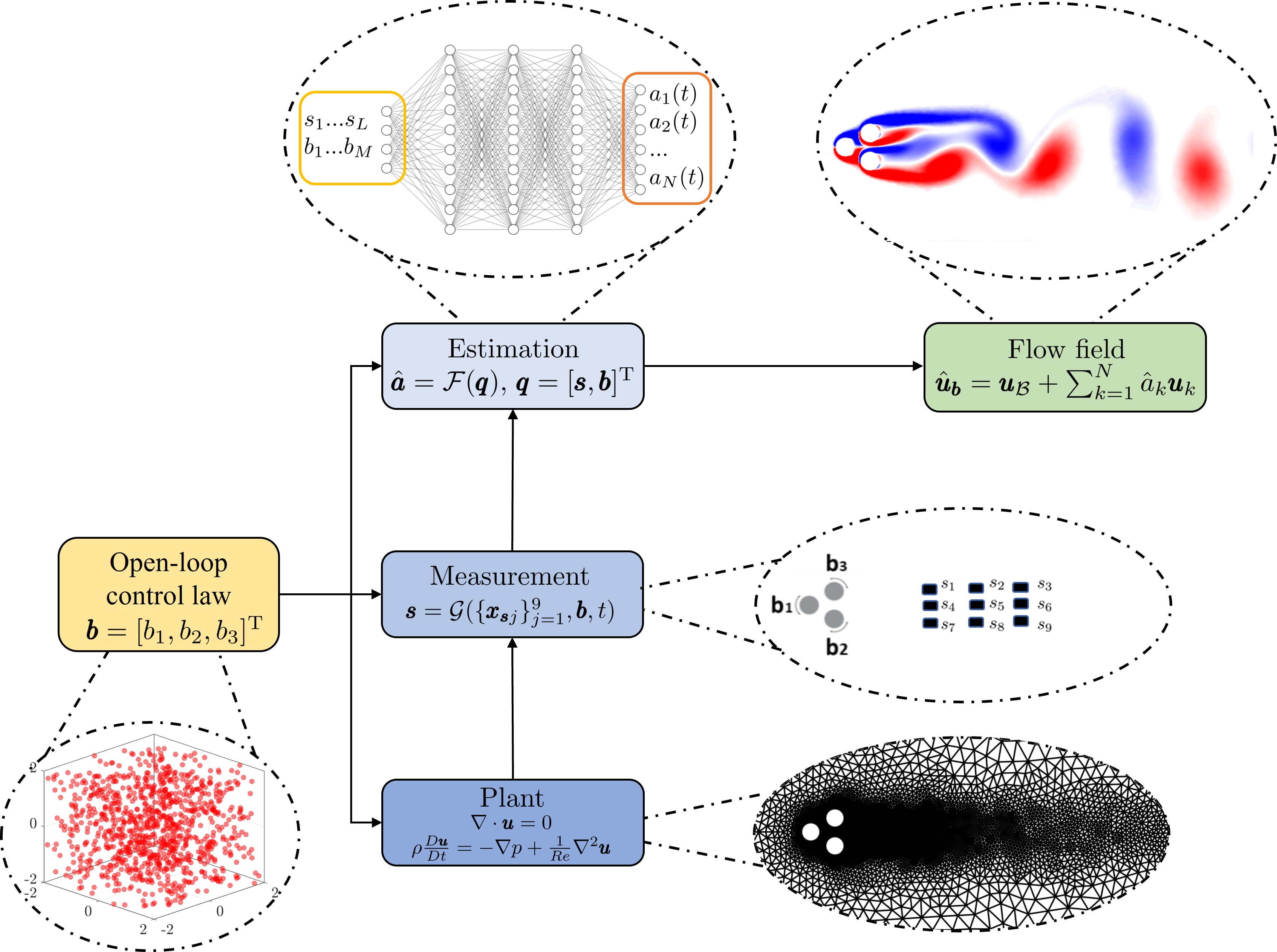}
    \caption{The flow chart of control-oriented full-state estimation exemplified for the fluidic pinball plant and the DNN estimator.}
    \label{fig:flowchart}
\end{figure}

To achieve this goal, we propose a data-driven procedure to estimate the flow state under different control laws.
\Cref{fig:flowchart} outlines the estimation procedure exemplified for the fluidic pinball, which is a benchmark problem which will be examined in this study.
This plant contains multiple control commands ($b_1$ to $b_3$) and multiple sensor measurements ($s_1$ to $s_9$), details of this plant will be introduced in \cref{sec:plant}.
For most sensor-based flow state estimation methods, only sensor signals are utilized as the input.
However, to include the influence of different control commands to the flow state,
in this study we employ a combined input vector $\pmb{q}$ which comprises the sensor signal $\pmb{s}$ and the control command $\pmb{b}$, such that:
\begin{equation}
    \pmb{q}=[\pmb{s},\pmb{b}]^{\text{T}}\in\mathbb{R}^{L+M}.
\end{equation}
\textcolor{black}{The effectiveness of this algorithmically easy extension of the estimator input in the parameter direction has been proved in \cite{Hasegawa2020}.
In this pioneering work, a CNN-LSTM-based reduced-order model is established for flow at different Reynolds numbers.
Here we extend this approach and include high-dimensional actuation parameters to the input vector.} 
For the purpose of establishing a data-driven mapping from the combined input $\pmb{q}$ to the flow state ${\pmb{u}}_{\pmb{b}}$, we first construct a database which contains massive input-output pairs under a set of $N_b$ randomly generated control commands. 
These control commands are denoted as $\mathcal{B}=\{\pmb{b}_j\}_{j=1}^{N_b}$.
For numerical plants, the flow snapshots are calculated from the numerical solver,
and the sensor signal $\pmb{s}$ are obtained by masking the flow field according to the sensor locations.
For experimental applications, the data can be acquired from synchronous measurements between Particle Image Velocimetry (PIV) and hotwire probes/pressure sensors (for example, see \citealt{Tu2013}).

Based on the well-established database described above, data-driven estimation can be performed to optimize and generalize the mapping function.
Nevertheless, given the high-dimensional nature of the flow state space, it is computationally expensive to estimate the velocity vector at every point inside the flow domain.
Hence a reduced-order representation of the flow field is required prior to the application of any estimation algorithm.
Different types of dimensionality reduction techniques have been proposed and are proven to be successful, including but not limited to deep autoencoder \citep{Fukami2020}, the proper orthogonal decomposition (POD, \citealt{Lumley1967}), and wavelet transform \citep{Bruce2002}, etc.
In this work a POD-based approach is employed to reduce the complexity of the flow snapshots.
The application of POD doesn't involve any pre-training which are typically computationally expensive, and the resulting POD eigenfunctions are human-interpretable.
In this work, the database which includes all flow snapshots are pre-processed with a nearly loss-less POD which compresses the high dimensional data onto an energy-optimum, low-dimensional subspace (see \Cref{ssec:pre-pod}).
Consequently, the estimation of instantaneous flow snapshots is equivalent to the estimation of the first $N$th POD modal coefficients $\pmb{a}$ which takes up the majority of fluctuating energy in the flow.

\begin{figure}
    \centering
    \includegraphics[width=.6\linewidth]{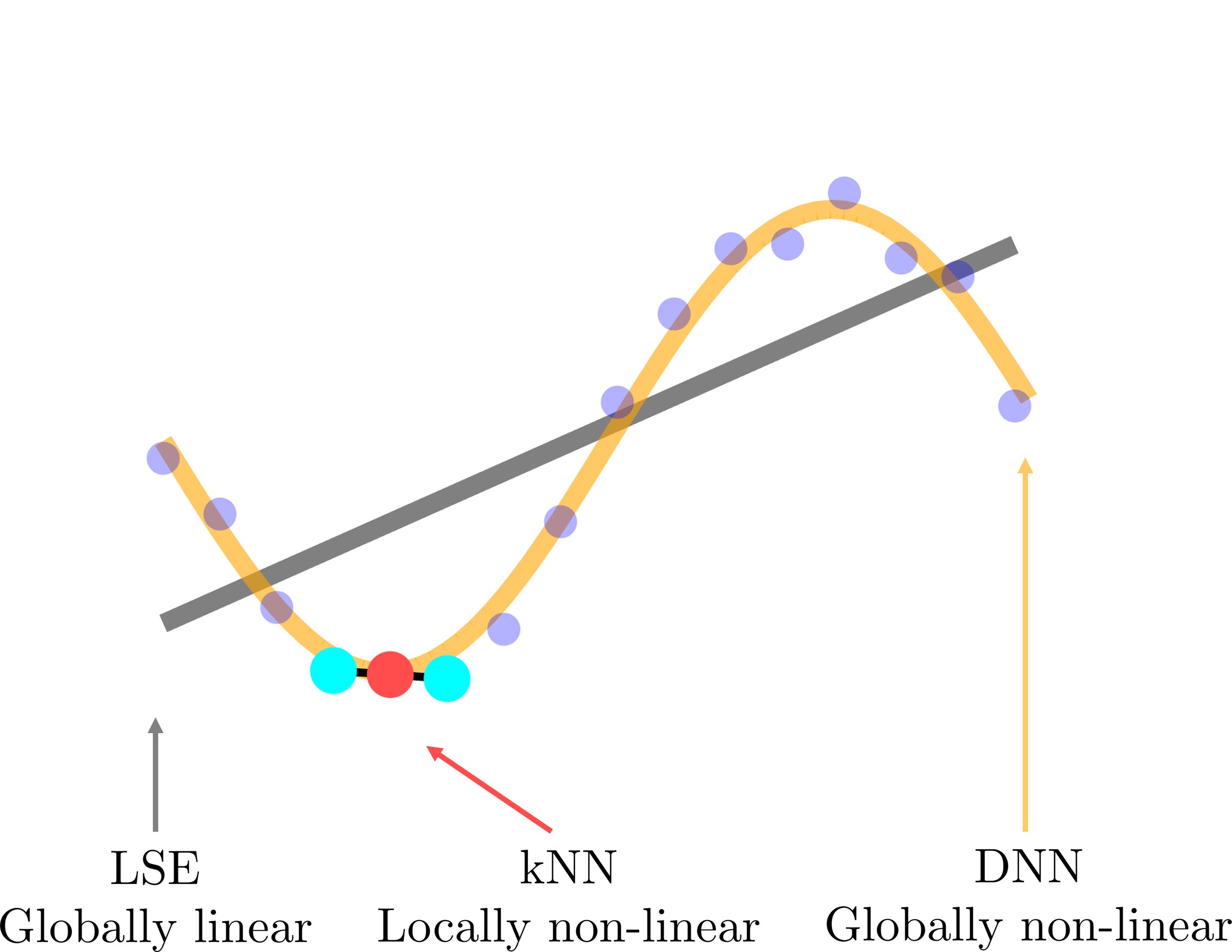}
    \caption{A graphical illustration of three estimation methods. Scattered points represents true data. LSE (gray) fits the data in a linear manner. $k$NN (red) estimates from the neighborhood observations \textcolor{black}{(light blue)}. DNN (yellow) optimizes the input-output relationship by a globally non-linear function.}
    \label{fig:methods}
\end{figure}

The generation of the database containing flow snapshots under various control laws, and the application of the dimensionality reduction technique, enable us to estimate the flow state via a spectrum of estimators which are capable to learn the mapping function $\mathcal{F}: \pmb{q}\rightarrow \pmb{{a}}$ from a total of $N_s$ input-output observation pairs $\mathcal{M} : \{(\pmb{q}^{(j)},\pmb{a}^{(j)})\}_{j=1}^{N_s}$ recorded under different control commands in the set $\mathcal{B}$.
In this work three representative estimators are proposed, and a graphical illustration of these data-driven methods can be found in \textcolor{black}{\cref{fig:methods}}.
The $k$ Nearest Neighbors ($k$NN) regression acts as a locally non-linear approach,
this method learns the mapping from its neighborhood observations via a non-parametric, distance-based interpolating process. 
The Deep Neural Network (DNN) establishes a globally non-linear model and optimizes the internal parameters of the model through an iterative optimization routine.
For Linear Stochastic Estimation (LSE), a linear relationship between inputs and outputs is first assumed, then a least-squares fit is performed to minimize the estimation error. 
These estimation methods, as well as the POD-based data compression technique, will be introduced in the following subsections.

\subsection{Data pre-processing: the proper orthogonal decomposition (POD)} \label{ssec:pre-pod}

In this study, the snapshot POD \citep{Sirovich1987} is employed as a data pre-processing technique, which reduces the computational cost of the estimation by compressing the high-dimensional flow snapshots onto a low-dimensional orthonormal subspace where energy is optimized.
\textcolor{black}{As we don't include the time coordinate for the control-oriented flow estimation in this study, Koopman-style decomposition techniques \citep{Chen2012} are not available for the data pre-processing.}
Given a set of flow snapshots under various control commands in $\mathcal{B}$, a velocity decomposition is first performed which splits the velocity vectors $\pmb{u}_{\pmb{b}}$ into the mean and fluctuating components:
\begin{equation}
    \pmb{u}_{\pmb{b}}(\pmb{x},t) = {\pmb{u}}_{\mathcal{B}} (\pmb{x}) + \pmb{u}'_{\pmb{b}}(\pmb{x},t).
    \label{eqn:decomposition}  
\end{equation}
Here ${\pmb{u}}_{\mathcal{B}}$ represents the mean velocity component and $\pmb{u}'_{\pmb{b}}$ the fluctuating counterpart. 
In this work the mean velocity ${\pmb{u}}_{\mathcal{B}}$ is defined as \textcolor{black}{the ensemble average over time $t$ and all control laws in the set $\mathcal{B}$}:
\begin{equation}
    {\pmb{u}}_{\mathcal{B}}(\pmb{x}) = \langle \pmb{u}_{\pmb{b}}(\pmb{x},t) \rangle_{\pmb{b},t},
\end{equation}
where $\langle\cdot\rangle_{\pmb{b},t}$ represents the ensemble average over both control commands and time instances.
For the fluctuating velocity $\pmb{u}'_{\pmb{b}}$, we apply the snapshot POD and expand the velocity field based on a series of orthonormal POD eigenfunctions.
As the first $N$th POD modes usually capture most of the energy and large-scaled coherent structures in the flow, an expansion of the fluctuating velocity can be formulated as:
\begin{equation}
    \pmb{u}'_{\pmb{b}}(\pmb{x},t)=\sum_{k=1}^{N} a_k(\pmb{b},t)\pmb{u}_{k}(\pmb{x}).
    \label{eqn:pod}
\end{equation}
Here $\pmb{u}_{k}$ represents the \textcolor{black}{orthonormal} eigenfunction of the $k$th POD mode and $a_k$ the corresponding modal coefficient \textcolor{black}{which contains the energy trace}.
Since the eigenfunctions $\pmb{u}_{k}$ are invariant with time and control command, the velocity expansion in \textcolor{black}{\cref{eqn:pod}} simplifies the full-state estimation problem to the estimation of the POD modal coefficients $\pmb{a}=[a_1,...,a_N]^{\text{T}}$. 
As a result, a significant reduction of the computational cost can be achieved. 
After the estimates of the POD coefficients are obtained, the flow field can be reconstructed according to~\textcolor{black}{\cref{eqn:decomposition,eqn:pod}}.

\subsection{$k$ nearest neighbors regression} \label{ssec:kNN}
\begin{figure}
    \centering
    \includegraphics[width=.55\linewidth]{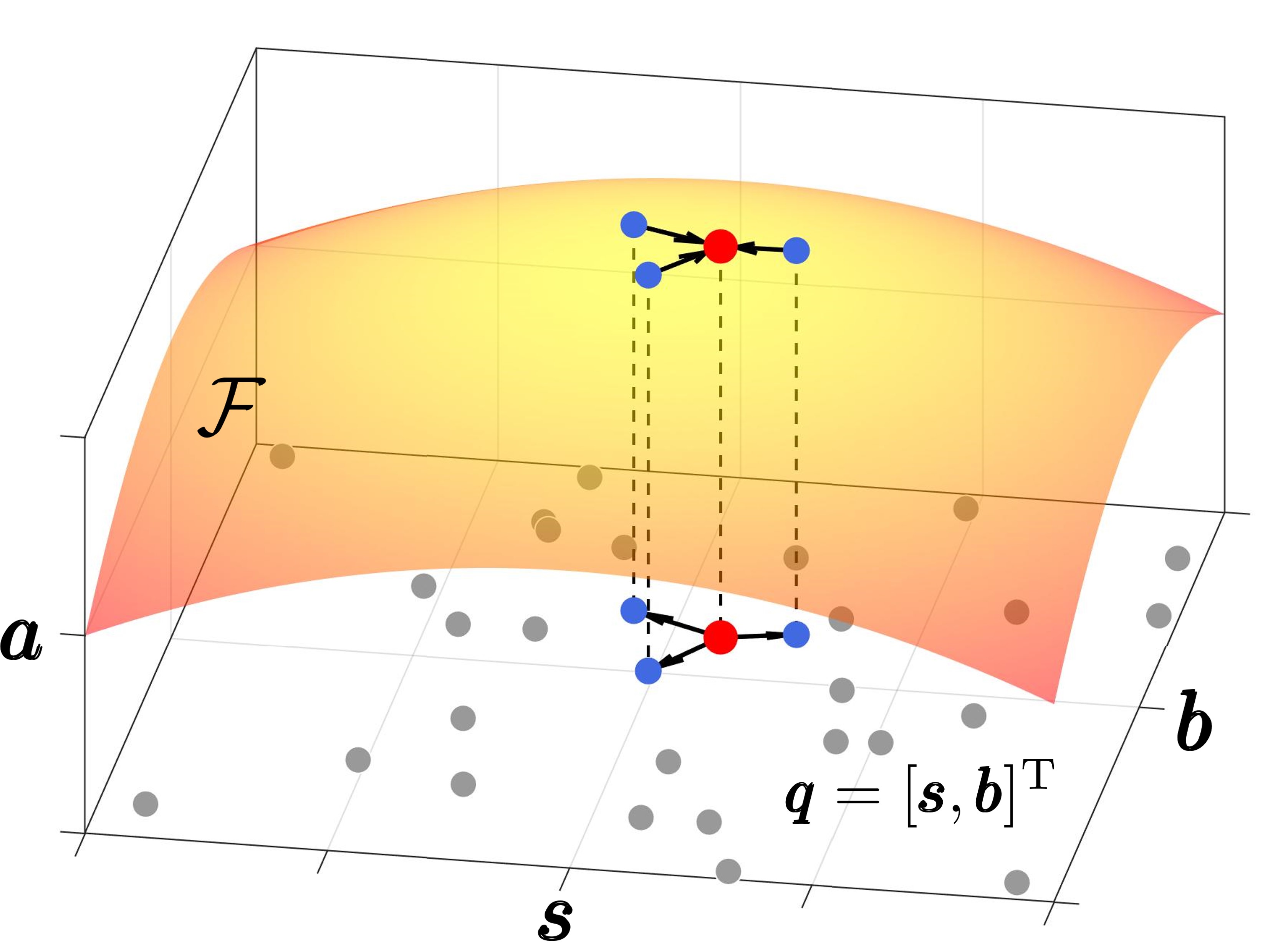}
    \caption{A schematic of $k$NN for the control-oriented flow estimation.}
    \label{fig:kNN}
\end{figure}

\begin{figure}
    \centering
    \includegraphics[width=.5\linewidth]{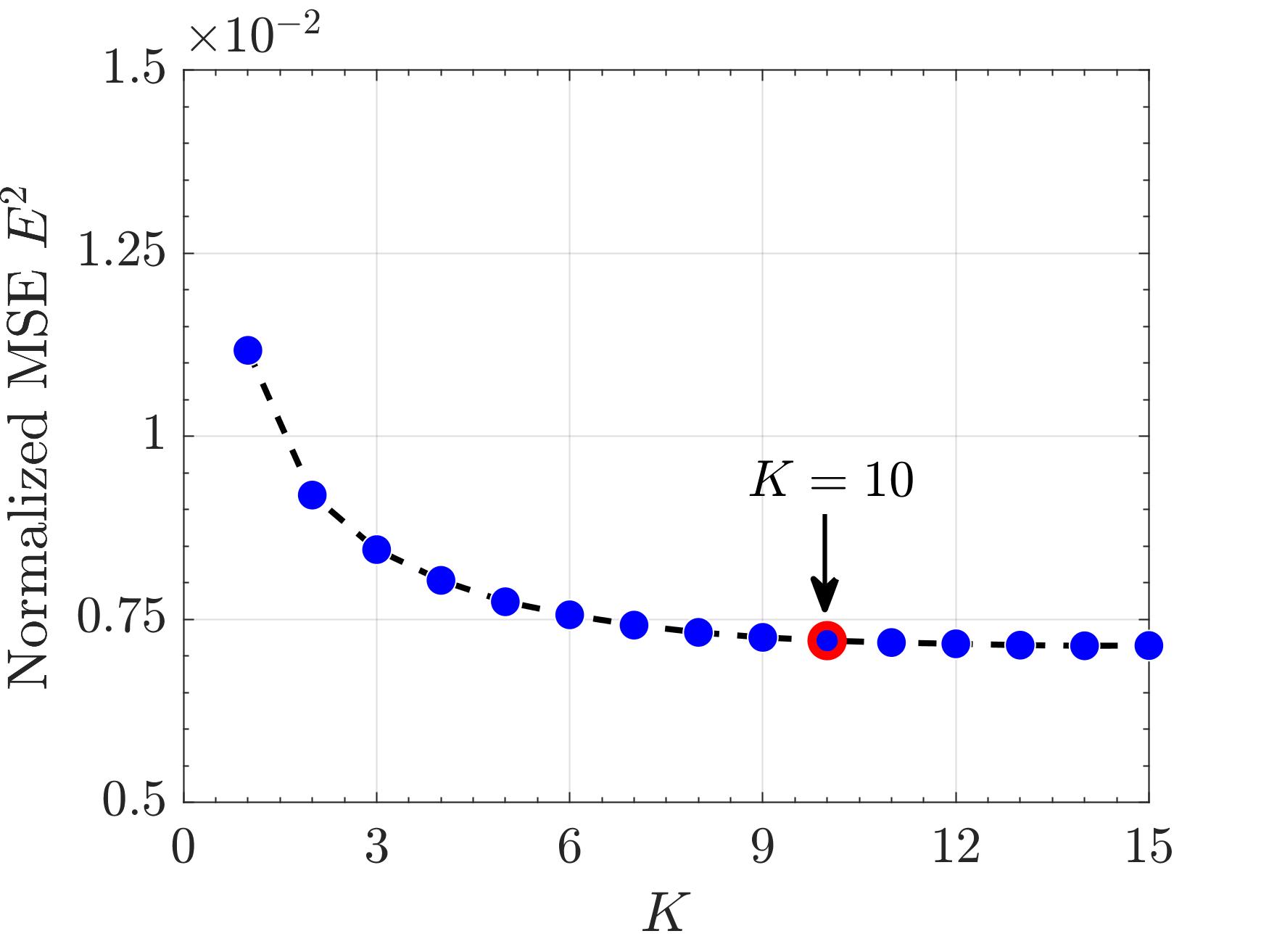}
    \caption{\textcolor{black}{Mean-squared error of $k$NN under different neighborhood numbers $K$ in \cref{eqn:kNN}.}}
    \label{fig:kNN_convergence}
\end{figure}

To establish a mapping function $\mathcal{F}:\pmb{q}\rightarrow\pmb{a}$ for the purpose of flow estimation, we first introduce the $k$ Nearest Neighbors ($k$NN, \citealt{Stone1977}) regression algorithm.
This method is a locally non-linear estimation approach, 
and a schematic diagram of $k$NN is presented in \textcolor{black}{\cref{fig:kNN}}.
$k$NN adopts a non-parametric approach where the mapping function $\mathcal{F}$ is locally approximated by averaging a subset of $K$ observations in which the input data resides in the same neighborhood of the observations.
In this study, a weighted average over the first $K$th nearest neighbors is carried out based on the distance between the estimation input and the neighborhood observations, such that the output can be approximated by the following equation:
\begin{equation}
{\hat{\pmb{a}}}(\pmb{q}) = \frac{\sum_{i=1}^K \pmb{a}(\pmb{q}^{(i)})1/d(\pmb{q},\pmb{q}^{(i)})}{\sum_{i=1}^K 1/d(\pmb{q},\pmb{q}^{(i)})}.
\label{eqn:kNN}
\end{equation}
In this equation $d(\pmb{\alpha_1},\pmb{\alpha_2})$ represents the distance between vectors $\pmb{\alpha}_1$ and $\pmb{\alpha}_2$, and the Euclidean distance $\lVert \pmb{\alpha_2}-\pmb{\alpha_1} \rVert_2$ is adopted throughout this study. 
$\pmb{a}(\pmb{q}^{(i)})$ denotes the output in the observation set which corresponds to the combined input $\pmb{q}^{(i)}$.
The superscript $^{(i)}$ represents the $i$th nearest neighbor in the observation set based on the measurement of distance in the input space, and the parameter $K$ stands for the number of neighborhood observations that are utilized during the averaging process.
In general, the value of $K$ in \textcolor{black}{\cref{eqn:kNN}} needs to be determined empirically. 
A small $K$ means the result can be greatly influenced by the outlying observation and random noise while a large value is usually computationally expensive. 
Therefore, a convergence study is carried out to determine the optimal value of $K$ used in this study.
\textcolor{black}{\Cref{fig:kNN_convergence}} shows the normalized mean-squared estimation error (i.e., $E^2$ in \textcolor{black}{\cref{eqn:error3}}) of $k$NN under different $K$ values for the fluidic pinball plant, from which $K=10$ is selected in an effort to balance the accuracy and the computational efficiency.

\subsection{Deep neural network} \label{ssec:dnn}
As the second approach we employ Deep Neural Network (DNN) to approximate the mapping function $\mathcal{F}$ in a globally non-linear manner.
In general, DNN is composed of several layers of artificial neurons that are connected to one another.
For each layer, a linear transformation of the input data $\pmb{y}_{\text{in}}$ is first performed with tunable weights $\pmb{W}$ and biases $\pmb{c}$, then a non-linear activation function $\sigma$ is exerted, such that:
\begin{equation}
\pmb{y}_{\text{out}} = \sigma(\pmb{W}^T\pmb{y}_{\text{in}} + \pmb{c}).   
\end{equation}
Here $\pmb{y}_{\text{in}}\in \mathbb{R}^{n_{\text{in}}}$, $\pmb{y}_{\text{out}}\in \mathbb{R}^{n_{\text{out}}}$, $W\in\mathbb{R}^{n_{\text{in}} \times n_{\text{out}}}$ and
$\pmb{c}\in \mathbb{R}^{n_{\text{out}}}$.
By stacking multiple layers of artificial neurons sequentially and assigning the output from one layer as the input to the next layer, DNN has the potential to accurately approximate complex non-linear functions by learning weights and biases from data \textcolor{black}{\citep{Hornik1989}}.

\begin{figure}
    \centering
    \includegraphics[width=.5\linewidth]{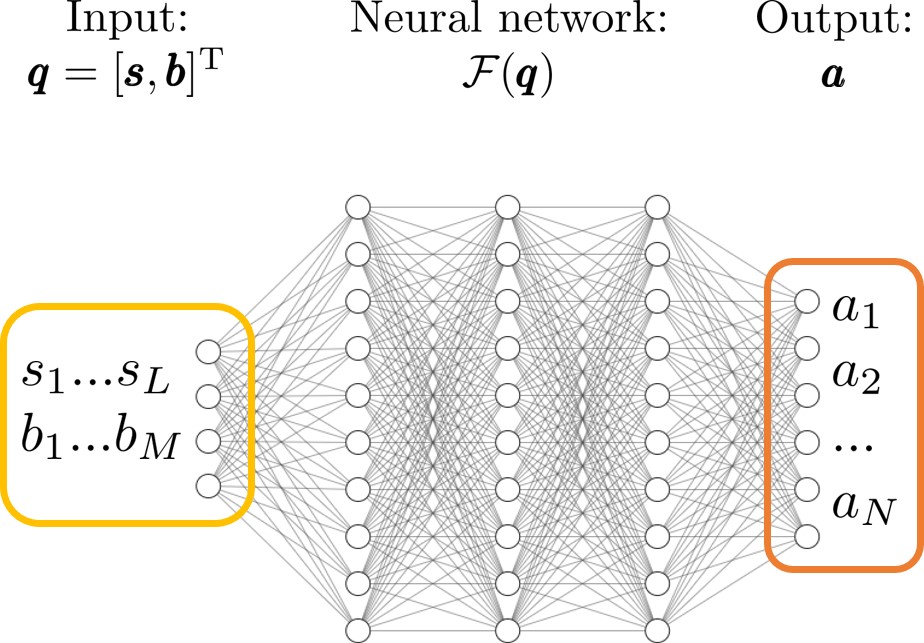}
    \caption{A schematic of deep neural network (DNN) for control-oriented flow estimation.}
    \label{fig:dnn}
\end{figure}

A schematic of the DNN architecture used for control-oriented flow estimation can be found in \textcolor{black}{\cref{fig:dnn}}.
Here the sensor signal components $s_1$ to $s_L$, as well as the control command parameters $b_1$ to $b_M$, are concatenated together and form the input of the neural network.
The output of the neural network contains the truncated POD modal coefficients $a_1$ to $a_N$.
\textcolor{black}{Both the combined input vector $\bm{q}$ and output vector $\bm{a}$ are re-scaled into the range of $[0,1]$ to facilitate convergence.
Note that in $k$NN and LSE,
the input vector is non-dimensionalized to an amplitude of $\mathcal{O}(1)$ by the free-stream velocity $U_{\infty}$, 
which is in accord to the recommendation in \citet{burkov2019}. }

A brief summary of the DNN architecture applied in this study can be found in \textcolor{black}{\cref{tab:dnn}}.
\textcolor{black}{The architecture of the neural network is determined empirically by balancing the estimation accuracy and the computational cost.}
\textcolor{black}{Based on the dimension of the combined input vector and the POD modal truncation (see \cref{ssec:pod})}, the sizes of the input and the output layers are 12 and 78, respectively. 
Three \textcolor{black}{256}-node hidden layers are fully connected in between, for each neuron the rectified linear activation function (ReLU, \citealt{Agarap2018}) is specified as the activation function. 

To minimize the loss function which indicates the performance of the neural network, all tunable weights and biases are optimized with gradient-based back-propagation via an iterative training approach \citep{LeCun2015}.
In this study, the Adam optimization method \citep{Kingba2014} is adopted, and the corresponding loss function is defined as $L_{\varepsilon}=\langle \lVert \pmb{a}-{\hat{\pmb{a}}}\rVert_2^2 \rangle$,
which represents the mean squared error between the desired output $\pmb{a}$ and the actual output $\hat{\pmb{a}}$ over the training samples.
To accelerate the training phase and avoid the convergence to local-minimum, the iterative back-propagation is achieved based on the evaluation of mini-batches which are randomly partitioned from the training data.
In the present work the batch size is set to 64, and a total number of 500 epochs were carried out to train the model.
\textcolor{black}{Other key settings related to the fluidic pinball dataset applied in this work, including the dataset partition and the iterative training process, will be detailed in \cref{ssec:partition+error}}.

\textcolor{black}{Before the iterative training, all weights are randomly initialized following the suggestion in \cite{Glorot2010}. 
In this method, the randomly generated initial weights yield Gaussian distributions and the initial biases are set to zero. 
To examine the influence of randomness from the network initialization to the estimation outcome, we trained the network 20 times, and for each time the initial weights are randomly generated. 
We calculate the averaged outputs and their variances based on the test set, and present these results in \cref{sec:results}.}
 
\begin{table}
\begin{center}
\def~{\hphantom{0}}
\begin{tabular}{ P{4cm} P{4cm} }
DNN Architecture & Parameters  \\ \hhline{==}
Input layer size & 12 ($\pmb{s} \in \mathbb{R}^9$, $\pmb{b} \in \mathbb{R}^3$)\\  
Output layer size & 78 ($  \pmb{a} \in \mathbb{R}^{78}$ )\\
Hidden layer sizes & \textcolor{black}{256, 256, 256}\\
Activation function & ReLU \textcolor{black}{\citep{Agarap2018}}\\
\textcolor{black}{Initialization} &  \textcolor{black}{Xavier \citep{Glorot2010}}\\
\textcolor{black}{Input-output scaling} & \textcolor{black}{Min-max $\rightarrow$ [0,1]}\\
Batch size & 64\\
Optimizer & Adam \citep{Kingba2014}\\
Loss & MSE ($L_{\varepsilon}$)\\
Number of epochs & 500\\
\textcolor{black}{Number of repetitions} &  \textcolor{black}{20}\\
\end{tabular}
\caption{The setup of DNN applied for the fluidic pinball plant.}
\label{tab:dnn}
\end{center}
\end{table}

\subsection{Linear stochastic estimation} \label{ssec:lse}
\begin{figure}
    \centering
    \includegraphics[width=.3\linewidth]{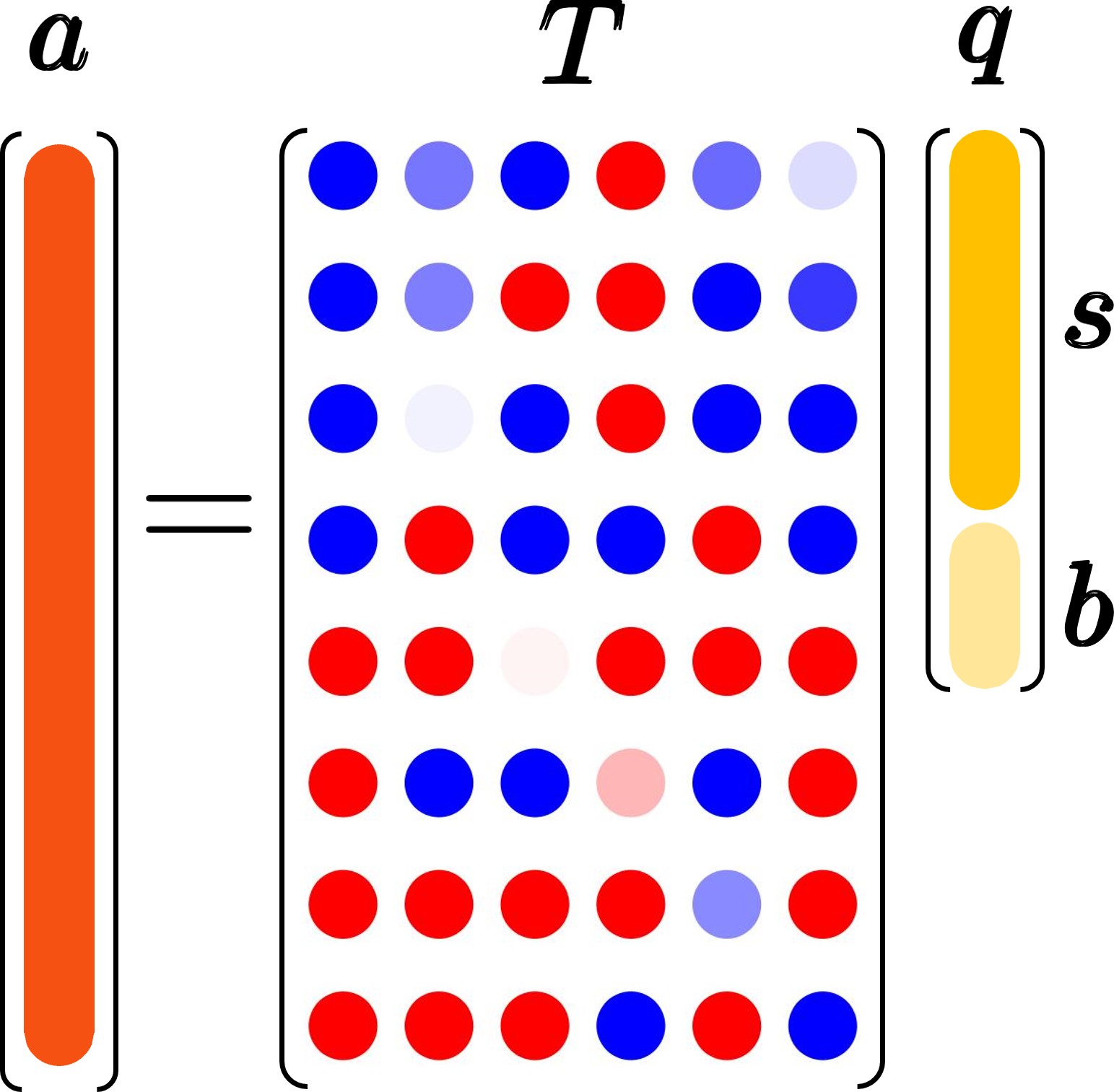}
    \caption{A schematic of the LSE algorithm for the control-oriented flow estimation.\textcolor{black}{The yellow column and light yellow column represent the sensor signal $\bm{s}$ and control command $\bm{b}$, respectively. The orange column represents the estimation of the POD modal coefficients $\bm{a}$. Dots in the transfer matrix corresponds to the weights of $\bm{T}$, which is visualized under a blue(negative)-white(0)-red(positive) color scheme.}}
    \label{fig:lse}
\end{figure}

Besides the non-linear, machine learning-based methods, we apply the Linear Stochastic Estimation (LSE, \citealt{Adrian1994}) as a reference point in this study.
\textcolor{black}{LSE represents the most general linear fit between fluctuating sensor signals and POD modal coefficients.
The solution of LSE is equivalent to the least-squared estimation error for the estimation of POD modal coefficients.}
As presented in \textcolor{black}{\cref{fig:lse}}, this method analytically optimize a liner input-output model, such that:
\begin{equation}
{a}_i(\pmb{b},t)=\sum\limits_{j=1}^N T_{ij} \> q_j(t).
\label{eqn:lse_1}
\end{equation}
Multiplying $q_k$ on both sides of \textcolor{black}{\cref{eqn:lse_1}} and taking the ensemble average over $\pmb{b}$ and $t$, one may get:
\begin{equation}
\sum\limits_{j=1}^N T_{ij} \> \langle q_jq_k\rangle_{\pmb{b},t}  = \langle  a_i q_k \rangle_{\pmb{b},t}.
\label{eqn:lse_2}
\end{equation}
Here $\langle q_jq_k\rangle_{\pmb{b},t}$ denotes the cross-correlation between inputs and $\langle a_i q_k  \rangle_{\pmb{b},t}$ represents the \textcolor{black}{auto-correlation} between inputs and outputs. 
Both matrices can be directly computed from data, and a least-squares solution of $T_{ij}$ can be obtained by solving \textcolor{black}{\cref{eqn:lse_2}}.
%

So far, we have described the control-oriented, sensor-based flow estimation problem, as well as a spectrum of machine learning flow estimation methods which are capable to tackle this problem.
In the following section, we will present the numerical plant, i.e., the fluidic pinball, which will be utilized in this study to evaluate the performance of these estimation methods.

\section{Numerical Plant: the Fluidic Pinball}
\label{sec:plant}

In this section we introduce the fluidic pinball as a benchmark plant to evaluate control-oriented flow estimation methods described in \cref{sec:estimation}.
An overview of the fluidic pinball is covered in \Cref{ssec:overview}, followed by the description of the flow configuration and the numerical solver in \Cref{ssec:config}.
The establishment of the fluidic pinball database that will be used in this study is detailed in \Cref{ssec:s+b}, including the control commands generation, data sampling, as well as sensor placement.
The partition of the database for the purpose of establishing data-driven mappings, as well as the criteria to evaluate the estimation performance, will be concluded in \Cref{ssec:partition+error}.

\subsection{An overview of the fluidic pinball plant} \label{ssec:overview}

\begin{figure}
    \begin{subfigure}{0.33\linewidth}
    \includegraphics[height=3cm]{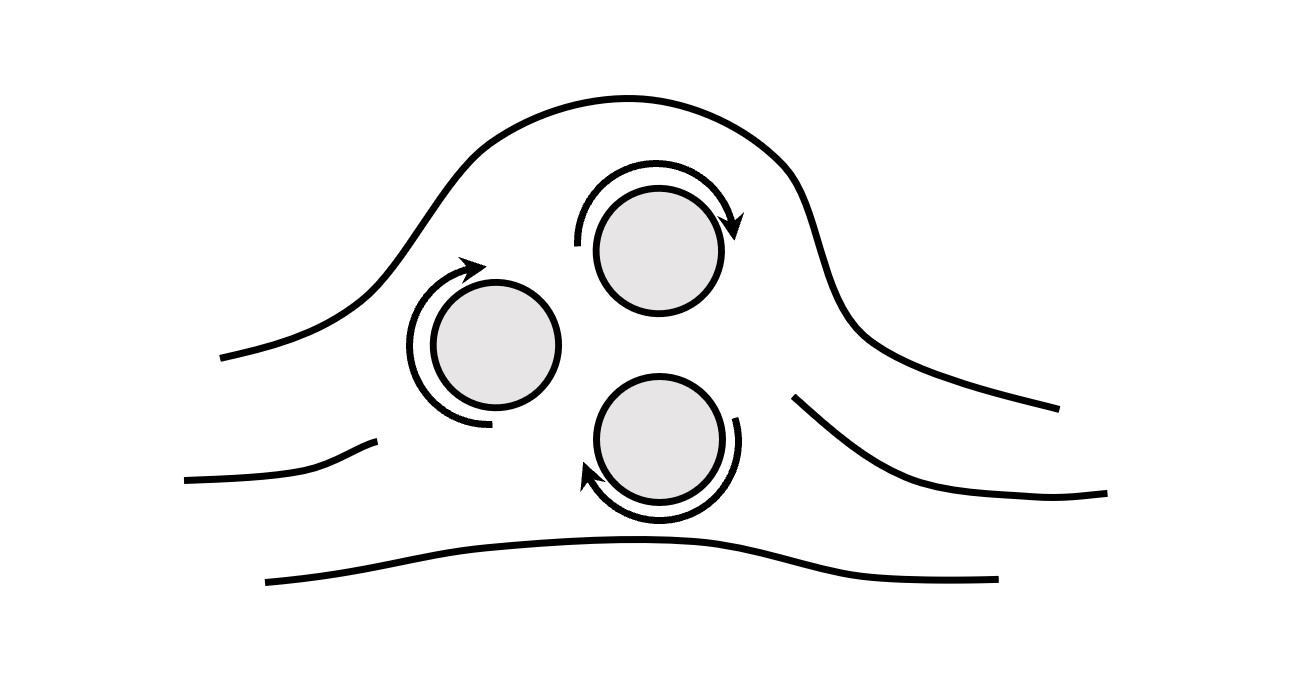}
    \subcaption{stagnation point control}
    \end{subfigure}
    \begin{subfigure}{0.33\linewidth}
    \includegraphics[height=3cm]{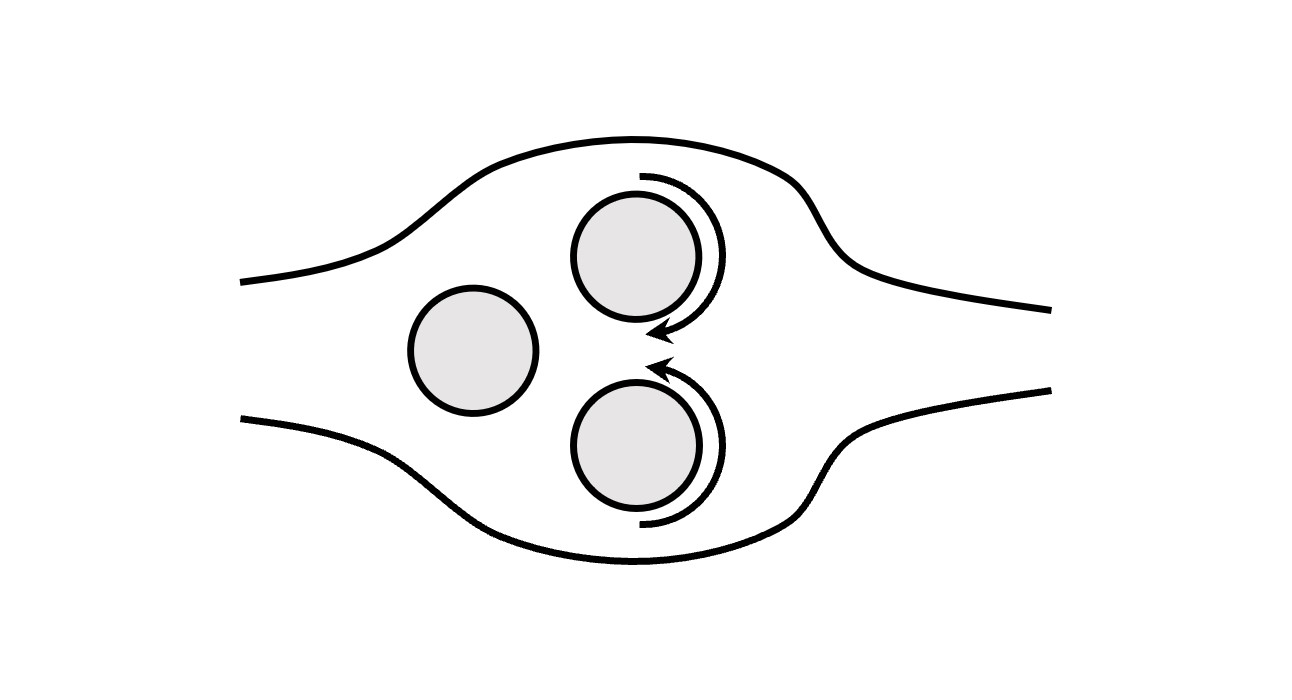}
    \subcaption{boat tailing}
    \end{subfigure}
    \vspace{1cm}
    \begin{subfigure}{0.33\linewidth}
    \includegraphics[height=3cm]{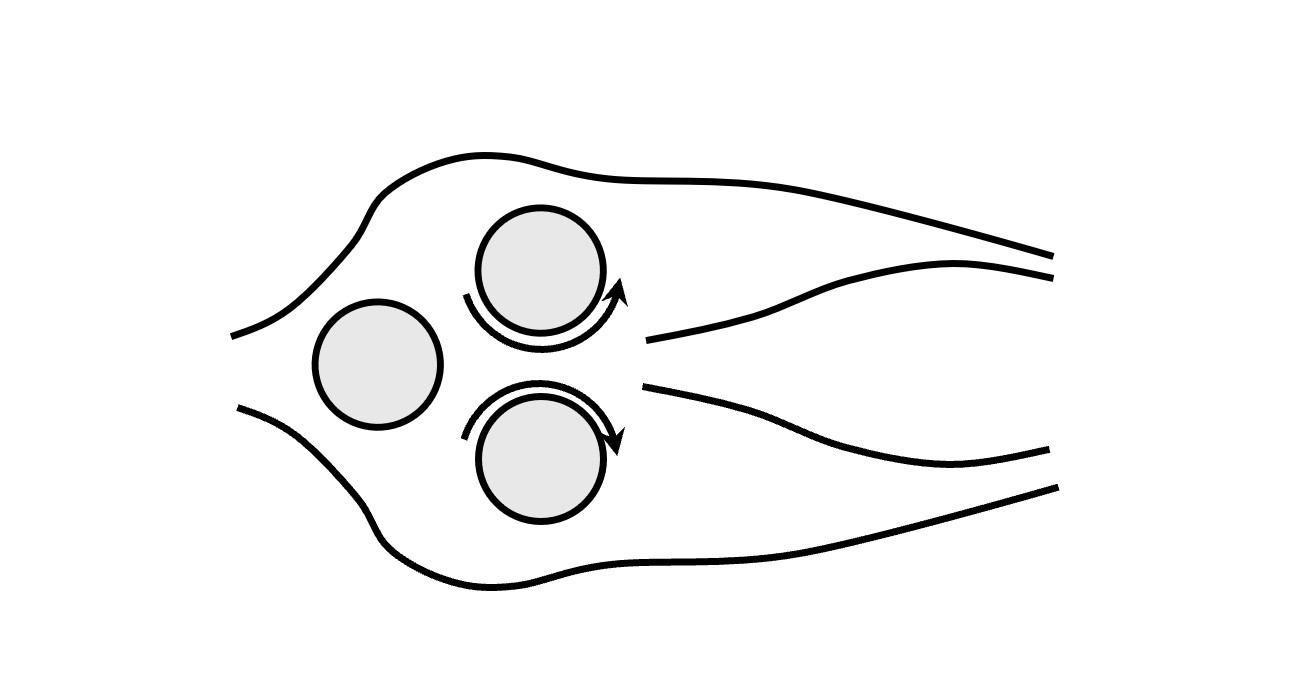}
    \subcaption{base bleed}
    \end{subfigure}
    \caption{A graphical illustration of three flow control mechanisms in the fluidic pinball plant. Figures from \citet{Lin2021}.}
    \label{fig:mechanisms}
\end{figure}

The fluidic pinball plant is a classical control problem which contains rich dynamical behaviors thanks to different flow control inputs \citep{Ishar2019}.
This plant is composed of a two-dimensional, uniform flow passing a set of three cylinders of the same diameter $D$.
Centers of the three cylinders form an equilateral triangle pointing towards the upstream direction, and the side length of the triangle is $1.5D$. 
The unforced flow exhibits 6 different Navier-Stokes solutions,
including two stable statistically asymmetric periodic vortex sheddings 
\textcolor{black}{\citep{Deng2020jfm,Deng2021}}.
The independent rotation of three cylinders allows the control of the wake formed downstream, and is capable to replicate most known control mechanisms, namely, phasor control, stagnation point control, boat tailing, base bleeding as well as high- and low-frequency forcing 
\citep{Ishar2019,Maceda2021}.
In this study, only steady, open-loop control commands are considered. 
Apart from the uncontrolled case with periodic vortex shedding, 
steady rotations of cylinders enables the realization of stagnation point control, boat tailing, and base bleeding.
The schematics of these control mechanisms can be found in \textcolor{black}{\cref{fig:mechanisms}}. 
With the stagnation point control, a unified rotation is exerted on the three cylinders, from which the incoming flow is deflected following the cylinder rotation direction thus attenuate periodic vortex shedding \citep{Seifert2012}.
Both boat tailing and base bleeding mechanisms starts from the counter rotation of the two rearward cylinders.
With boat tailing, the top cylinder rotates clockwise and the bottom rotates counter-closewise.
This will guide the boundary layer flowing towards the center region downstream of the cylinders, which delays the flow separation and leads to significant drag reduction
\citep{Geropp2000,Evrard2016}.
When the top cylinder rotates counter-clockwise and the bottom rotates closewise, the base bleeding mechanism takes place which generates a fluid jet on the centerline.
The jet actuation will limit the interaction between upper and lower shear layers hence impair the development of large scale vortices downstream. 
\textcolor{black}{The ability of base bleeding mechanism to reduce bluff body aerodynamic drag has been reported in \citet{Wood1964,Bearman1967,Howell2003}, among many others.} 
With the deployment of sparse sensors in the flow, the fluidic pinball forms a multi-input multi-output (MIMO) system.
Here, we focus on the application of sensor signal, along with the control command, to estimate the instantaneous flow field under steady control. 

\subsection{Flow configuration and numerical solver} \label{ssec:config}

\begin{figure}
    \centering
    \includegraphics[width=.6\linewidth,trim={1cm 5.5cm 0cm 9cm},clip]{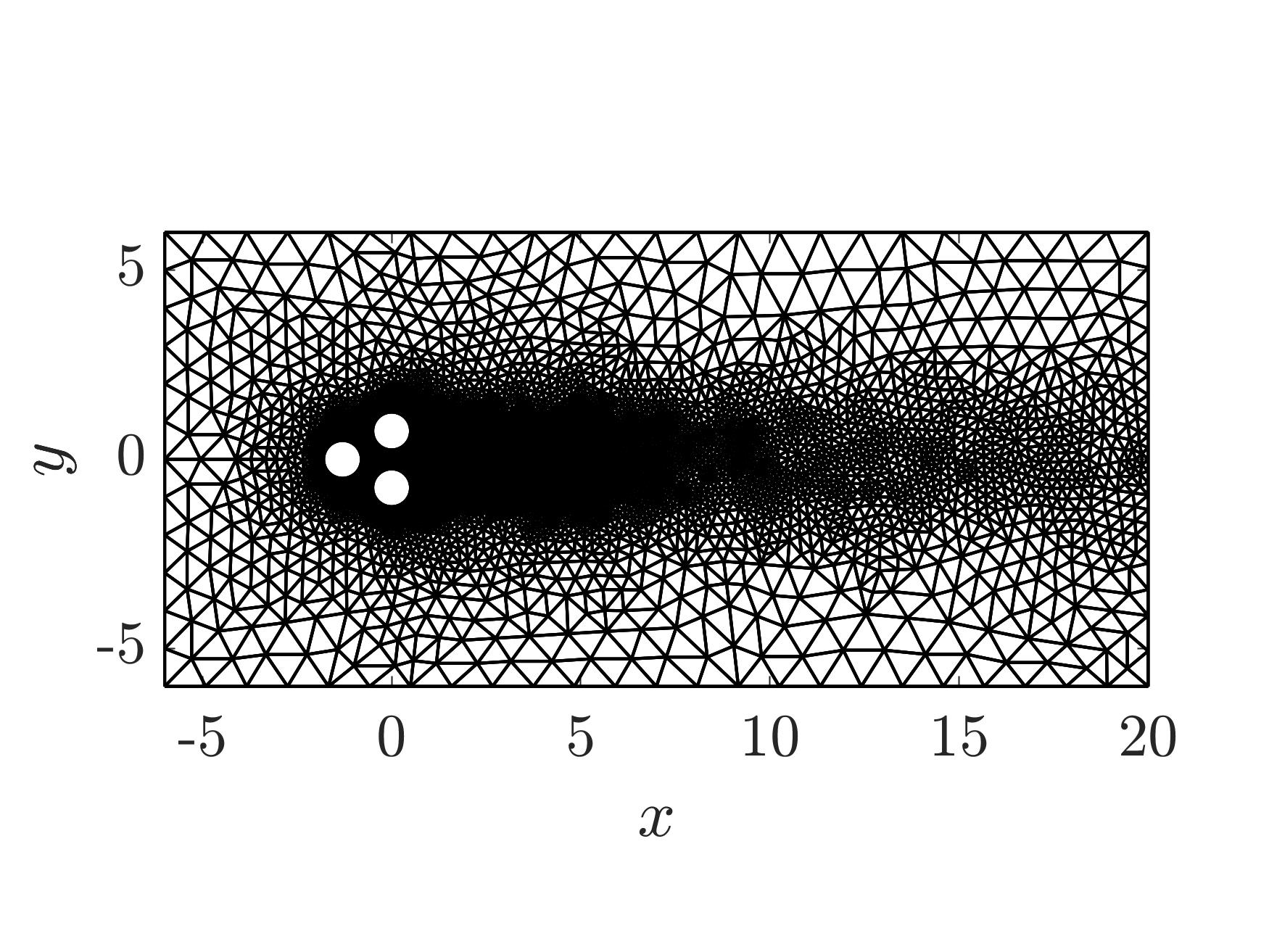}
    \caption{Computational grid for the fluidic pinball simulation.}
    \label{fig:grid}
\end{figure}

The flow is depicted in a two-dimensional Cartesian coordinate system, where the origin is located on the midpoint of the two rear cylinders.
The $x$ axis represents the streamwise direction and the $y$ axis points perpendicular to the freestream direction.
\textcolor{black}{Both axes are scaled by the diameter of the cylinders.
The velocity vector is scaled by the freestream velocity $U_{\infty}$, and is denoted by $\pmb{u}=(u,v)$ where $u$ represents the streamwise velocity and $v$ the transverse velocity.}
In this study we consider an incompressible Newtonian fluid with constant density $\rho$ and kinematic viscosity $\nu$. 
The Reynolds number based on the cylinder diameter $D$ is defined as $Re_{D} = U_{\infty}D/\nu$. 
Throughout this study the Reynolds number is set to 100, which corresponds to the asymmetric periodic vortex shedding according to \citet{Deng2020jfm}.
The computational domain $\Omega$ is bounded by a rectangular box $[-6,20]\times[-6,6]$ externally and the three cylinders internally:
\begin{minipage}[c]{.99\linewidth}
\begin{equation*}
\Omega=\{[x,y]^{\text{T}}\in \mathcal{R}^2:[x,y]^{\text{T}}\in[-6,20]\times[-6,6]\cap(x-x_i)^2+(y-y_i)^2\geqslant 1/4,i=1,2,3\}.
\end{equation*}
\end{minipage}
Here $x_i$ and $y_i$ represents the centers of the three cylinder positioned in the flow, where:
\begin{center}
\begin{minipage}[c]{.99\linewidth}
\begin{center}
\begin{tabular}{m{4cm} m{4cm}}
$x_1=-3/2\text{cos}(30^\circ)$ & $y_1=0$,\\
$x_2=0$ & $y_2=-3/4$,\\
$x_3=0$ & $y_3=3/4$.\\
\end{tabular}
\end{center}
\end{minipage}
\end{center}
The computational domain $\Omega$ is discretized by the unstructured mesh shown in \textcolor{black}{\cref{fig:grid}}. This mesh contains 4225 triangles and 8633 vertices. 
The grid independence of the direct Navier-Stokes solutions has been established by \citet{Deng2020jfm}.

The boundary conditions for the inflow, upper and lower boundaries are specified as the freestream velocity $U_{\infty}$ and a stress-free condition is assumed at the outflow boundary. 
To model the steady rotation of the cylinders, a non-slip condition is enforced on the cylinder boundaries.
The flow adopts the circumferential velocities on the boundary of the front, bottom and
top cylinders, which is specified by $b_1 = U_F$, $b_2 = U_B$, and $b_3 = U_T$. 
\textcolor{black}{Here $b_1$ to $b_3$ are scaled with $U_{\infty}$.}
For each cylinder, the rotation speed on the boundary is within the range of [-2, 2].
A positive value corresponds to counter-clockwise rotation of the cylinders and vice versa. 
The control command can then be written as $\pmb{b}=[b_1,b_2,b_3]^{\text{T}}\in[-2,2]^3$. 
The initial condition for the numerical simulations is the symmetric steady solution calculated from the steady Navier-Stokes equations.
We use an in-house implicit finite-element method solver UNS3 \citep{Noack2017} which employs implicit finite element method to perform numerical integration of Navier-Stokes equations with third-order accuracy in space and time. 
In this study, only the post-transient data are collected for the purpose of constructing the fluidic pinball database that will be used to model the input-output relationship between the sparse sensors and the flow state.
\textcolor{black}{We non-dimensionalize the time with $U_{\infty}$ and $D$.
For every control command, 100 snapshots are recorded with a time interval $\Delta t = 1$.}
\textcolor{black}{The numerical simulation under a single control command takes up around 1 core hour and the data storage for 100 snapshots is about 78 MB.}
For the unforced flow, the vortex shedding period is about 8.5 times the non-dimensionalized time \textcolor{ForestGreen}{unit}.
We choose $\Delta t=1$ in an attempt to balance the computational cost and the overall sampling range in time.
The generation of control commands, as well as the placement of virtual sensors, will be detailed in the following subsection.

\subsection{Control command generation and sensor placement} \label{ssec:s+b}

\begin{figure}
    \centering
    \includegraphics[width=.99\linewidth]{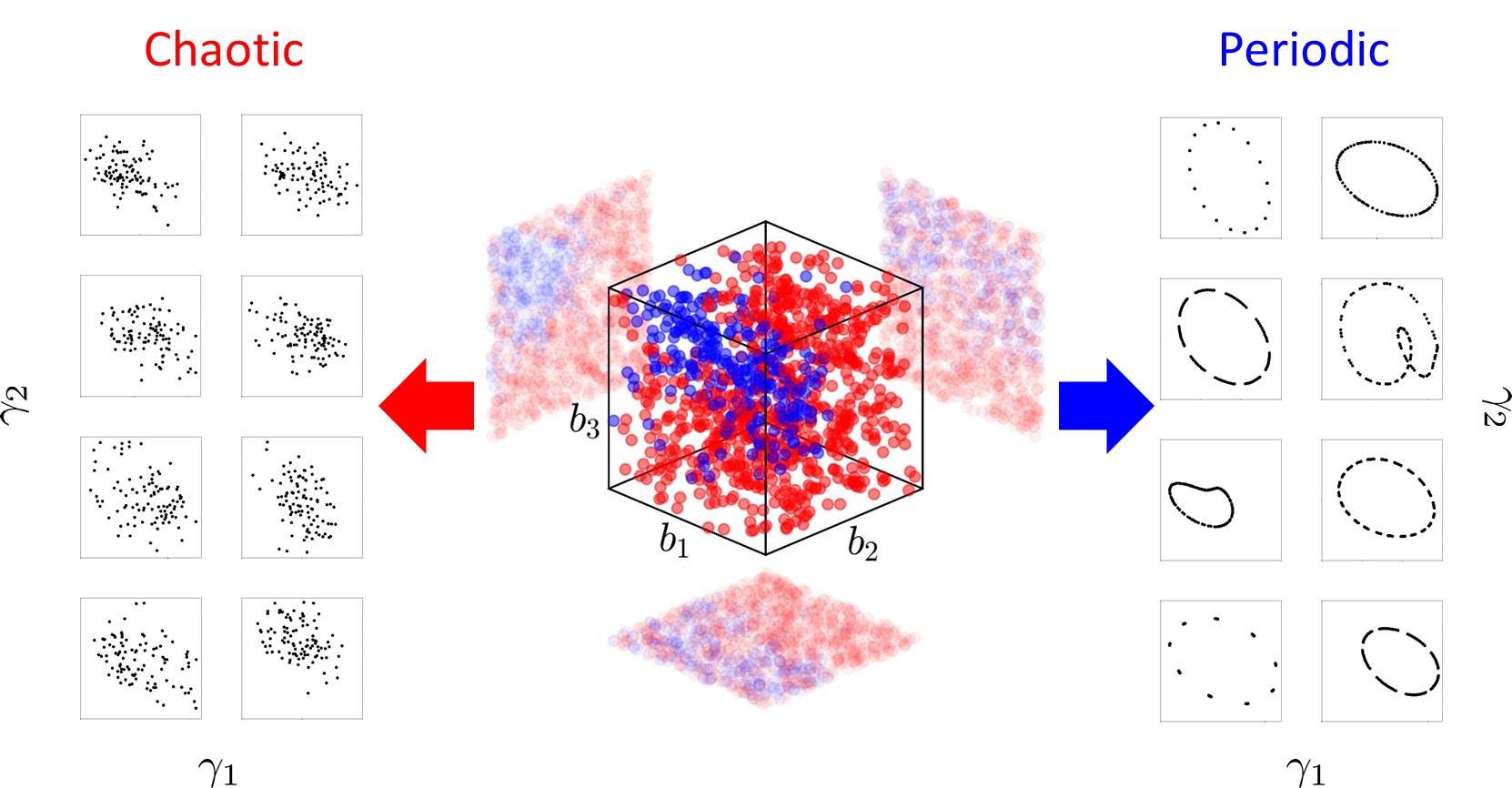}
    \caption{Randomly sampled control commands $\mathcal{B}$ in the control space $[-2, 2]^3$ and their projections on the three orthogonal coordinate planes. Periodic and chaotic cases are rendered by red and blue dots, respectively. \textcolor{black}{Representative patterns of $\gamma_1$ versus $\gamma_2$ in each category are also presented in this figure. See text for the definition of $\gamma_1,\gamma_2$.}}
    \label{fig:LHS}
\end{figure}

To establish control-oriented flow estimators, a set of 1,000 different control commands $\mathcal{B}=\{\pmb{b}_j\}_{j=1}^{1000}$ are randomly generated inside the control parameter space $[-2,2]^3$. 
As mentioned in \cref{ssec:config}, the control parameter $\pmb{b}$ represents the circumferential rotation speed of three cylinders in the flow.
Here we employ the Latin Hypercube Sampling (LHS) \citep{McKay1979} to obtain a set of randomly distributed control commands $\mathcal{B}$.  
As presented in \textcolor{black}{\cref{fig:LHS}}, this sampling method ensures a nearly uniform distribution of samples inside the multidimensional control parameter space. 
For each control command, we categorize the flow state into periodic and chaotic types after the flow snapshots are obtained from the numerical solver.
\textcolor{black}{The classification criteria is based on the relationship between two variables $\gamma_1$ and $\gamma_2$ obtained from the sampled snapshots.
Here $\gamma_1 = C_L(t)$ and $\gamma_2 = (C_L(t+\Delta t)-C_L(t))/\Delta t$, and $C_L$ represents the lift coefficient.}
The lift coefficient is computed by integrating pressure over the surfaces of three cylinders.
\textcolor{black}{
From these plots different characteristics between periodic and chaotic flows can be clearly distinguished.
Representative periodic and chaotic phase pattern approximations are displayed in \textcolor{black}{\cref{fig:LHS}}.
For periodic flows, the discrete points in the plots will form circular or periodic trajectories.
In contrast, the patterns in chaotic flows are disperse and disordered.}
The classification of the flow state will enable us to evaluate the performance of the flow estimators introduced in \Cref{sec:estimation} under different types of the flow states.

\begin{figure}
    \centering
    \includegraphics[width=.6\linewidth,trim={1cm 7cm 0cm 10cm},clip]{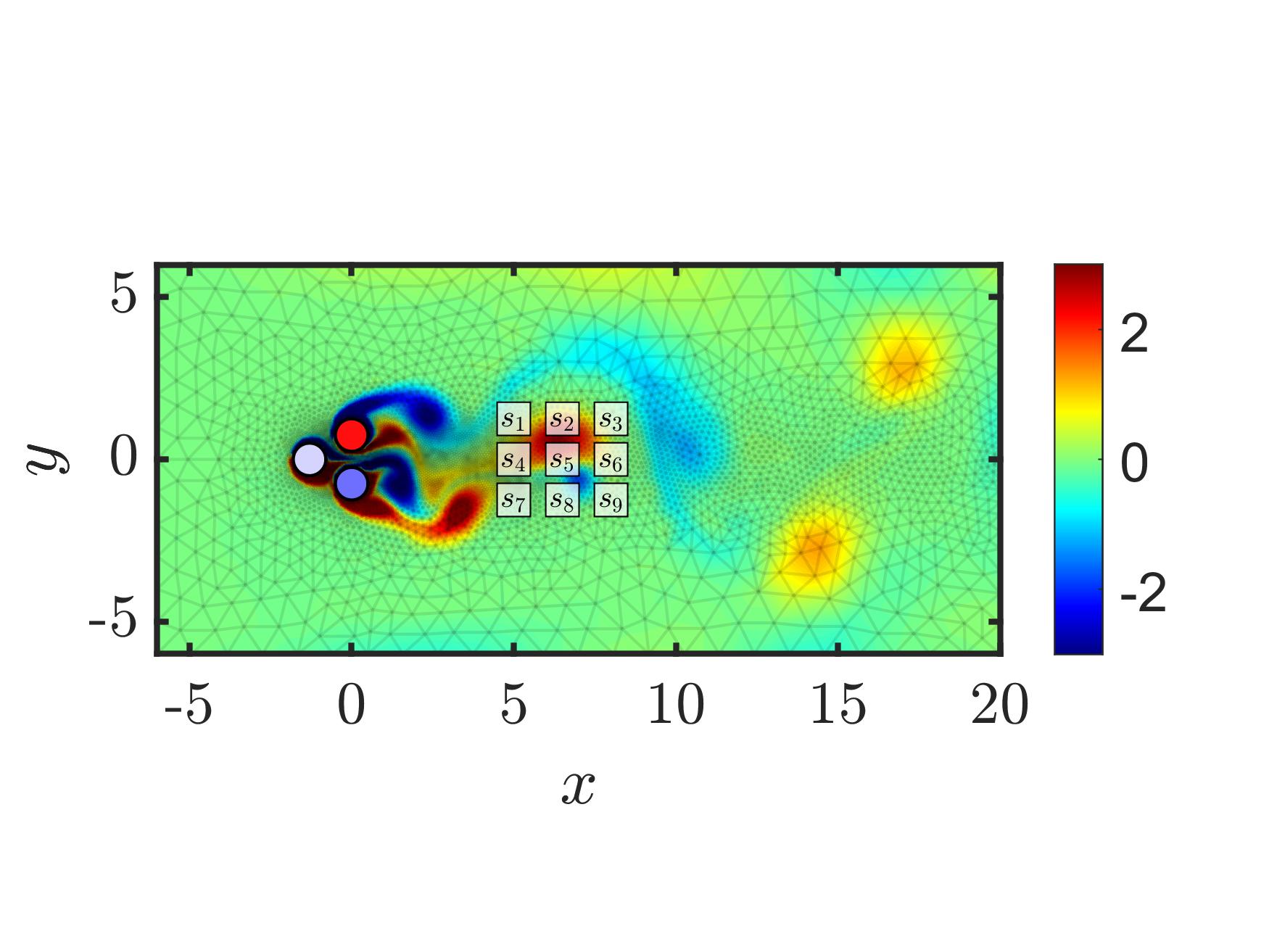}
    \caption{Visualization of sensor placement in the flow domain. The background is rendered by an instantaneous vorticity field. The corresponding control command $\pmb{b}$ is visualized by the colors on three cylinders, under a blue(negative)-white(0)-red(positive) color scheme.}
    \label{fig:sensor}
\end{figure}

To monitor the flow state, 9 virtual probes are placed downstream to measure the discrete velocity signal, either in streamwise or spanwise component.
The velocity probes are deployed as a $3\times3$ rectangular grid symmetric about the centerline, where the upper and lower sensors measure the streamwise velocity and the middle ones on the centerline measure the spanwise velocity.
The streamwise locations of the sensors are $x = 5, 6.5, 8$, and the spanwise locations are $y = -1.5,0,1.5$.
\textcolor{black}{This sensor placement has been validated by \citet{Maceda2021}, in a sense that such arrangement is capable to detect large-scaled vortex shedding while encompassing phase information between sensors in its feedback flow control experiment.
In other words, this arrangement can help infer the flow state in the dynamically controlled flow with a wide range of control commands. 
Hence in this work, we continue to use this promising sensor placement to obtain flow state estimation.}
A summary of the sensor placement can be found in \textcolor{black}{\cref{tab:sensors}}, and a comprehensive visualization regarding to the flow state, control parameters, and the downstream sensing is presented in \textcolor{black}{\cref{fig:sensor}}.

\begin{table}
	\begin{center}
	\def~{\hphantom{0}}
	\begin{tabular}{P{2cm} P{2cm} P{2cm} P{2cm}}
		Sensors & $x$ & $y$ & Velocity\\ \hhline{====}
		$s_1, s_2, s_3$ & 5, 6.5, 8 & 1.25 & streamwise\\
		$s_4, s_5, s_6$ & 5, 6.5, 8 & 0 & spanwise\\
		$s_7, s_8, s_9$ & 5, 6.5, 8 & -1.25 & streamwise\\
	\end{tabular} 
	\caption{Virtual Sensor locations and the corresponding velocity components.}
	\label{tab:sensors}
	\end{center}
\end{table}

\subsection{Data partition and estimation performance evaluation} \label{ssec:partition+error}
According to the steady control command and virtual probe placement described above, the fluidic pinball database is established aiming for sensor-based, control-oriented flow estimation.
To preserve the generality of the data-driven estimation and prevent over-fitting, the collected input-output data pairs are partitioned into a training set, a validation set, and a test set based on a random split of the control commands $\mathcal{B}$.
The partition of the control commands follows a 80\%---10\%---10\% ratio and each set is composed of all time instances generated from the corresponding control commands.
The training and validation sets compose the observation data $\mathcal{M}$, and will be utilized for the neighborhood searching in $k$NN and the calculation of $T_{ij}$ in LSE.
\textcolor{black}{
To prevent overfitting the training data in DNN, the iterative training and gradient-based back-propagation are only performed on the mini-batched training set. 
After every epoch, we evaluate the model performance on the validation set. 
The network parameters corresponding to the minimum validation loss over the 500 epochs are outputted after training. 
The number of epochs are empirically determined by observing the fact that the training loss will hardly be minimized after 500 epochs.
}
For all estimation methods, the test set is not accessible during the training stage. 
Therefore with this separate set, we can evaluate the generalized performance of the proposed methods. 
All key parameters regarding the fluidic pinball database has been concluded and presented in \textcolor{black}{\cref{tab:database}}.

To evaluate the performance of each estimation method from various perspectives, different types of errors are defined and calculated based on the data from the test set.
In this work different types of errors will be discussed and the definitions are presented in the following.
To evaluate the estimation performance under each control command, the normalized estimation error of the POD modal coefficients is defined as:
\begin{equation}
\varepsilon_{\pmb{a}}(\pmb{b}) 
= \sqrt{\frac{\langle \lVert \pmb{a}-\hat{\pmb{a}} \rVert_2^2 \rangle_t}{\langle \lVert \pmb{a} \rVert_2^2 \rangle_t}}.
\end{equation}
Similarly, the normalized velocity estimation error $\varepsilon_{\pmb{u}}(\pmb{x})$ evaluates the spatial distribution of the velocity estimation error which accounts for all time instances and control commands and is normalized by the overall fluctuating energy, such that:
\begin{equation}
\varepsilon_{\pmb{u}}(\pmb{x})
= \sqrt{\frac{\langle \lVert \pmb{u}'_b(\pmb{x})-\hat{\pmb{u}}'_b (\pmb{x})  \rVert_{2}^2 \rangle_{\pmb{b},t}}{\langle \lVert \pmb{u}'_b(\pmb{x}) \rVert_{2}^2 \rangle_{\pmb{b},t}}}.
\end{equation}
Furthermore, we define the normalized overall estimation error $E$ which is a scalar representing the overall accuracy of each estimation method.
The value of $E$ is determined as the cumulative estimation error over the spatial domain, all control commands, and all temporal instances in the test set normalized by energy, such that:
\begin{equation}
E = \sqrt{\frac{\langle \int_{\Omega} \lVert  \pmb{u}(\pmb{x})-\hat{\pmb{u}}(\pmb{x}) \rVert_{2}^2  d\pmb{x}\rangle_{\pmb{b},t}}{\langle \int_{\Omega} \lVert \pmb{u}(\pmb{x}) \rVert_2^2  d\pmb{x}\rangle_{\pmb{b},t}}}.
\label{eqn:error3}
\end{equation}

\begin{table}
	\begin{center}
	\def~{\hphantom{0}}
	\begin{tabular}{ P{4cm} P{4cm} }
		Parameters & Value\\
		\hhline{==}
		Control parameters & $\pmb{b} \in [-2,2]^3$\\
		Number of control commands & 1,000\\
        Number of snapshots per case & 100\\
		Sampling time interval & $1$\\
		Training set & 80\%\\
		Validation set & 10\%\\
		Test set & 10\%\\
	\end{tabular}
	\caption{A conclusion of the fluidic pinball database.}
	\label{tab:database}
	\end{center}
\end{table}


\section{Results and Discussion} \label{sec:results}

In this section, we apply the three estimation methods which are introduced in \cref{sec:estimation} to the fluidic pinball plant.
First, we present the results from the snapshot POD in \cref{ssec:pod}, which includes the energy distribution and the physical interpretation of the leading POD modes.
Then, the performance of the estimation methods are compared under three representative flow control commands which are selected from the test cases, and the results are presented in \cref{ssec:case1}, \cref{ssec:case2}, and \cref{ssec:case3}, respectively. 
Here each control command represents a specific flow control mechanism introduced in \cref{ssec:overview},
and the comparison includes the estimation of reduced-order POD modal coefficients as well as the full-state flow field.
Furthermore, the overall performance of each estimation method will be evaluated and discussed in \cref{ssec:evaluation}, 
and this section is closed by a concluding diagram which contains a thorough assessment of each estimation methods from various perspectives.
\textcolor{black}{Throughout this section, we adopt the non-dimensionalized form of the data and spatial dimensions, as introduced in \Cref{ssec:config}}.

\subsection{The snapshot POD}\label{ssec:pod}
\begin{figure}
\centering
\begin{subfigure}{0.45\linewidth}
\includegraphics[width=.99\linewidth,trim={1cm 8cm 0cm 13cm},clip]{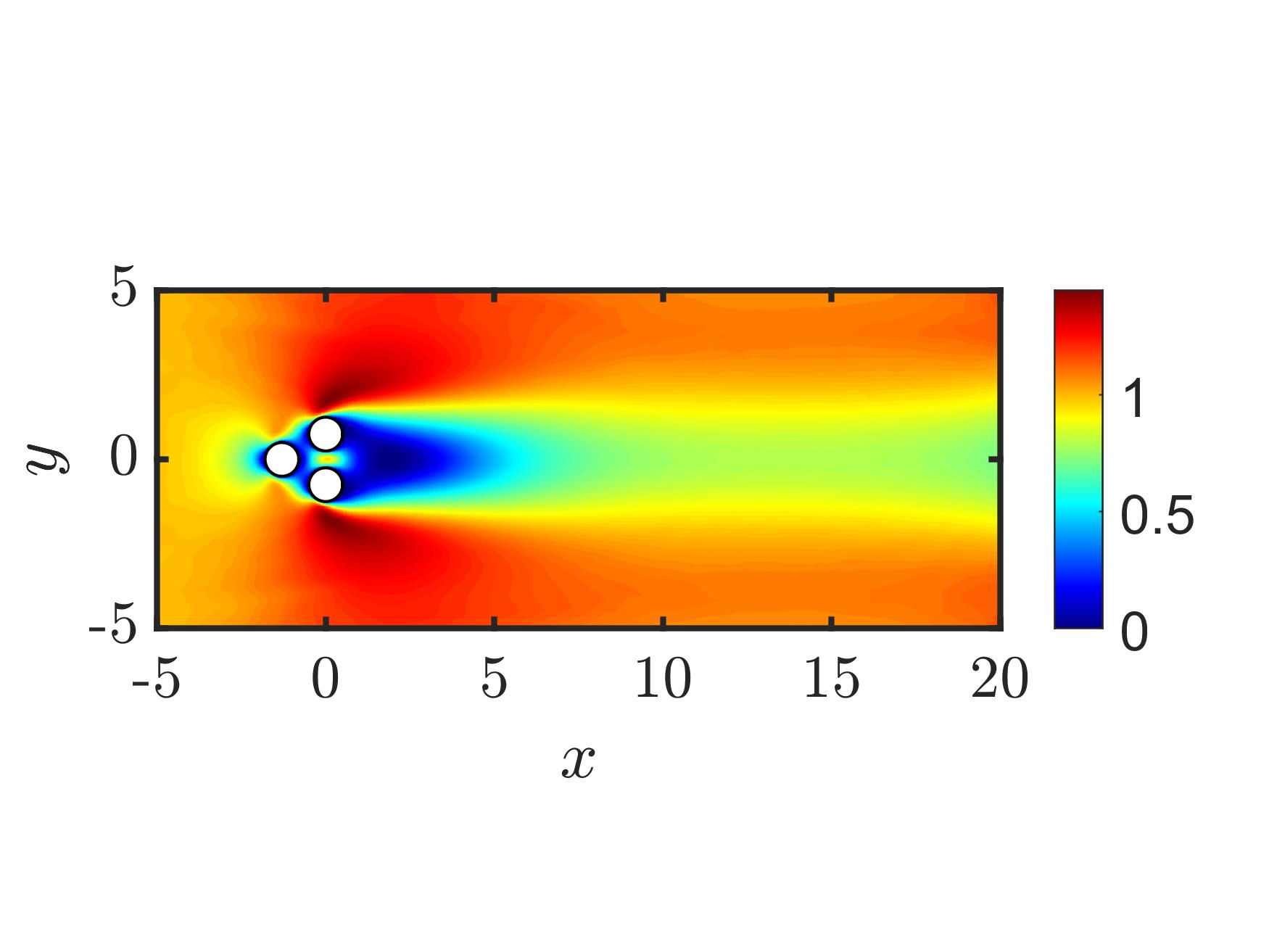}
\caption{}
\label{fig:mean}
\end{subfigure}
\begin{subfigure}{0.45\linewidth}
\includegraphics[width=.99\linewidth,trim={1cm 8cm 0cm 13cm},clip]{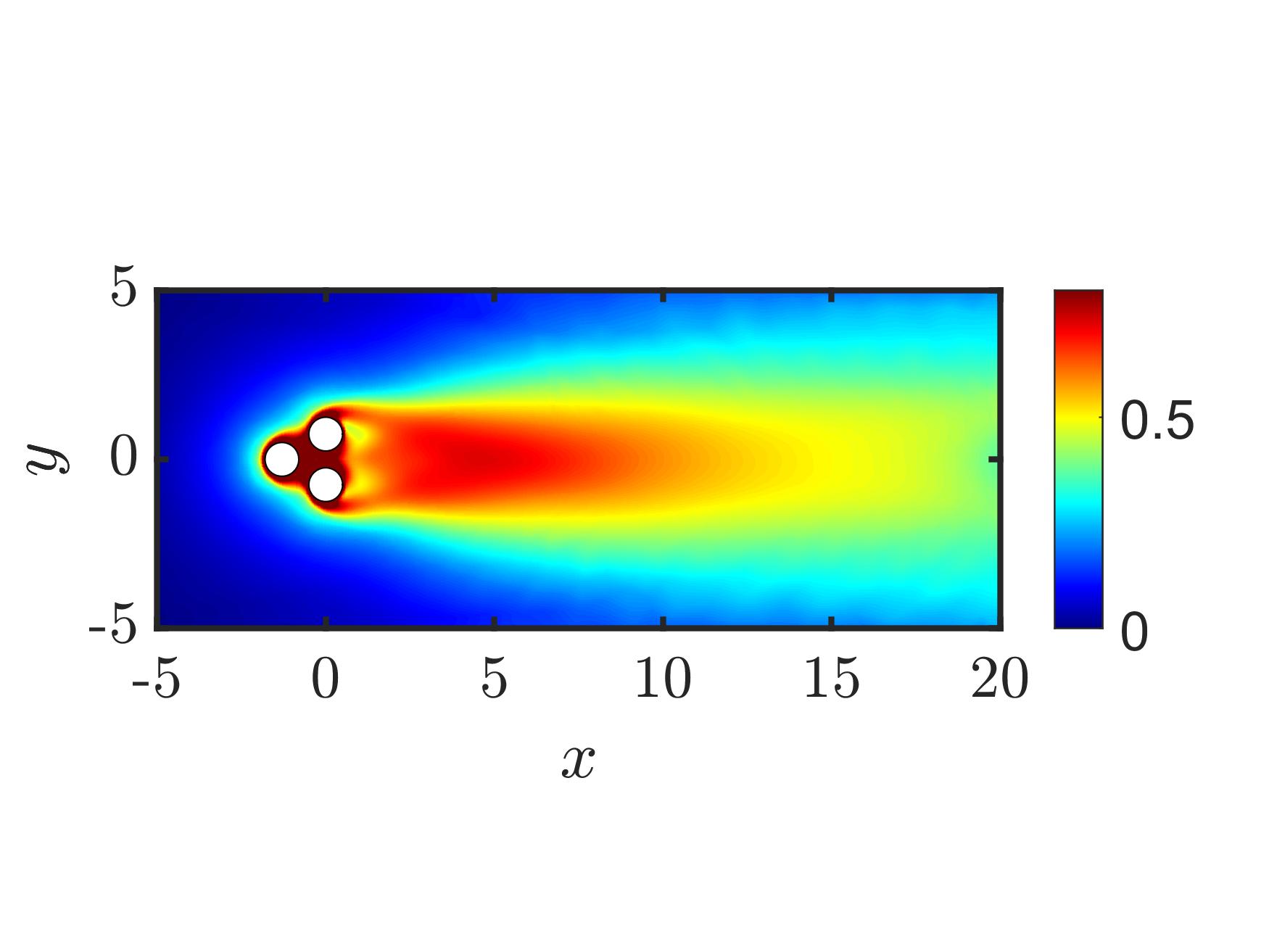}
\caption{}
\label{fig:rms}
\end{subfigure}\\
\begin{subfigure}{0.45\linewidth}
\includegraphics[width=.99\linewidth,trim={1cm 8cm 0cm 13cm},clip]{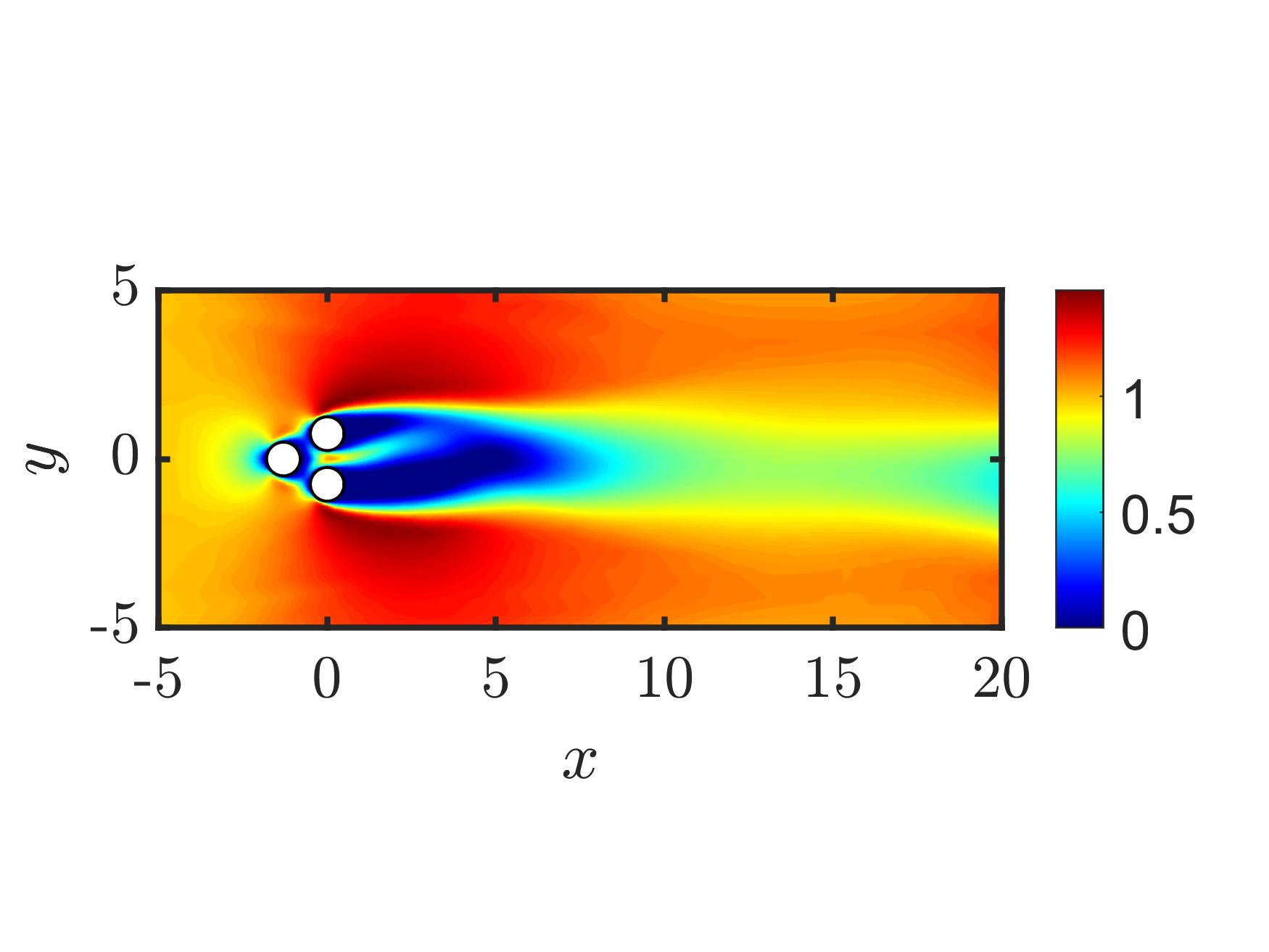}
\caption{}
\label{fig:unforced_mean}
\end{subfigure}
\begin{subfigure}{0.45\linewidth}
\includegraphics[width=.99\linewidth,trim={1cm 8cm 0cm 13cm},clip]{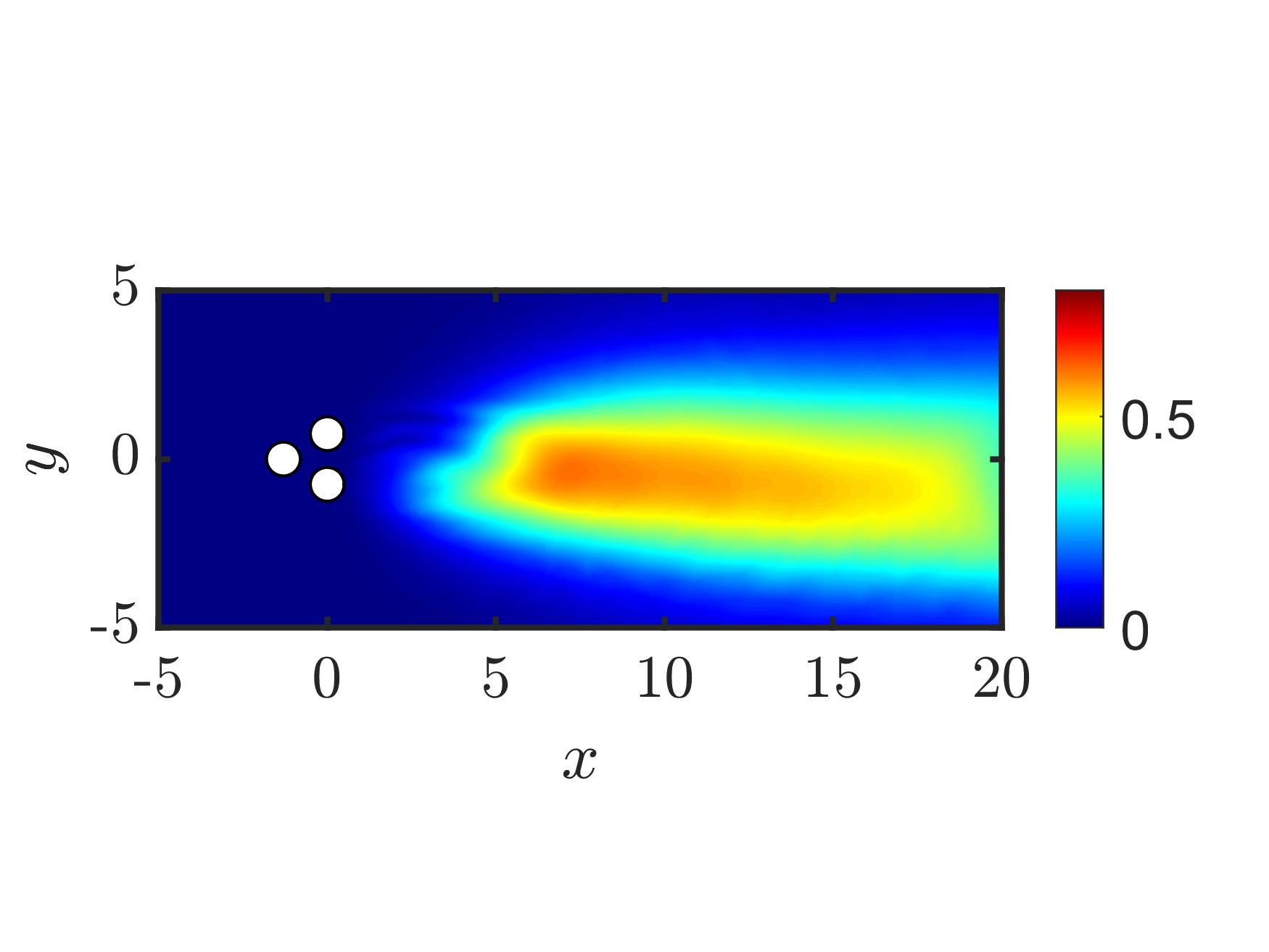}
\caption{}
\label{fig:unforced_rms}
\end{subfigure}
\caption{(a) Ensemble-averaged streamwise velocity field over 1000 different control commands, and (b) intensity of the fluctuating velocity; (c) uncontrolled mean streamwise velocity field with pitchfork bifurcation \citep{Deng2020jfm}, and (d) the corresponding fluctuating velocity intensity.}
\label{fig:statistics}
\end{figure}

To perform sensor-based flow estimation, a necessary step is to find a reduced-order representation of the high dimensional flow snapshots. 
In this work we decompose the velocity into mean and fluctuating components according to \textcolor{black}{\cref{eqn:decomposition}}, then apply the snapshot POD to describe the velocity field using the linear combination of the most energetic POD modes.
\Cref{fig:mean,fig:rms} presents the ensemble-averaged mean streamwise velocity field and the intensity of the fluctuating velocity component $\sqrt{\langle u^2+v^2 \rangle_{\pmb{b},t}}$, respectively.
For comparison the corresponding profiles of the uncontrolled case $\pmb{b} = [0,0,0]^{\text{T}}$ are also displayed in \textcolor{black}{\cref{fig:unforced_mean,fig:unforced_rms}}.  
Unlike the unforced mean profiles which are biased to the negative-$y$ direction due to the pitchfork bifurcation described in \citet{Deng2020jfm}, the profiles in \textcolor{black}{\cref{fig:statistics}} are seen to be symmetric about the centerline $y=0$ thanks to the effect of control. 
In addition, one can clearly capture a reduced length of the separation region behind the cylinders, as well as the expansion of the most energetic region (in the sense of fluctuating velocity), when the flow is under random control.

\begin{figure}
    \centering
    \includegraphics[width=.5\linewidth]{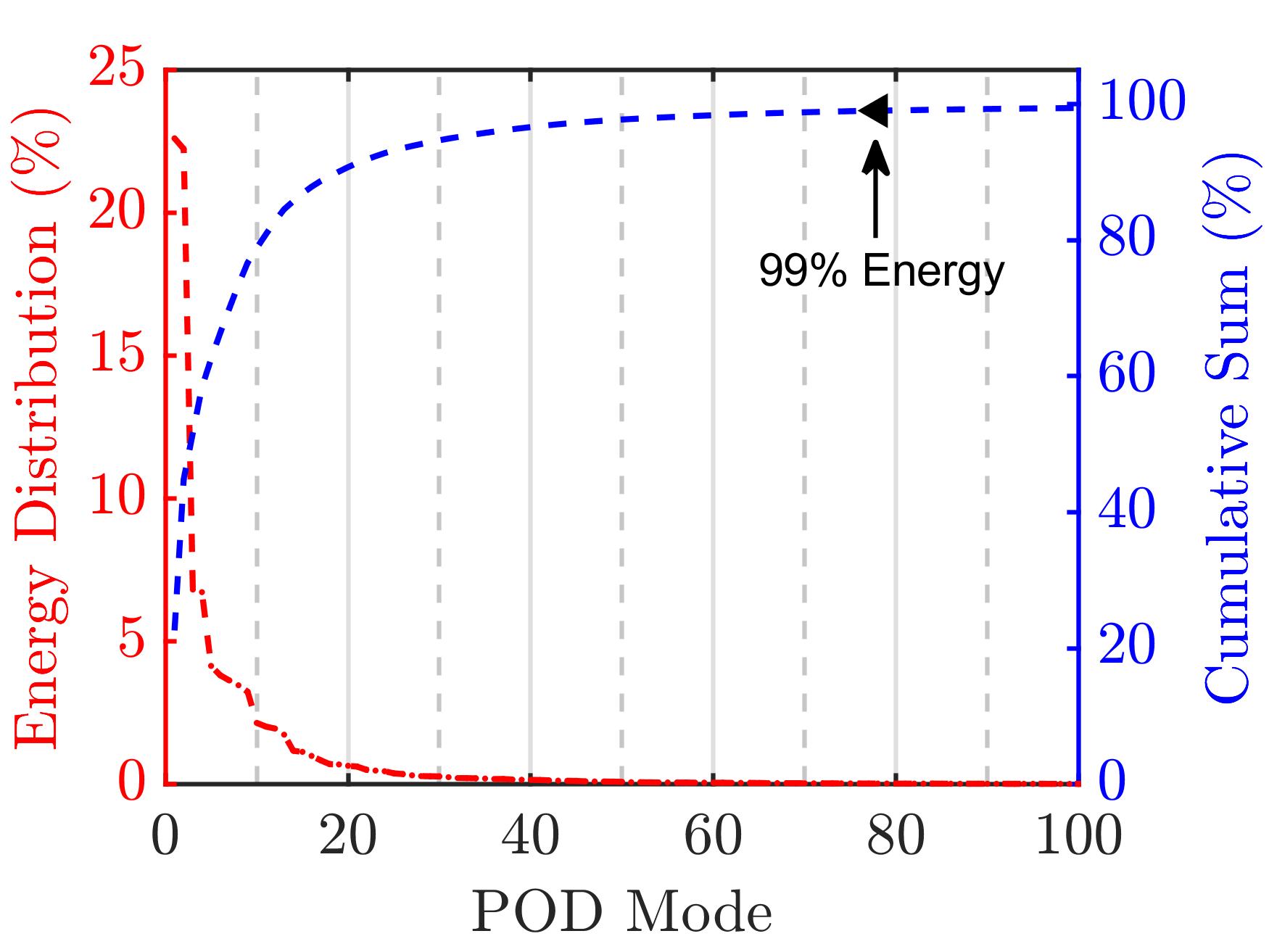}
    \caption{Energy distribution of the leading POD modes and the cumulative sum.}
    \label{fig:pod_energy}
\end{figure}

Following the velocity decomposition, we employ the snapshot POD to set up a reduced-order representation of the fluctuating velocity $\pmb{u}'_{\pmb{b}}$.
\Cref{fig:pod_energy} presents the energy distribution as well as the cumulative sum of the first 100 POD modes. 
The leading POD modes are seen to possesses most of the fluctuating energy, in which the first two and the first ten modes takes up 44.9\% and 78.9\% of the overall energy, respectively.
A fast decay of the energy in lower order POD modes is also observed.
In an attempt to reduce the computational cost while preserving the most energetic flow structures, the modal truncation criteria is defined as 99\% of the total energy, which corresponds to the first 78 POD modes in this database.
These modes will serve as the expected outputs to investigate the performance of the estimation methods.

\begin{figure}
    \centering
    \includegraphics[width=.8\linewidth]{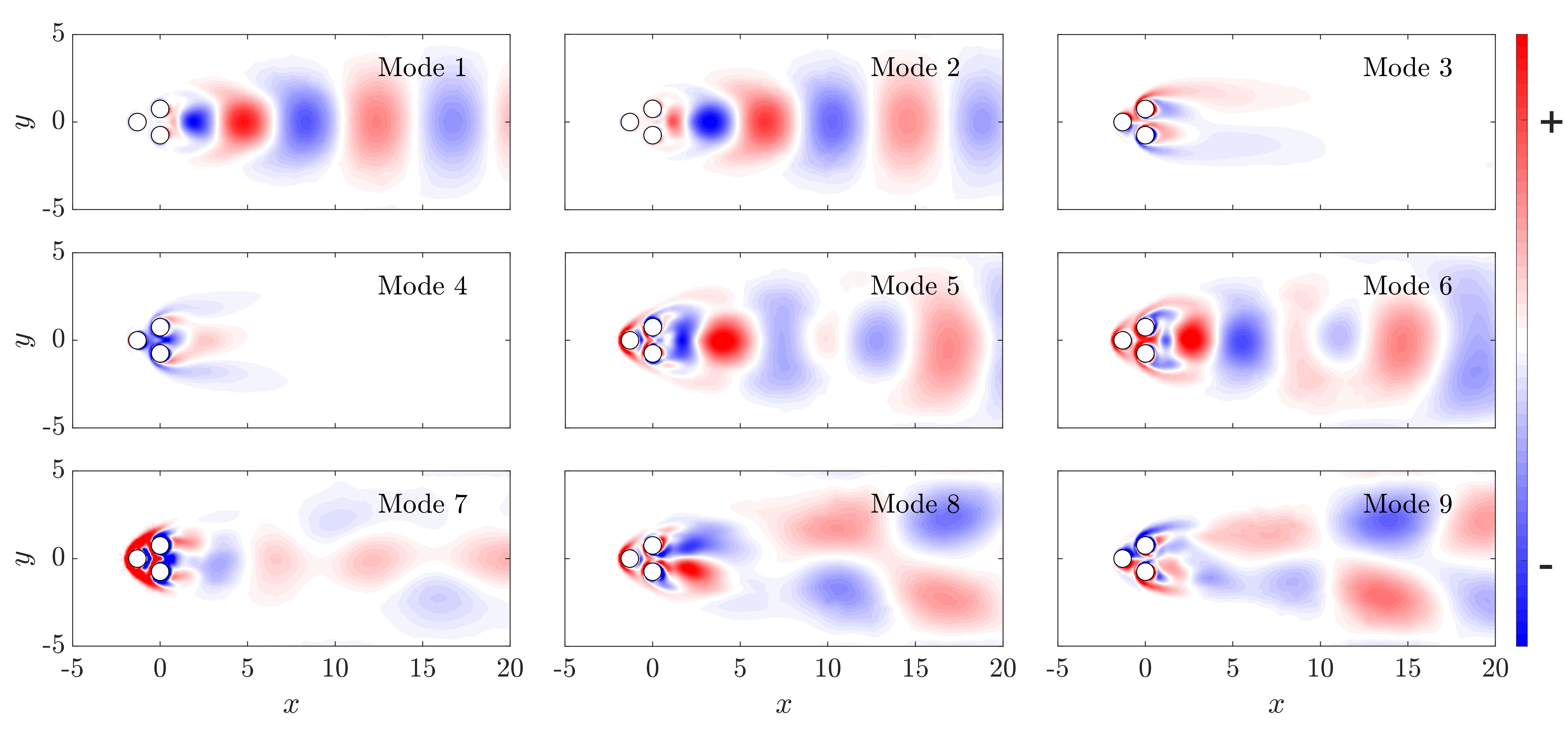}
    \caption{Modal shapes of the first 9 POD modes visualized in vorticity ($\nabla\times\pmb{u}_k$).}
    \label{fig:pod_modes}
\end{figure}

\begin{figure}
    \centering
    \includegraphics[width=.6\linewidth]{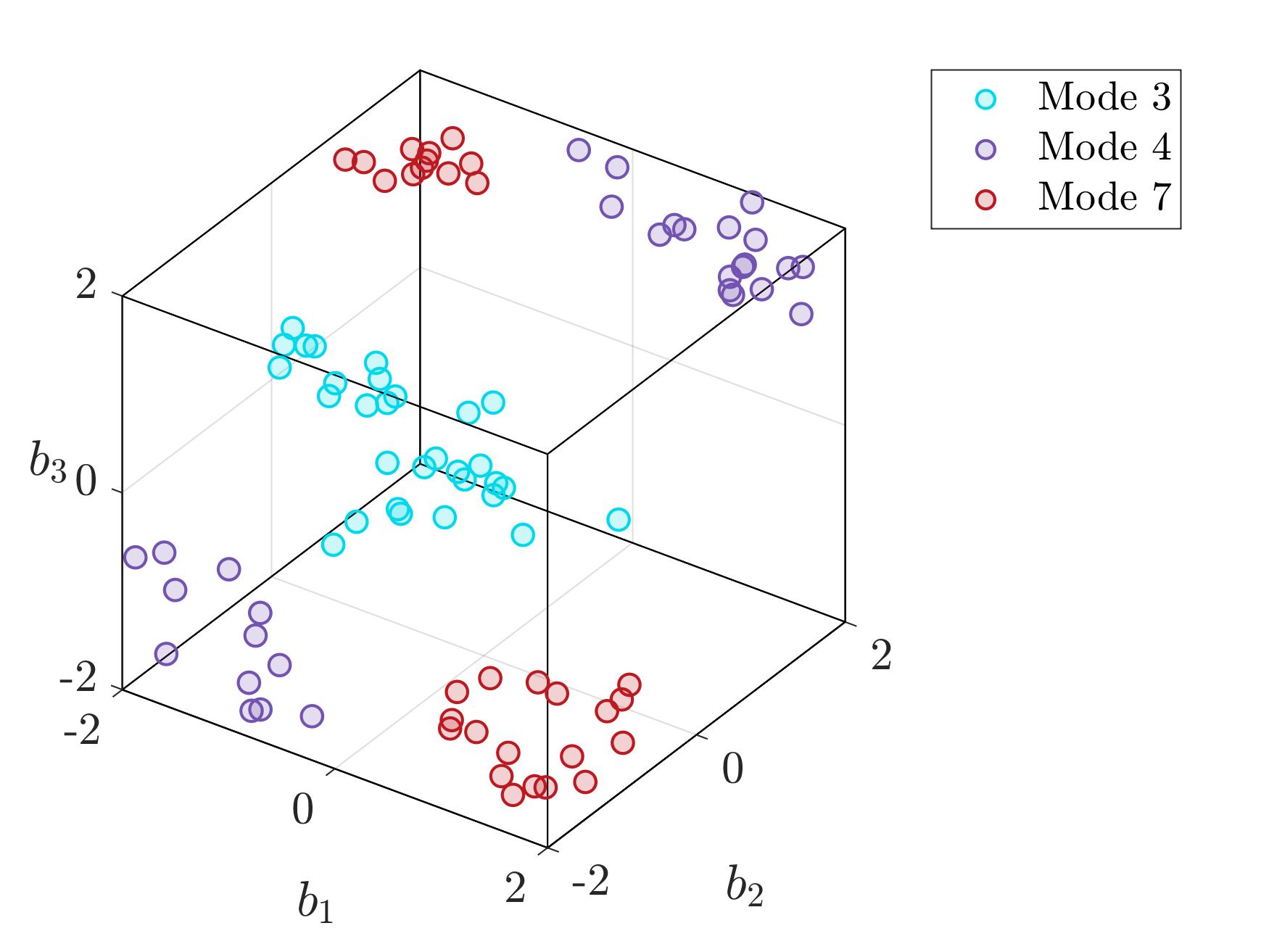}
    \caption{Distribution of the first 30 control commands which possess the highest overall amplitudes of the POD modal coefficients for modes 3, 4, and 7.}
    \label{fig:mode_amp}
\end{figure}

\Cref{fig:pod_modes} presents the modal shapes of the first 9 dominant POD modes. 
Each contour represents the curl of the corresponding POD eigenfunction ($\nabla\times\pmb{u}_k$).
The combination of POD modes 1 and 2 can effectively depict the spatial structures of the von K\'arm\'an vortex street appearing downstream of the three cylinders.
Modes 4 and 5 also represent the vortex shedding structures in the flow, and modes 8 and 9 describe the second harmonics of the dominant coherent structures. 
Although these structures can be clearly recognized from visualization, the shapes of these modes are no longer strictly symmetric/anti-symmetric about the centerline thanks to the influence of cylinder rotations.
\textcolor{black}{To advance the physical understanding of POD modes 3, 4, and 7, we calculate the root-mean-square of these POD modal coefficients over time ($\sqrt{\langle a_i^2(\bm{b},t) \rangle_t}$ for $i=3,4,7$) for all $\bm{b}\in\mathcal{B}$.
For each mode, we pick the first 30 control laws possessing the highest r.m.s. value, and visualize them in \cref{fig:mode_amp}.
In other words, these POD modes become more pronounced under the actuation commands displayed in the figure.}
 Mode 3 plays the role of  the shift mode between steady and time-averaged periodic solution \citep{Noack2003}.
Furthermore, as the control commands in \textcolor{black}{\cref{fig:mode_amp}} mostly locate around the boundary lines ($b_2=-2$ and $b_3=2$) and ($b_2=2$ and $b_3=-2$), we find this shift mode also represents the effect when the two rear cylinders rotates in opposite directions.
Mode 4 is more energetic when the control commands approach the boundary points $[-2, -2, -2]$ and $[2, 2, 2]$, 
hence this mode is related to the stagnation point control in which all cylinders rotate in a uniform manner.
This causal relationship can be further confirmed by its modal shape, where a symmetrical pattern emerges from the separation points of the top and bottom cylinders.
Finally, the control commands for mode 7 in \textcolor{black}{\cref{fig:mode_amp}} aggregates near $[2,-2,-2]$ and $[-2,2,2]$. 
From this characteristic we can  infer that this mode is highly related to the control when the two rear cylinders rotate in the same direction and the front cylinder rotates oppositely.

The results from the snapshot POD provide a basic understanding of the dominant flow structures under different steady control commands.
Based on the information above, in the following we will present the application of estimation methods proposed in \cref{sec:estimation} to three representative steady control cases, i.e., boat tailing, stagnation control, and base bleeding.
\textcolor{black}{To investigate the influence of random initialization on the DNN estimation results, we train the same network architecture 20 times with different randomly-generated initial weights, as mentioned in \cref{ssec:dnn}. 
In the following discussions, we use the averaged output from all trained networks as the DNN estimation result and discuss the sensitivity to the initial condition using the corresponding standard deviation.}

\subsection{Case \RNum{1}: Boat tailing}\label{ssec:case1}
\begin{figure}
    \centering
    \begin{subfigure}{0.3\linewidth}
    \includegraphics[width=.99\linewidth]{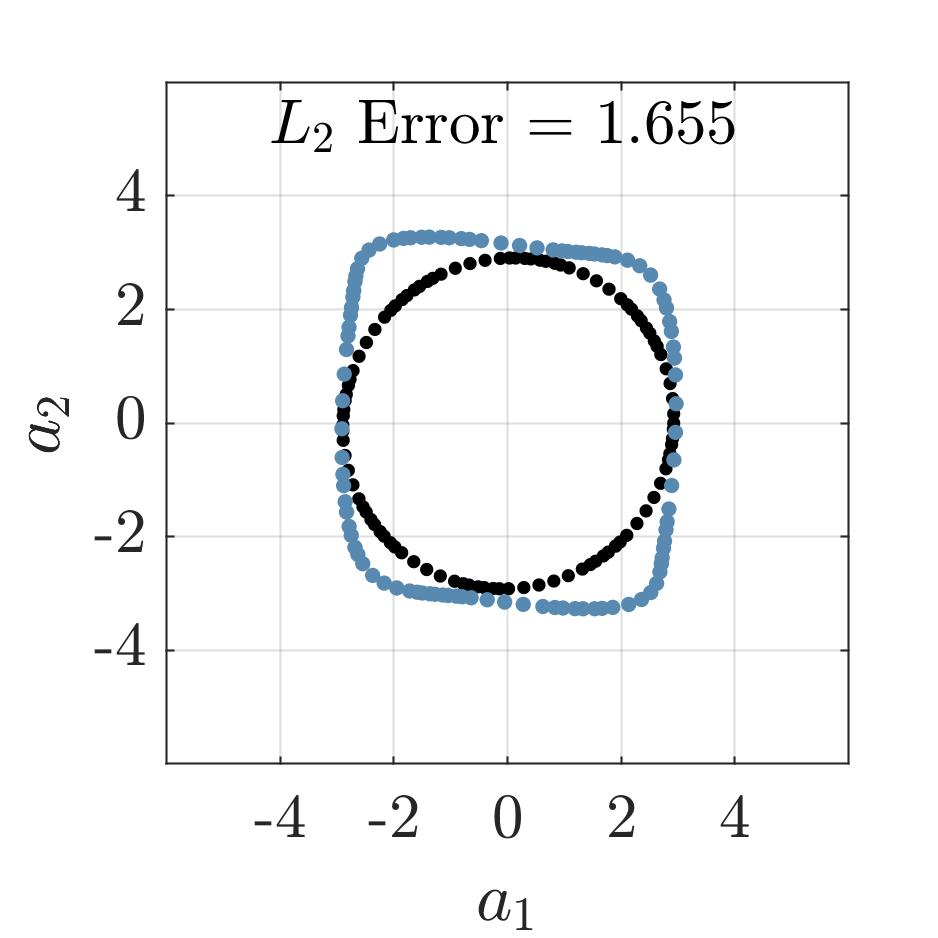}
    \subcaption{LSE}
    \end{subfigure}
    \begin{subfigure}{0.3\linewidth}
    \includegraphics[width=.99\linewidth]{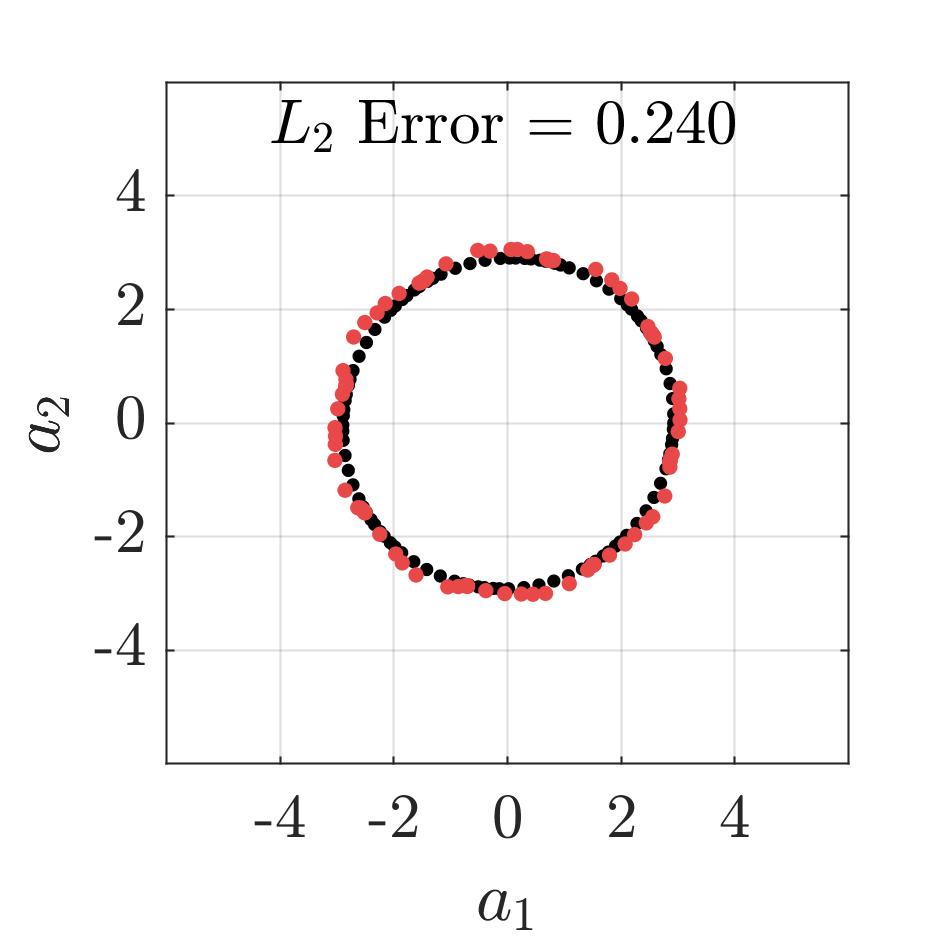}
    \subcaption{$k$NN}
    \end{subfigure}
    \begin{subfigure}{0.3\linewidth}
    \includegraphics[width=.99\linewidth]{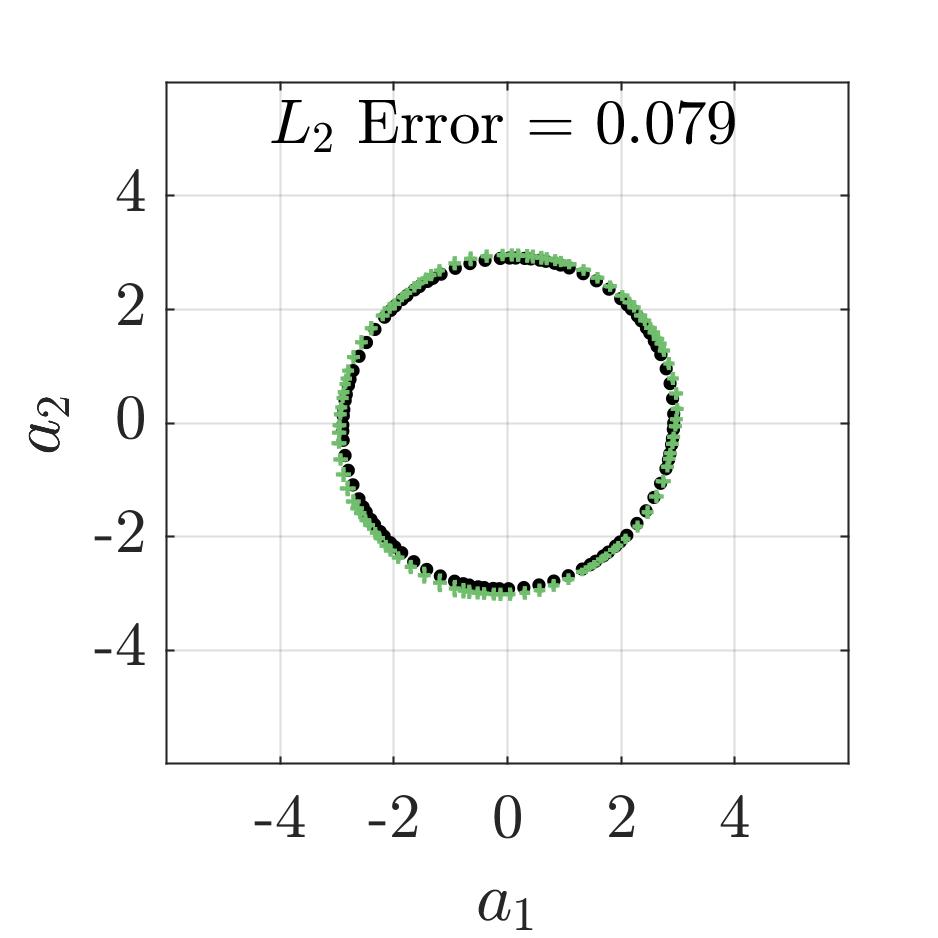}
    \subcaption{DNN}
    \end{subfigure}
    \caption{Phase relationship ($a_1$ versus $a_2$) obtained from three estimation methods \textcolor{black}{and the $L_2$ error associated with the first two modes} in the boat-tailing case. Black dots represents the target outputs from the DNS data. \textcolor{black}{The center, horizontal and vertical lines of the DNN marker represents the mean, standard deviation of $a_1$ and $a_2$ estimates from 20 trained neural networks with different initializations.}}
    \label{fig:ID82_phase}
\end{figure}

To start with, we apply the flow estimation methods to a boat tailing case where $\pmb{b}= [-0.0196, 0.7078,  -0.7431]^{\text{T}}$.
By rotating the top cylinder in the clockwise direction and the bottom cylinder in the opposite direction, the flow control is realized by the delay of separation on the rearward cylinders in this case.
As discussed in \cref{ssec:pod}, the combination of the first two POD modes represents the streamwise propagation of the dominant coherent structures in the flow.
Hence, to evaluate the estimation performance we start with the examination of the phase relationship between $a_1$ and $a_2$ obtained from each method.
\Cref{fig:ID82_phase} presents the estimated phase relationship in comparison to the ground truth provided by the DNS simulation.
\textcolor{black}{The the $L_2$ norm of the estimation error from the first two POD modes
is also displayed in the figure}.
The temporal evolution of $a_1$ and $a_2$ from the simulation data forms a circular pattern in \textcolor{black}{\cref{fig:ID82_phase}}, which indicates the occurrence of periodic vortex shedding downstream of the cylinders.
The phase relationships of $k$NN and DNN can follows the target outputs in principal, and the estimation errors are relatively small.
\textcolor{black}{In comparison, DNN produces a lower estimation error than $k$NN in this case.}
However, the LSE misinterprets the circular phase relationship and generates a nearly squared pattern. 
As a result, a relatively large estimation error can be observed from the LSE estimates of the first two POD modes.

\begin{figure}
    \centering
    \includegraphics[width=.8\linewidth,trim={7cm 0cm 8cm 2cm},clip]{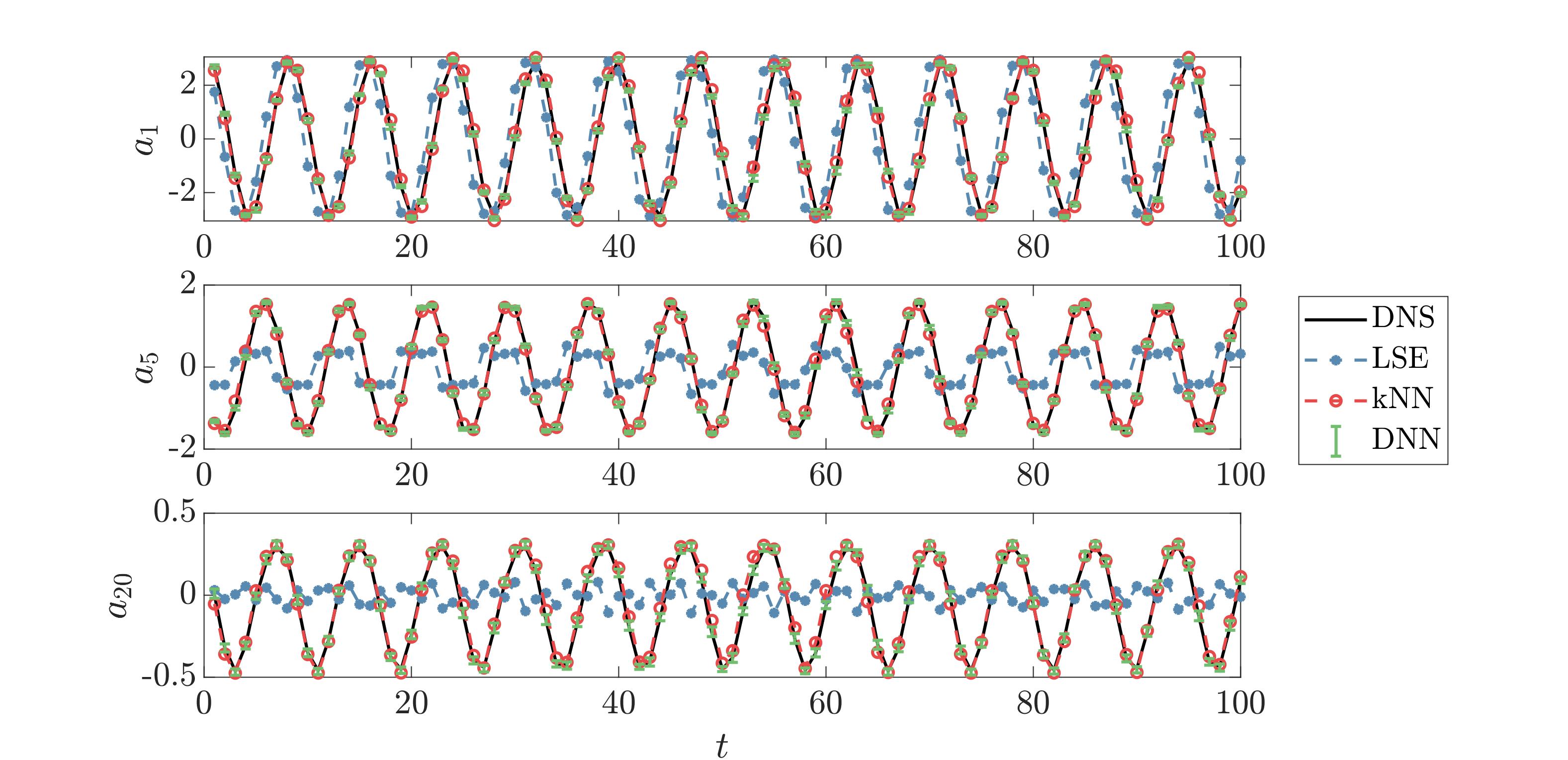}
    \caption{A comparison between estimated POD modal coefficients and the ground truth from DNS data in the boat tailing case. \textcolor{black}{Error bars of the DNN estimates represent the standard deviation from 20 trained neural networks with different initializations.}}
    \label{fig:ID82_pred}
\end{figure}

\begin{figure}
    \centering
    \begin{subfigure}{0.45\linewidth}
    \includegraphics[width=.99\linewidth,trim={0cm 8cm 1cm 12cm},clip]{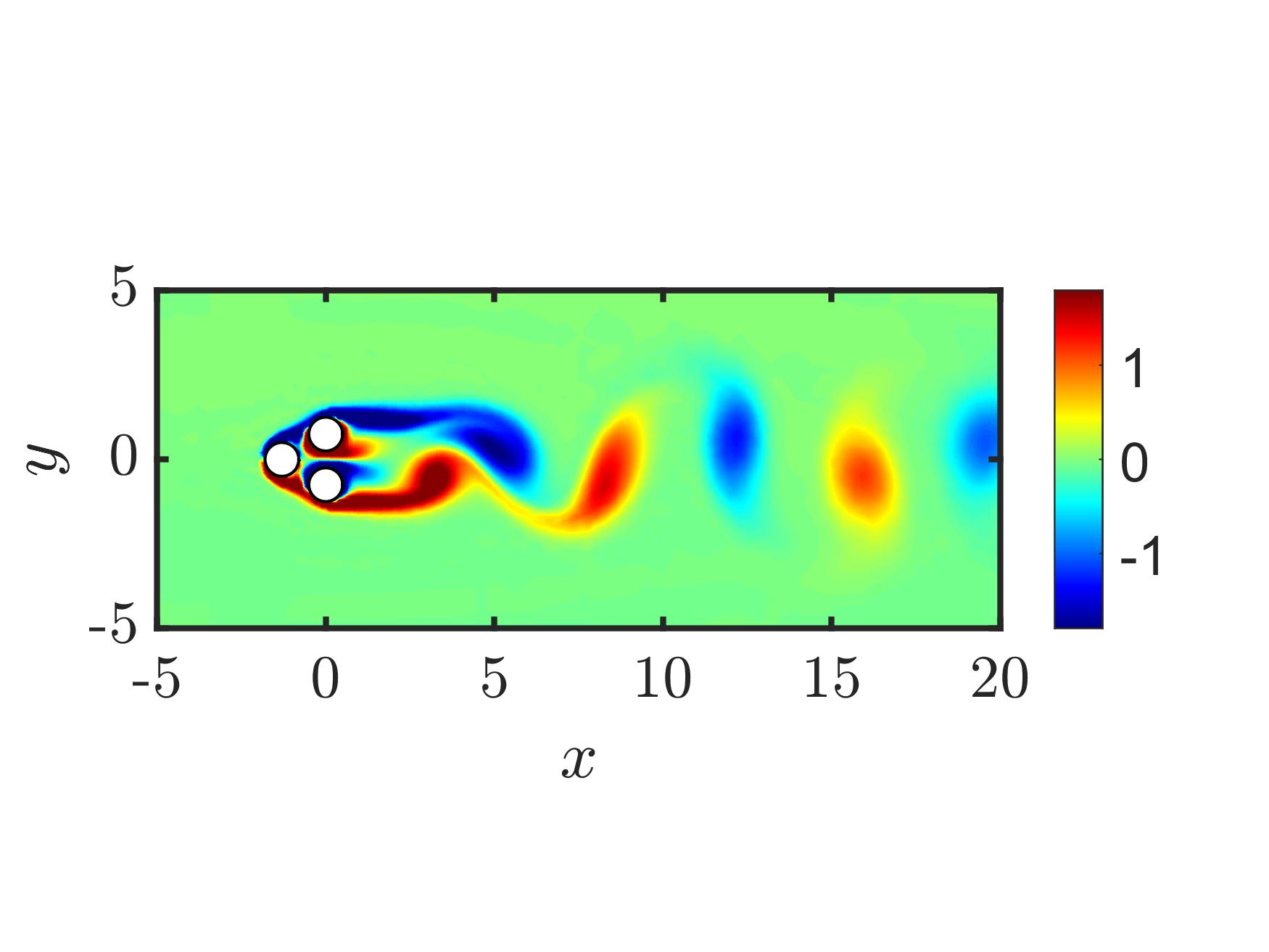}
    \subcaption{DNS}
    \end{subfigure}    
    \begin{subfigure}{0.45\linewidth}
    \includegraphics[width=.99\linewidth,trim={0cm 8cm 1cm 12cm},clip]{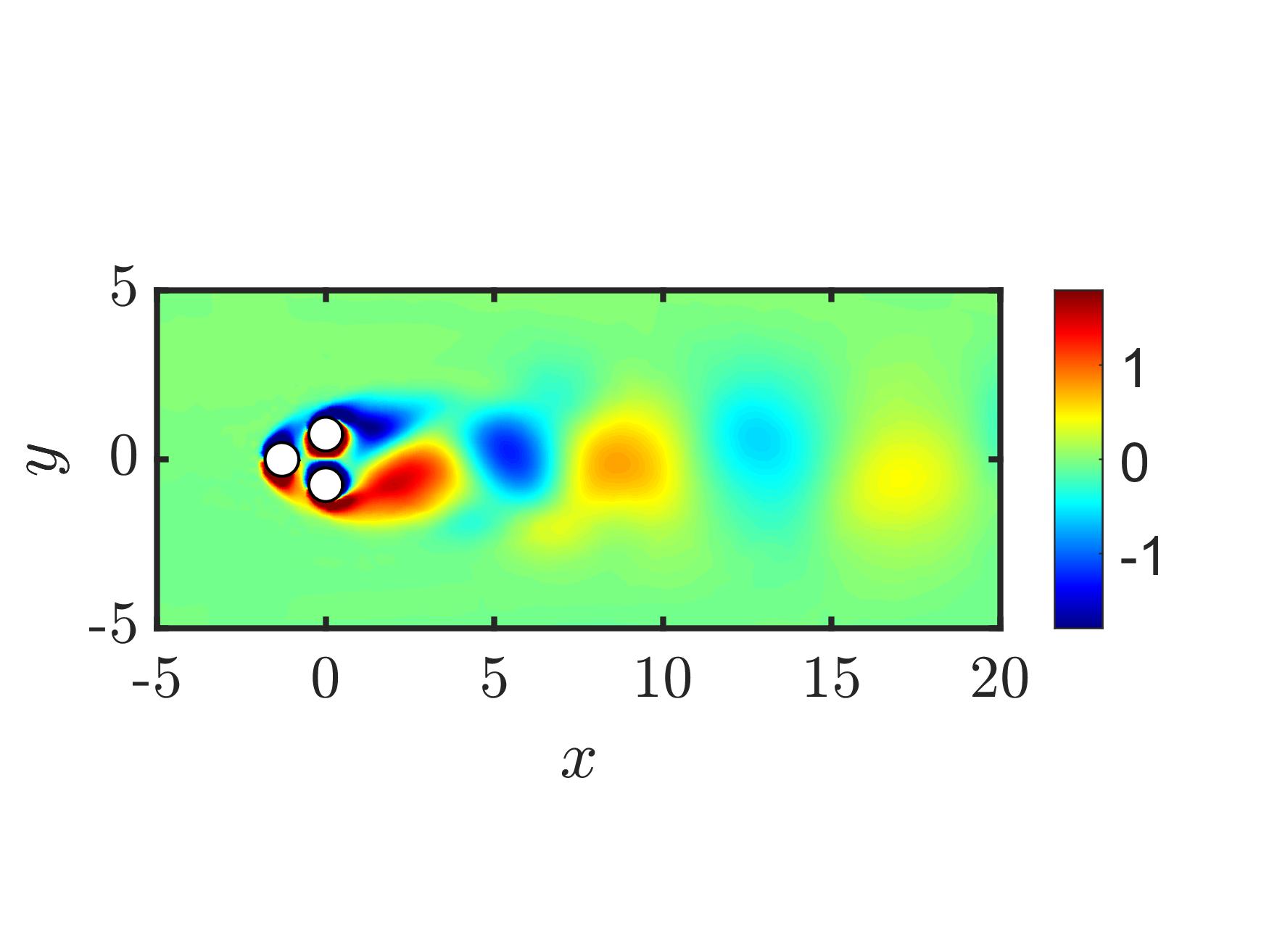}
    \subcaption{LSE}
    \end{subfigure}
    \begin{subfigure}{0.45\linewidth}
    \includegraphics[width=.99\linewidth,trim={0cm 8cm 1cm 12cm},clip]{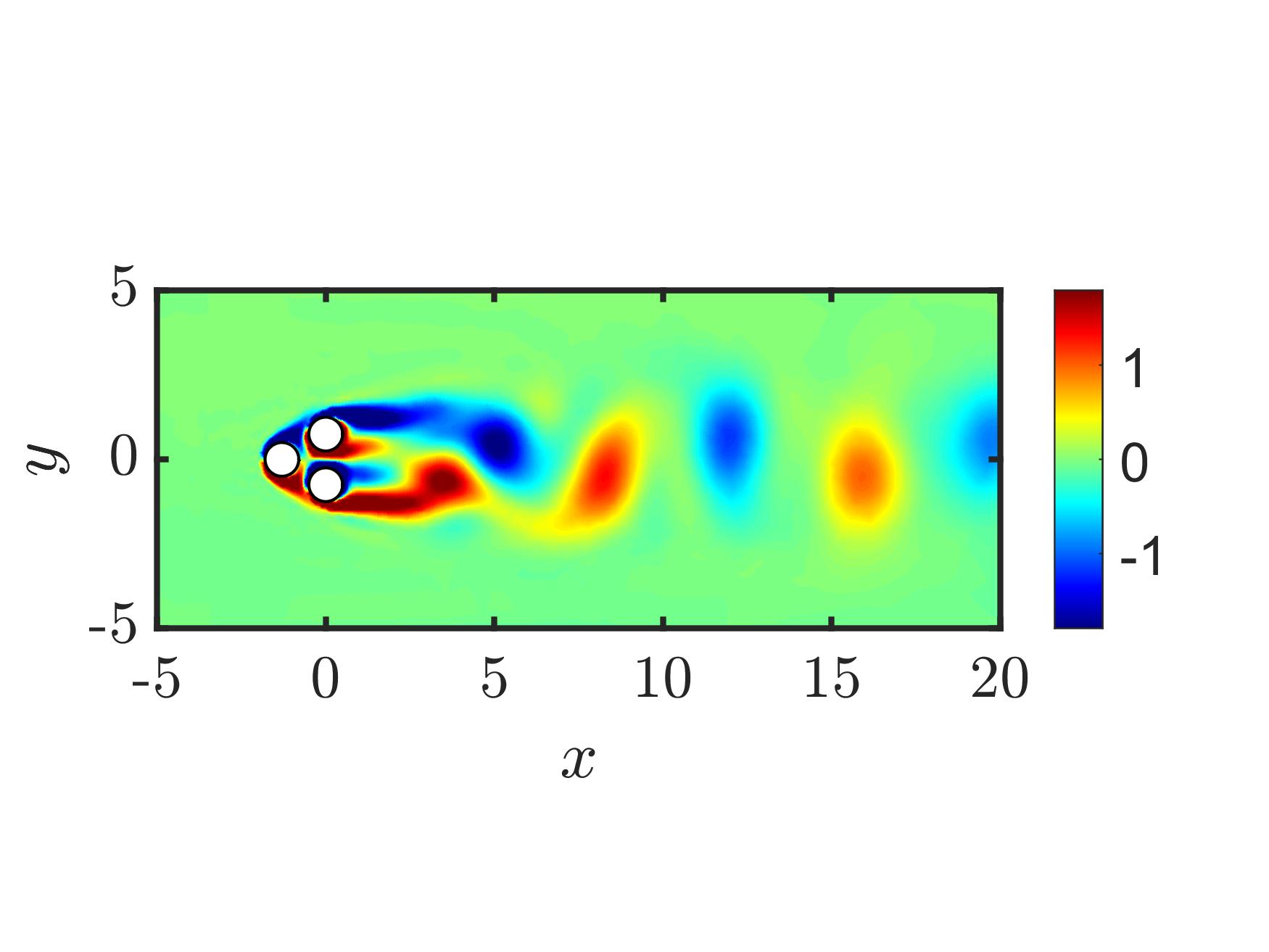}
    \subcaption{$k$NN}
    \end{subfigure}
    \begin{subfigure}{0.45\linewidth}
    \includegraphics[width=.99\linewidth,trim={0cm 8cm 1cm 12cm},clip]{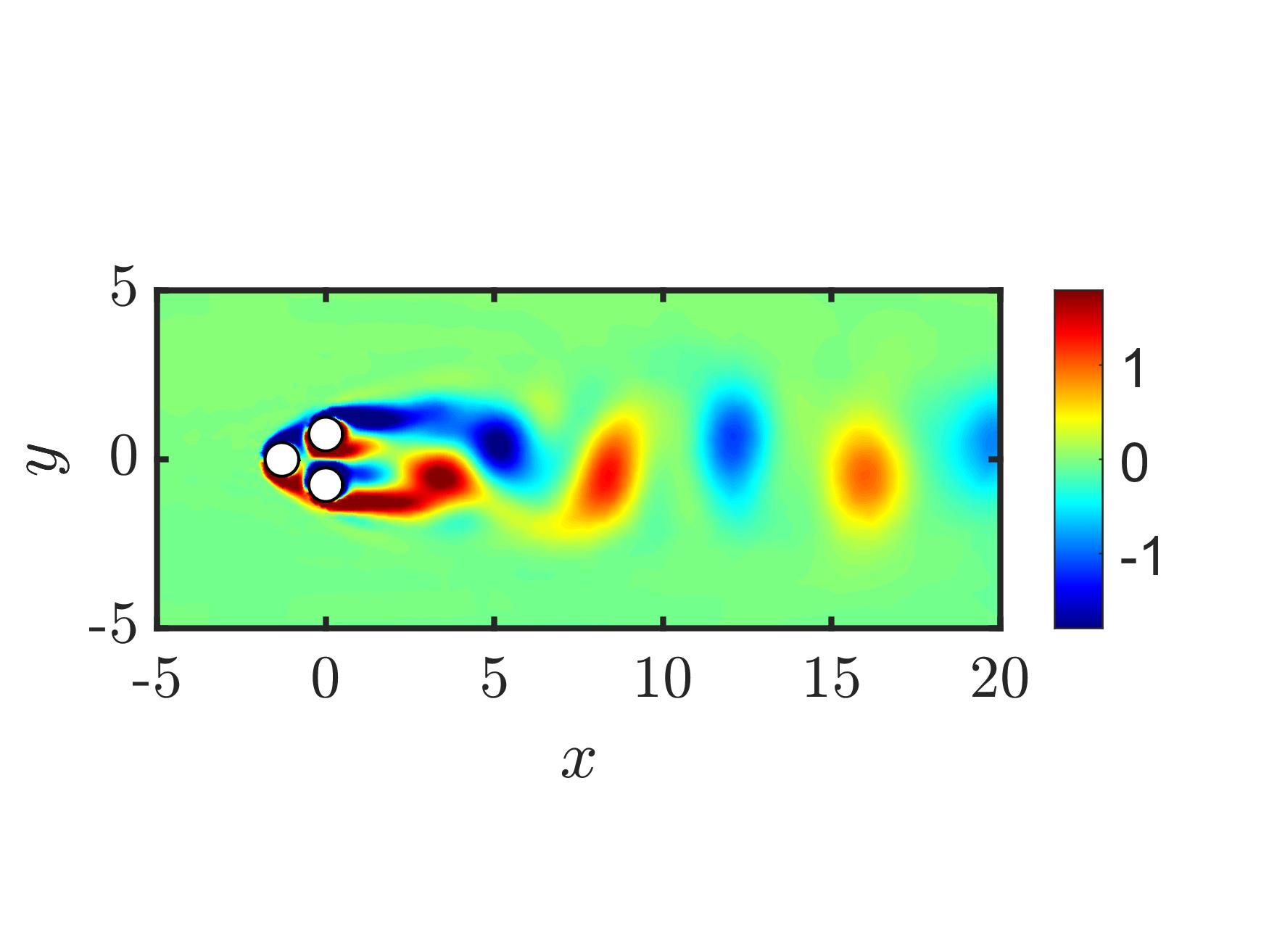}
    \subcaption{DNN}
    \end{subfigure}
    
    \caption{A comparison between the estimated instantaneous vorticity fields and the DNS data in the boat tailing case. \textcolor{black}{The DNN estimate is averaged from 20 trained neural networks with different initializations.}}
    \label{fig:ID82_recon}
\end{figure}

\Cref{fig:ID82_pred} compares the estimated POD modal coefficients and the target outputs from DNS data over all time instances in this boat tailing case.
Here the estimates of three representative POD modes 1, 5, and 20 are presented in an attempt to evaluate estimation performance at different energy levels.
The $k$NN and DNN estimates are observed to be consistent to the DNS data, and they are able to accurately describe the periodic behavior of the POD modal coefficients.
On the contrary, LSE is observed to generate a relatively large error in the estimation of $a_1$, 
and it can hardly follow the DNS data at higher order modes $a_5$ and $a_{20}$.
\textcolor{black}{Regarding to the variation of the DNN estimates with different initializations, the standard deviations of modal coefficients are hardly recognizable either for leading or higher order POD modes, as shown in \cref{fig:ID82_phase,fig:ID82_pred}.}
\Cref{fig:ID82_recon} displays a comparison between the estimated flow fields and the target DNS data at a randomly selected time instance.
The vorticity fields obtained from $k$NN and DNN can be seen to remain consistent to the DNS snapshot,
while LSE fails to accurately describe the flow state regarding to the eddy locations and intensities. 

\subsection{Case \RNum{2}: Stagnation point control}\label{ssec:case2}
\begin{figure}
    \centering
    \begin{subfigure}{0.3\linewidth}
    \includegraphics[width=.99\linewidth]{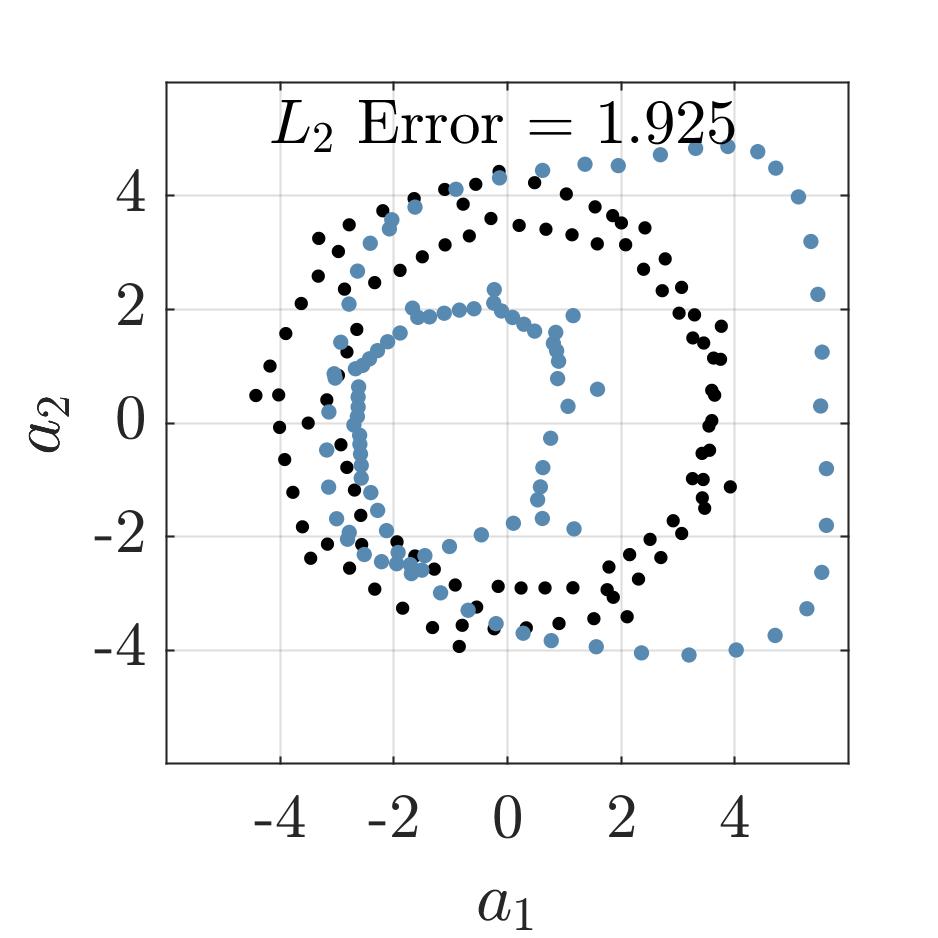}
    \subcaption{LSE}
    \end{subfigure}
    \begin{subfigure}{0.3\linewidth}
    \includegraphics[width=.99\linewidth]{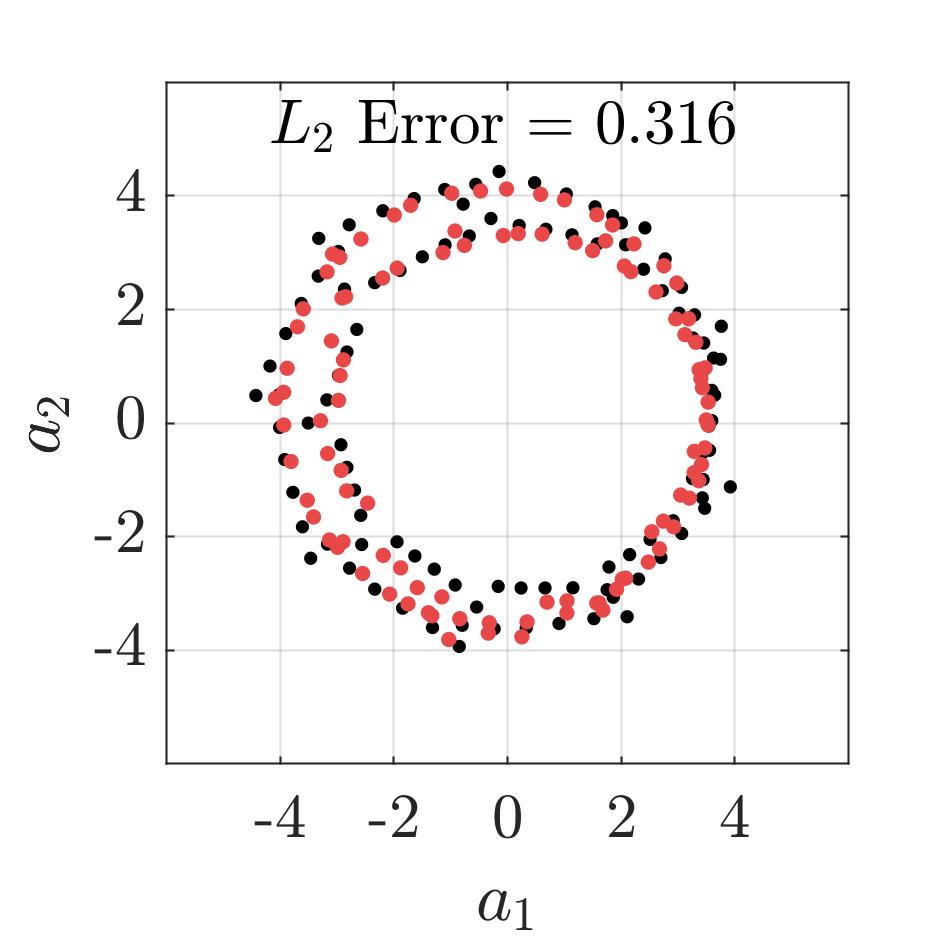}
    \subcaption{$k$NN}
    \end{subfigure}
    \begin{subfigure}{0.3\linewidth}
    \includegraphics[width=.99\linewidth]{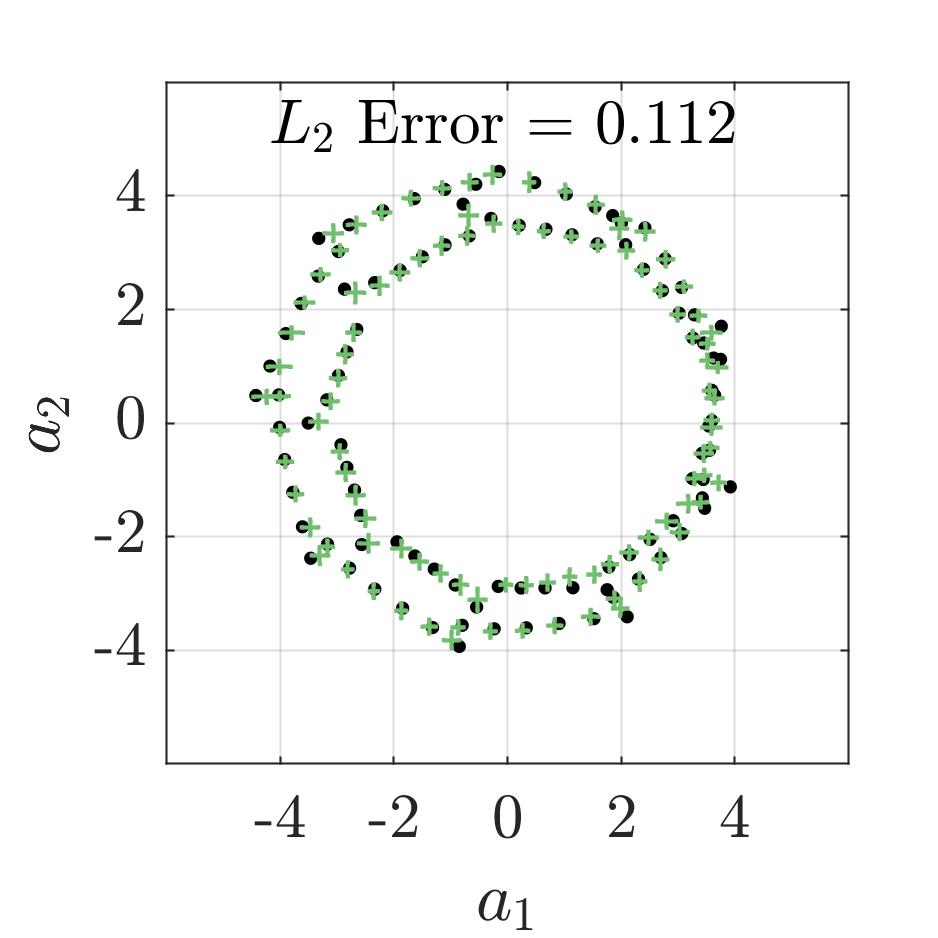}
    \subcaption{DNN}
    \end{subfigure}
    \caption{Phase relationship ($a_1$ versus $a_2$) obtained from three estimation methods \textcolor{black}{and the $L_2$ error associated with the first two modes} in the stagnation point control case. Black dots represents the target outputs from the DNS data.
    \textcolor{black}{The center, horizontal and vertical lines of the DNN marker represents the mean, standard deviation of $a_1$ and $a_2$ estimates from 20 trained neural networks with different initializations.}}
    \label{fig:ID63_phase}
\end{figure}

\begin{figure}
    \centering
    \includegraphics[width=.8\linewidth,trim={7cm 0cm 8cm 2cm},clip]{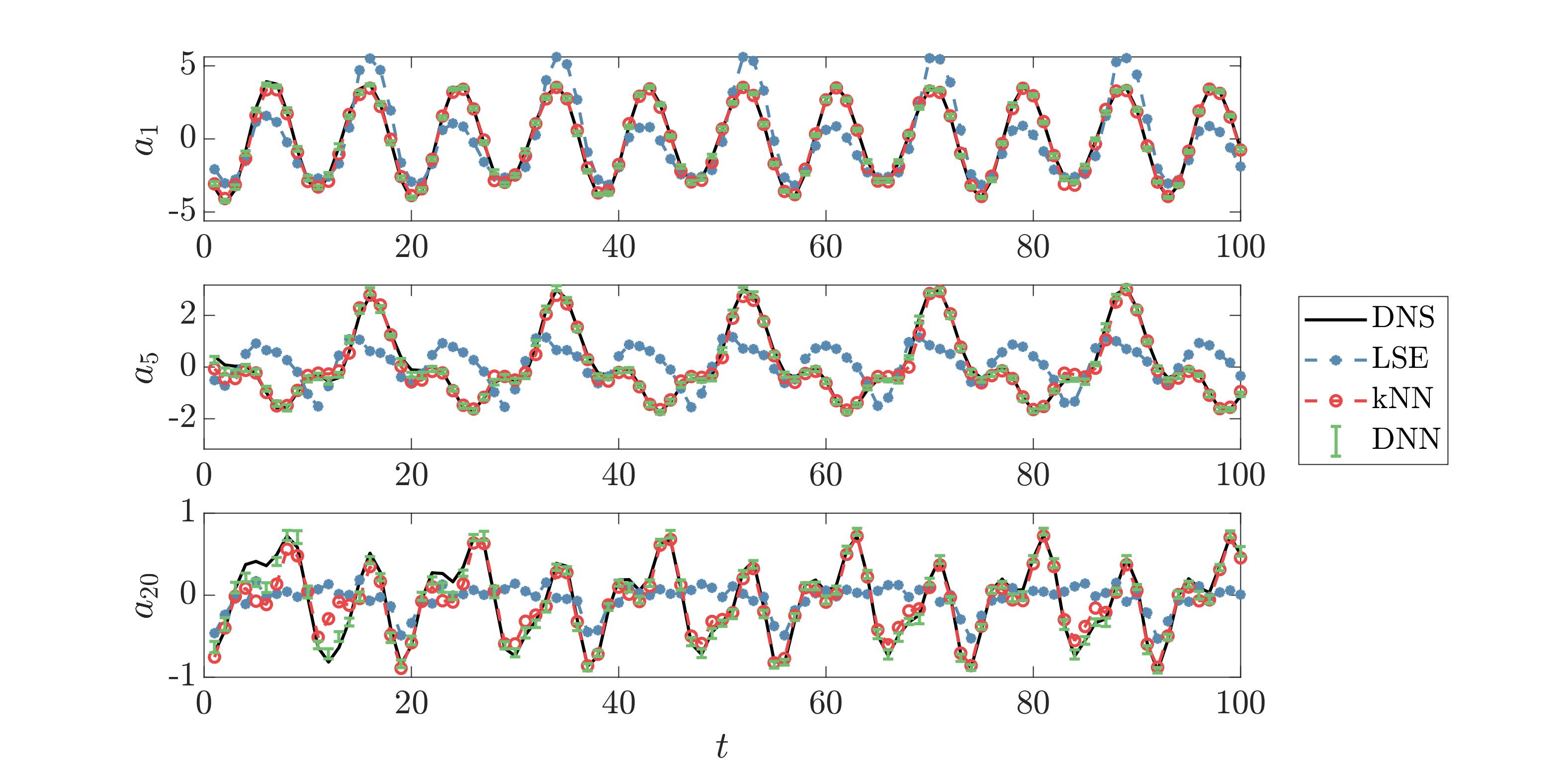}
    \caption{A comparison between estimated POD modal coefficients and the ground truth from DNS data in the stagnation point control case. 
    \textcolor{black}{Error bars of the DNN estimates represent the standard deviation from 20 trained neural networks with different initializations.}}
    \label{fig:ID63_pred}
\end{figure}

\begin{figure}
    \centering
    \begin{subfigure}{0.45\linewidth}
    \includegraphics[width=.99\linewidth,trim={0cm 8cm 1cm 12cm},clip]{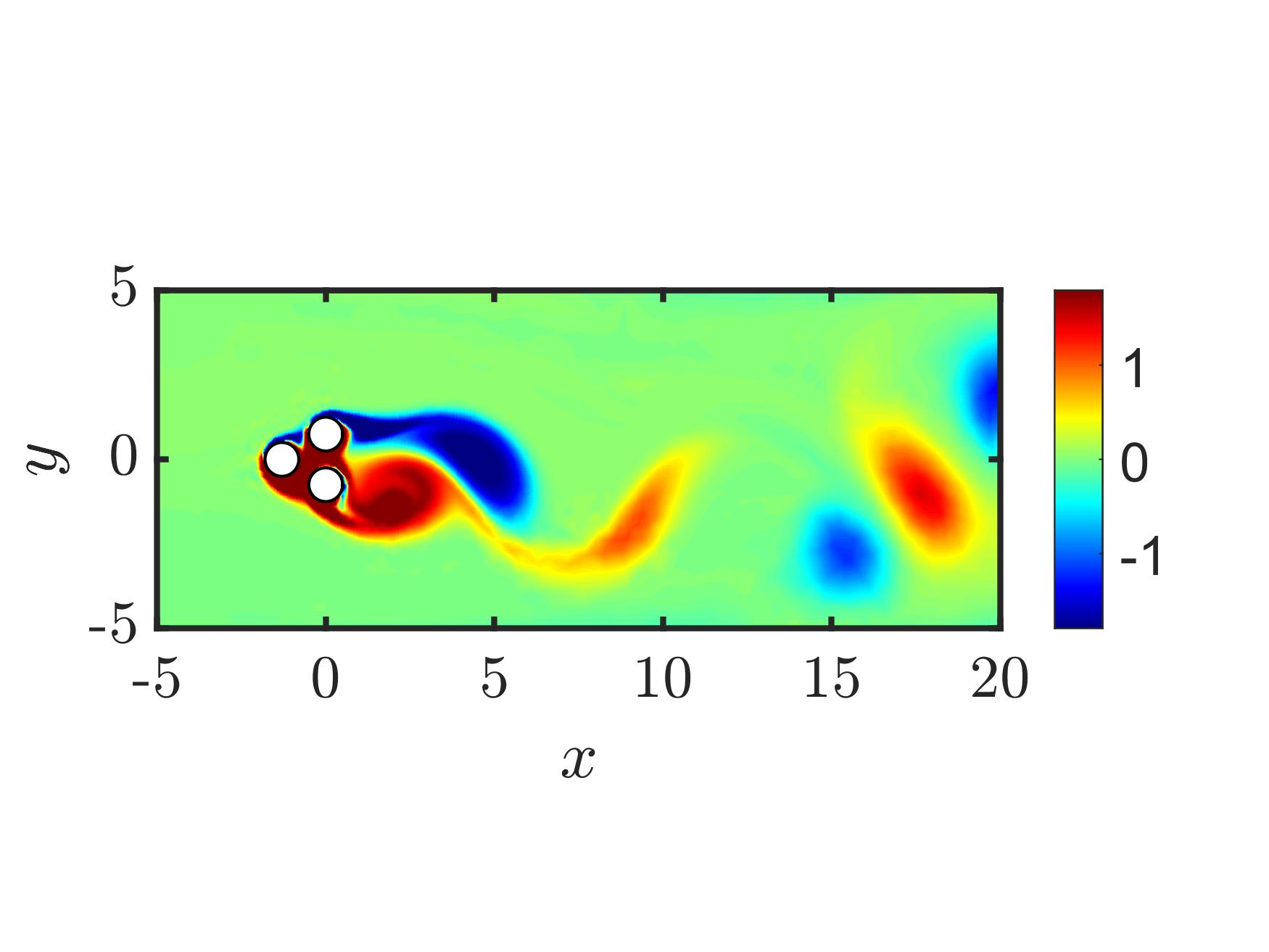}
    \subcaption{DNS}
    \end{subfigure}
    \begin{subfigure}{0.45\linewidth}
    \includegraphics[width=.99\linewidth,trim={0cm 8cm 1cm 12cm},clip]{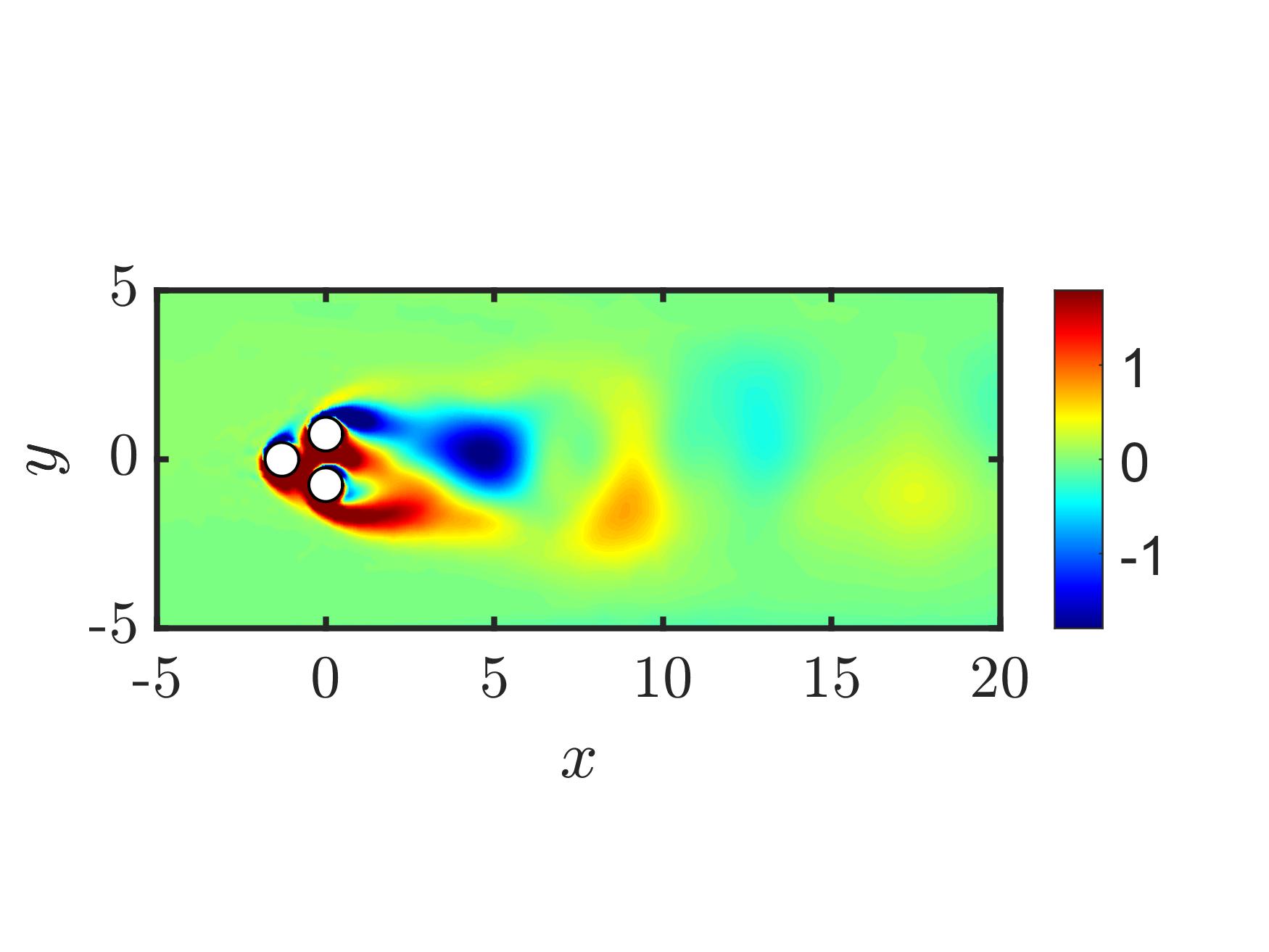}
    \subcaption{LSE}
    \end{subfigure}
    \begin{subfigure}{0.45\linewidth}
    \includegraphics[width=.99\linewidth,trim={0cm 8cm 1cm 12cm},clip]{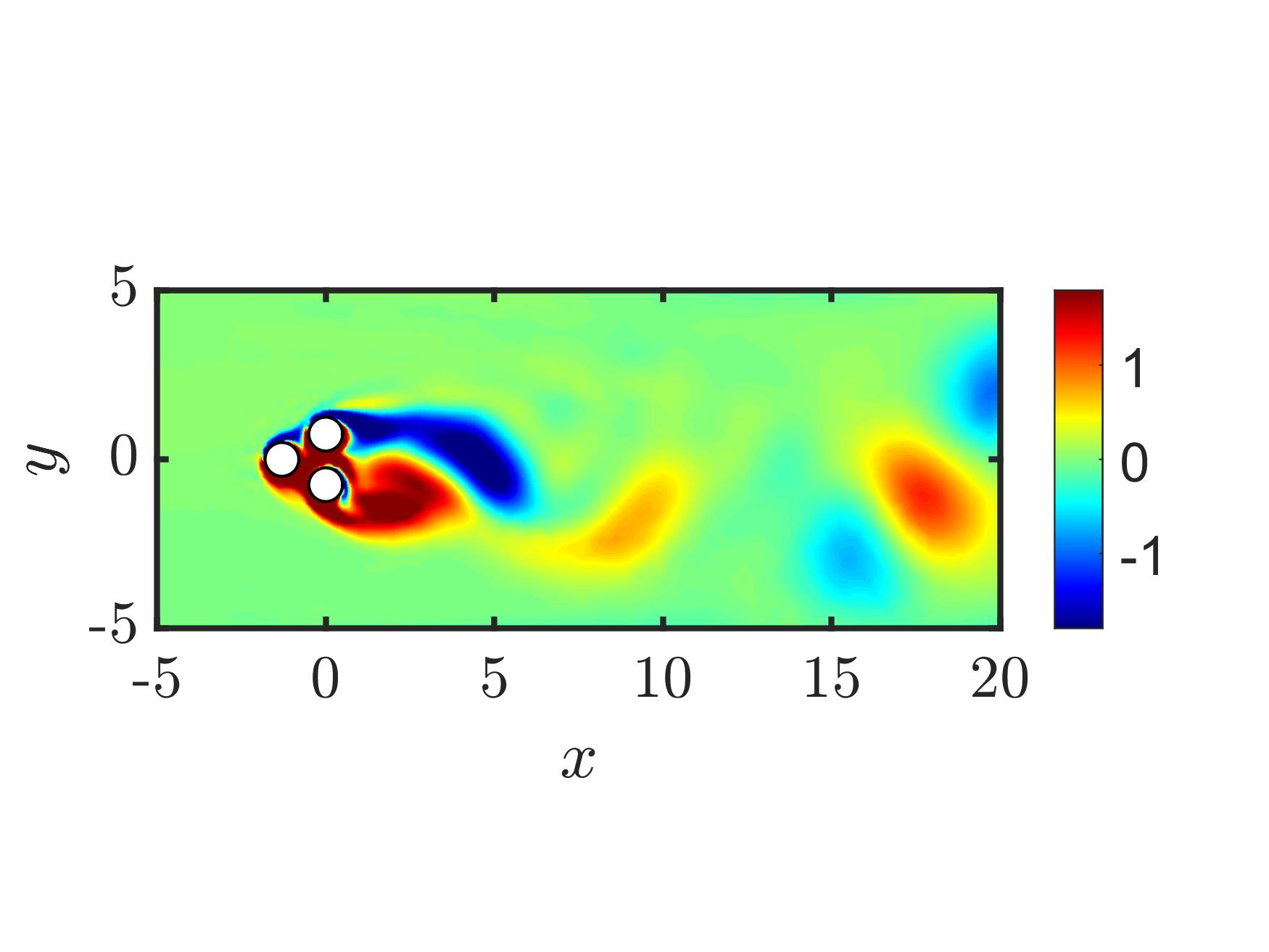}
    \subcaption{$k$NN}
    \end{subfigure}
    \begin{subfigure}{0.45\linewidth}
    \includegraphics[width=.99\linewidth,trim={0cm 8cm 1cm 12cm},clip]{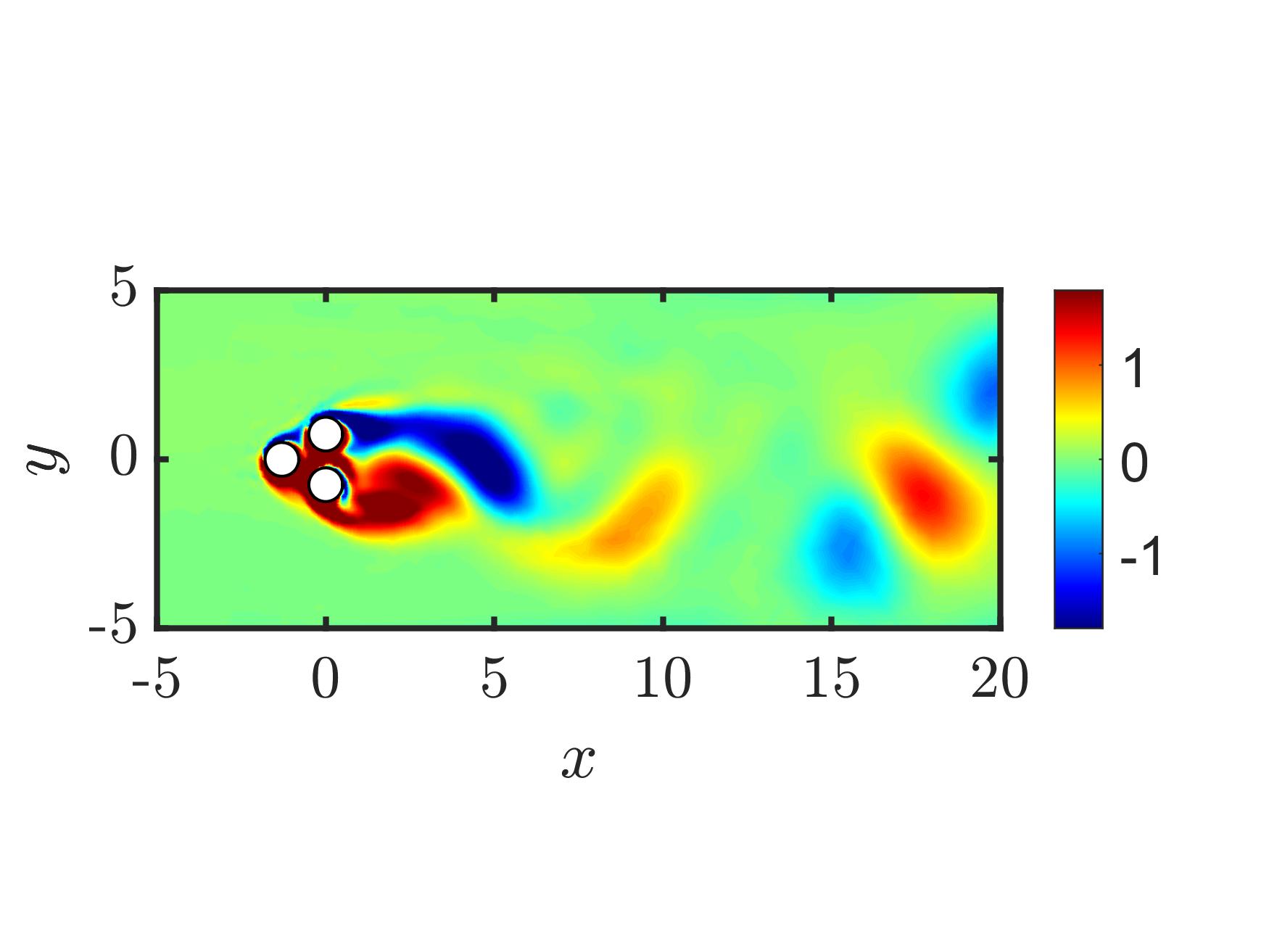}
    \subcaption{DNN}
    \end{subfigure}
    
    \caption{A comparison between the estimated instantaneous vorticity fields and the DNS data in the stagnation point control case.
    \textcolor{black}{The DNN estimate is averaged from 20 trained neural networks with different initializations.}}
    \label{fig:ID63_recon}
\end{figure}

In the second case we investigate the estimation performance when the flow field is under stagnation point control.
This strategy represents the uniform rotation of all cylinders, and the control command for this representative case is $ \pmb{b} = [-1.3932 ,  -1.5026,   -1.5952]^{\text{T}}$.
\textcolor{black}{\cref{fig:ID63_phase}} presents the phase relationship from the three estimation methods as well as the expected outcome from DNS.
The phase portrait between $a_1$ and $a_2$ from DNS is seen to be quasi-periodic, in a sense that the recurring pattern is no longer circular.
In this exemplary case, the LSE estimates completely deviate from the true phase pattern, and result in significant estimation error.
On the other hand, the two non-linear methods can faithfully reflect the expected phase relationship.
Comparatively $k$NN performs slightly better than DNN regarding to the $L_2$ error which are shown on the figure.
\Cref{fig:ID63_pred} compares the estimated modal amplitudes to the ground truth for POD modes 1, 5, and 20.
Besides the accurate estimation of the first order modal coefficient, 
$k$NN and DNN also shows promising estimation performance at $a_5$ and $a_{20}$.
On the contrary, LSE estimates are seen to deviate much more from the ground truth at $a_1$, 
and can hardly follow the DNS curve at higher order modes.
\textcolor{black}{Regarding to the sensitivity to random initialization from DNN estimates, only insignificant level of uncertainty is observed in \cref{fig:ID63_phase,fig:ID63_pred}.}
The estimation results of an instantaneous flow snapshot are displayed in \textcolor{black}{\cref{fig:ID63_recon}}, and are compared to the target output from DNS.
In the DNS snapshot, the dominant features in the flow are the vortex shedding near $x=5$, as well as the downstream eddies appearing after $x=15$.
The estimated flow fields from LSE and DNN are capable to capture these dominant flow features from the combined input vector.
However, in the LSE estimate, the flow structures before $x=10$ can only be roughly captured with significant inconsistency in shape.
At the same time, vortices appearing further downstream can no longer be recognized from this linear estimation approach.

\subsection{Case \RNum{3}: Base bleeding}\label{ssec:case3}

\begin{figure}
    \centering
    \begin{subfigure}{0.3\linewidth}
    \includegraphics[width=.99\linewidth]{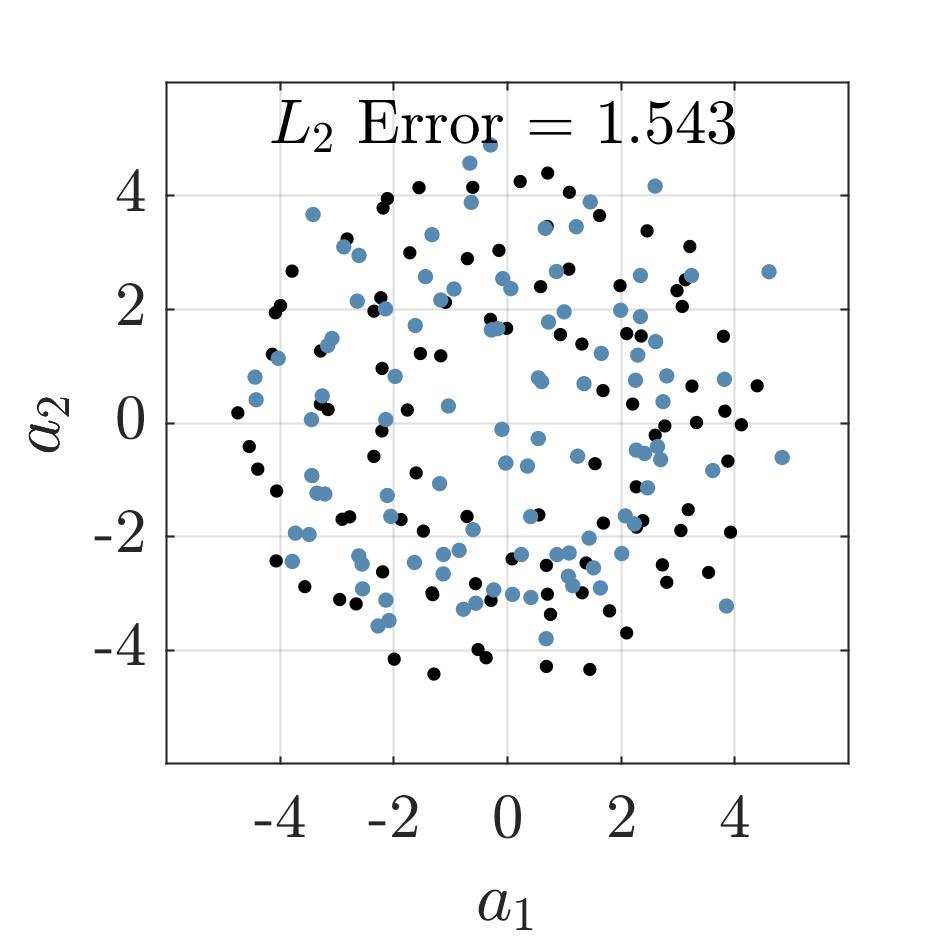}
    \subcaption{LSE}
    \end{subfigure}
    \begin{subfigure}{0.3\linewidth}
    \includegraphics[width=.99\linewidth]{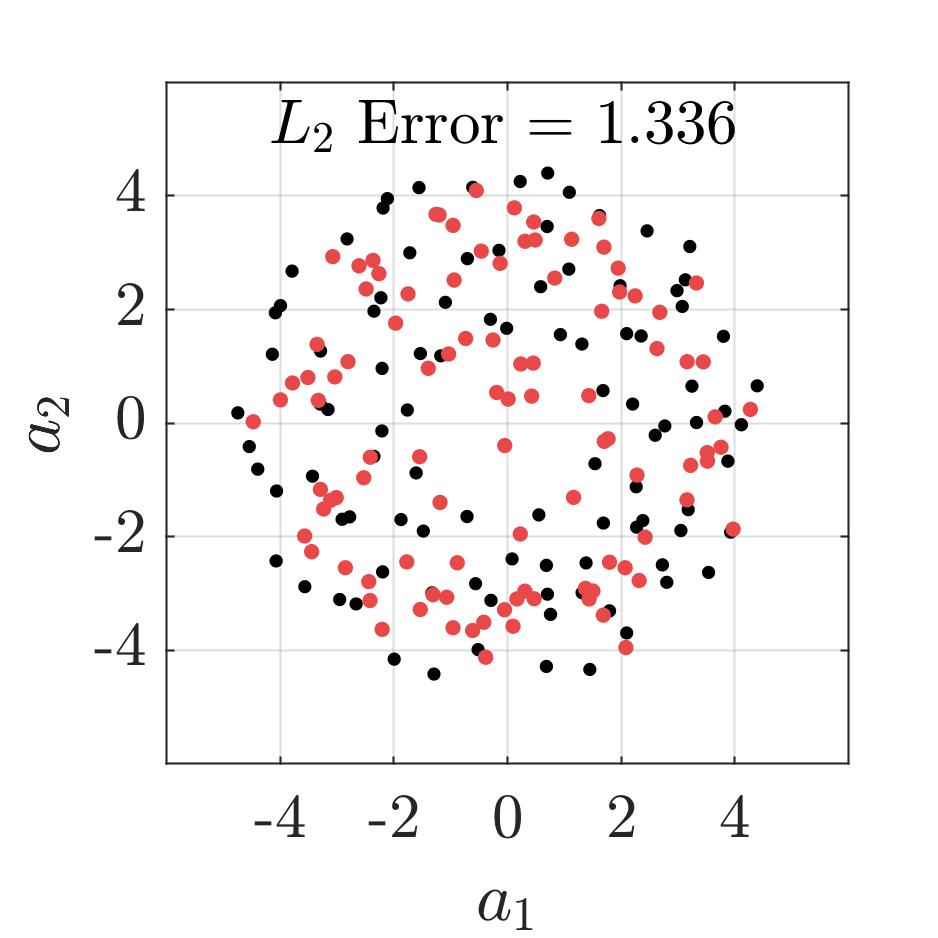}
    \subcaption{$k$NN}
    \end{subfigure}
    \begin{subfigure}{0.3\linewidth}
    \includegraphics[width=.99\linewidth]{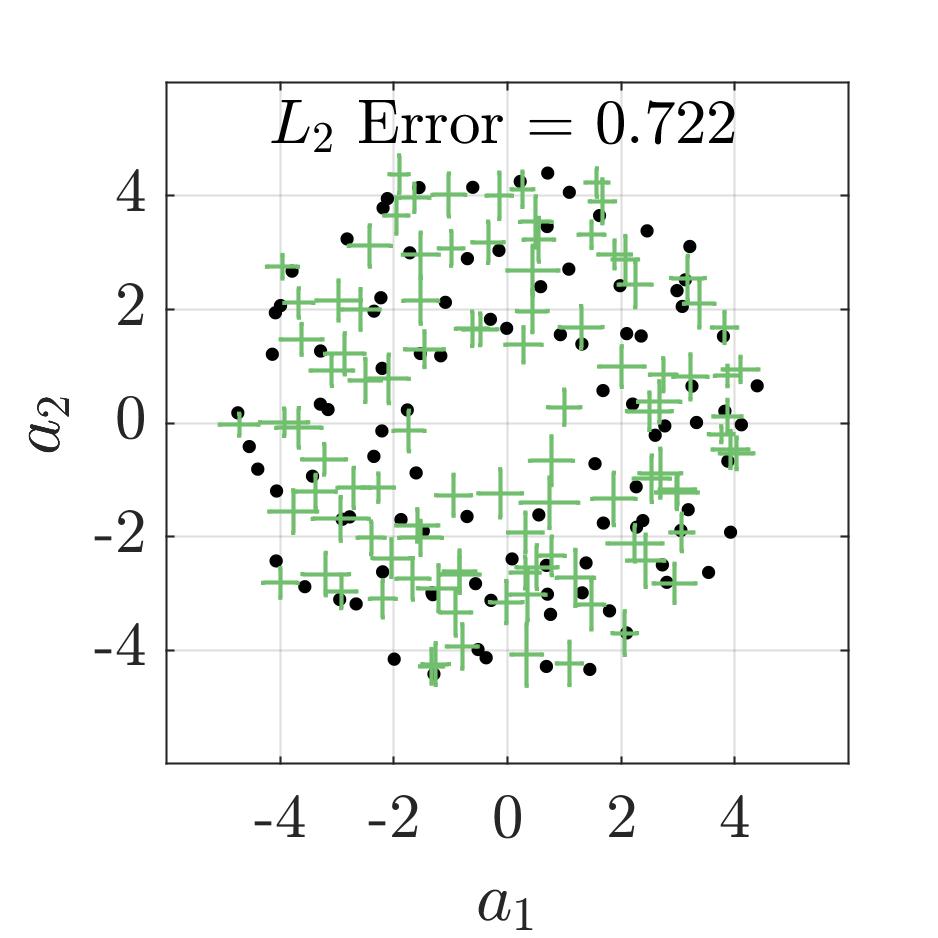}
    \subcaption{DNN}
    \end{subfigure}
    \caption{Phase relationship ($a_1$ versus $a_2$) obtained from three estimation methods \textcolor{black}{and the $L_2$ error associated with the first two modes} in the base-bleeding case. Black dots represents the target outputs from the DNS data.
    \textcolor{black}{The center, horizontal and vertical lines of the DNN marker represents the mean, standard deviation of $a_1$ and $a_2$ estimates from 20 trained neural networks with different initializations.}}
    \label{fig:ID75_phase}
\end{figure}

\begin{figure}
    \centering
    \includegraphics[width=.8\linewidth,trim={7cm 0cm 8cm 2cm},clip]{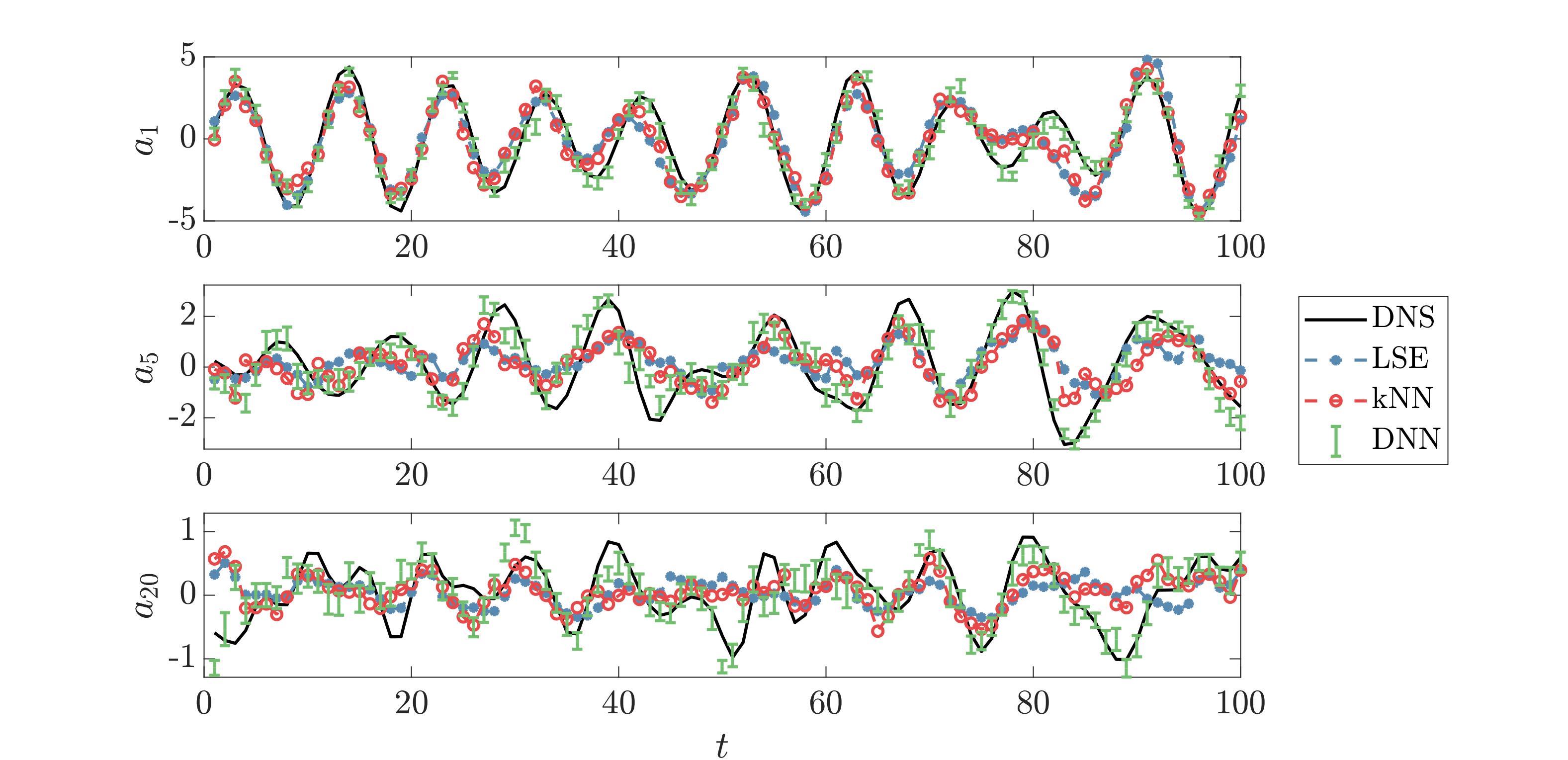}
    \caption{A comparison between estimated POD modal coefficients and the ground truth from DNS data in the stagnation control case.
    \textcolor{black}{Error bars of the DNN estimates represent the standard deviation from 20 trained neural networks with different initializations.}}
    \label{fig:ID75_pred}
\end{figure}

\begin{figure}
    \centering
    \begin{subfigure}{0.45\linewidth}
    \includegraphics[width=.99\linewidth,trim={0cm 8cm 1cm 12cm},clip]{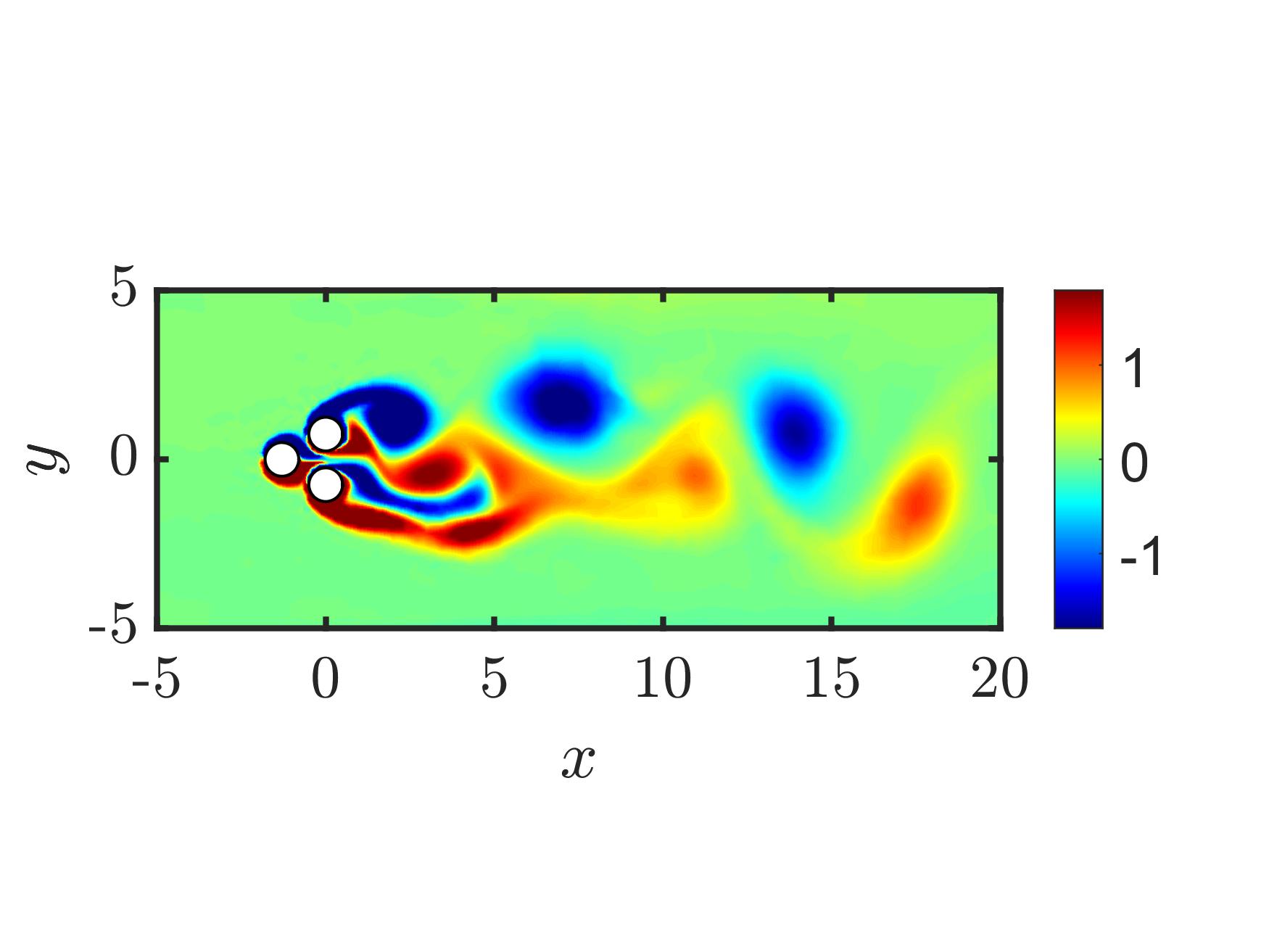}
    \subcaption{DNS}
    \end{subfigure}
    \begin{subfigure}{0.45\linewidth}
    \includegraphics[width=.99\linewidth,trim={0cm 8cm 1cm 12cm},clip]{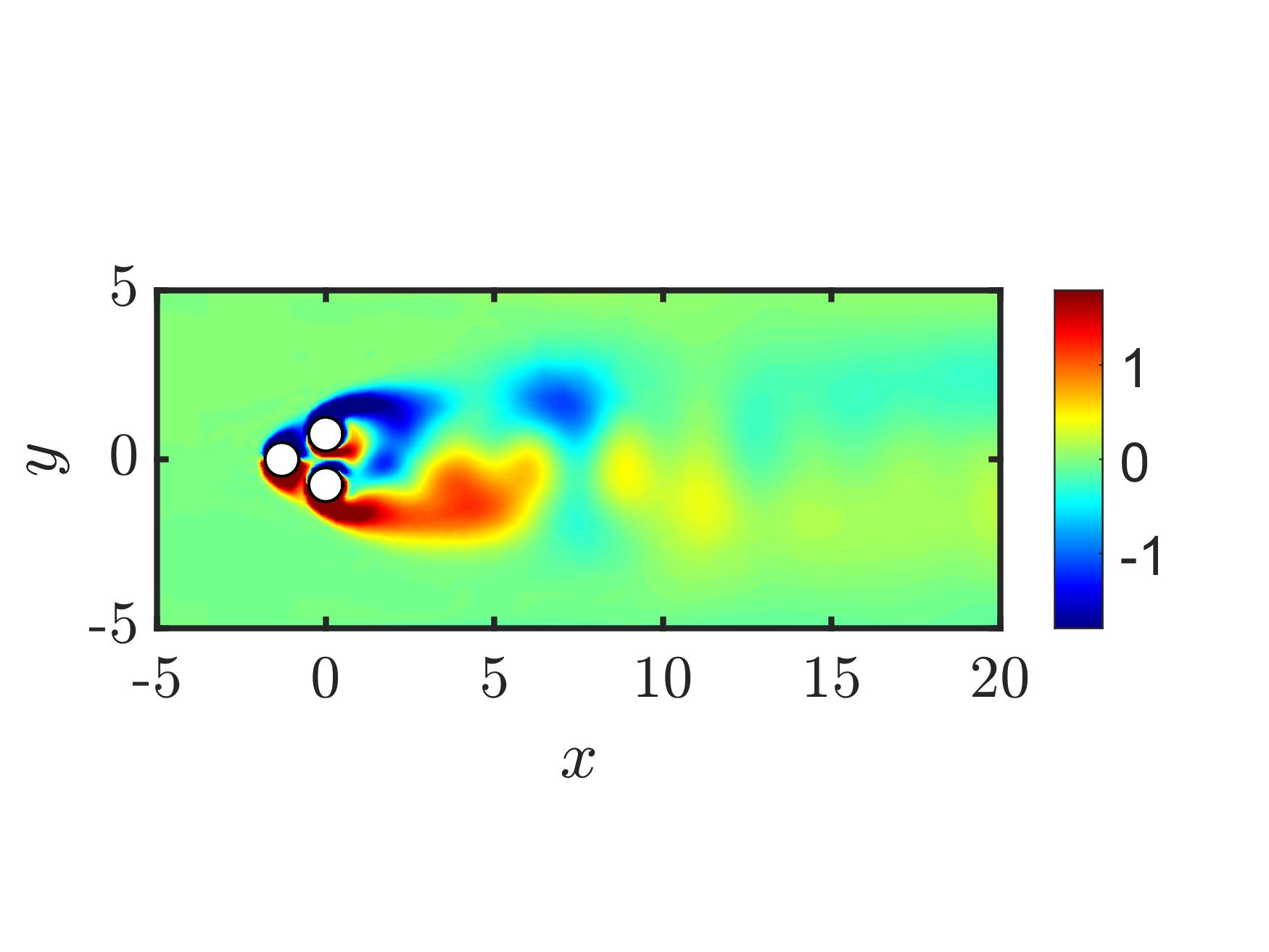}
    \subcaption{LSE}
    \end{subfigure}
    \begin{subfigure}{0.45\linewidth}
    \includegraphics[width=.99\linewidth,trim={0cm 8cm 1cm 12cm},clip]{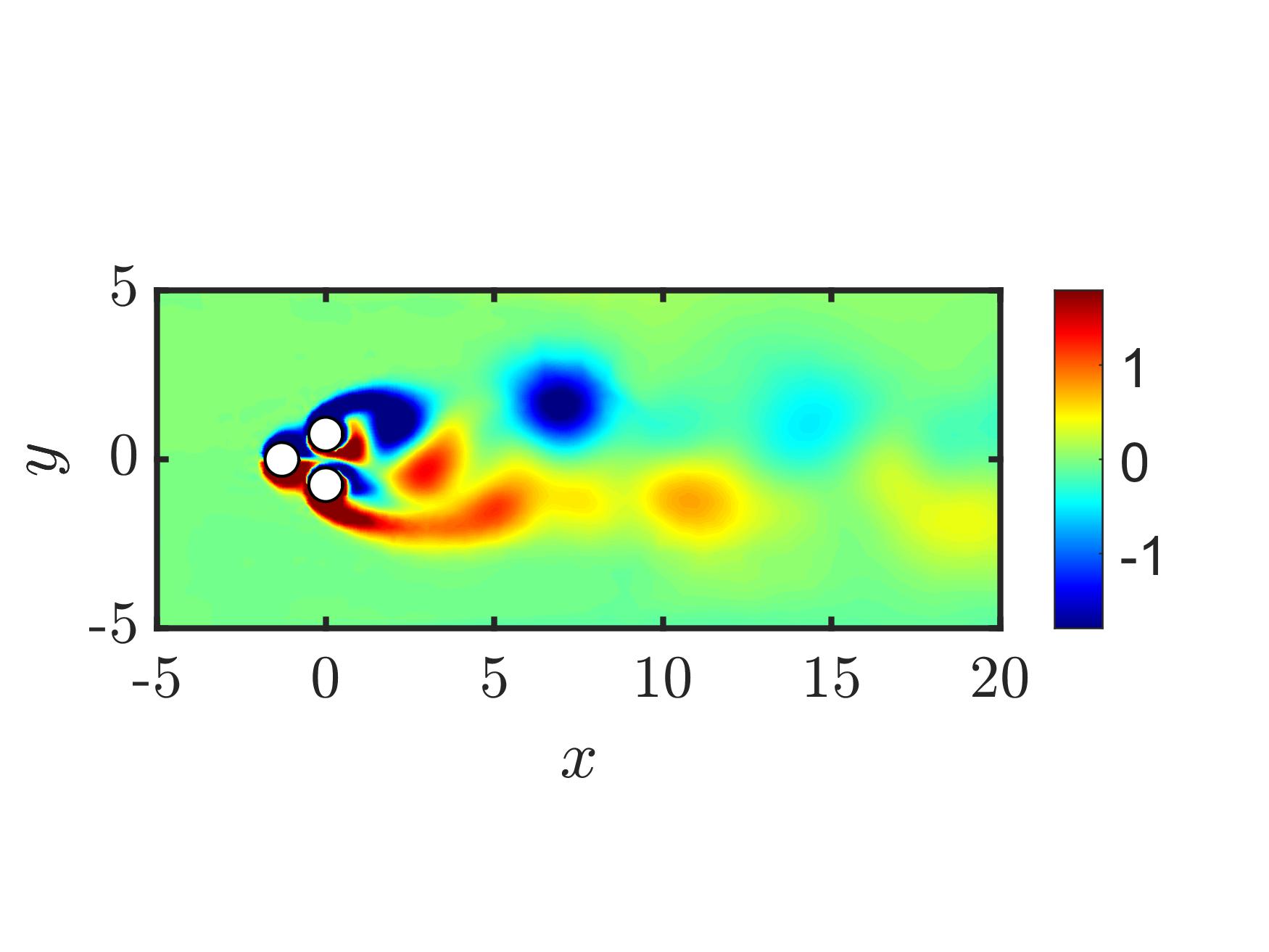}
    \subcaption{$k$NN}
    \end{subfigure}
    \begin{subfigure}{0.45\linewidth}
    \includegraphics[width=.99\linewidth,trim={0cm 8cm 1cm 12cm},clip]{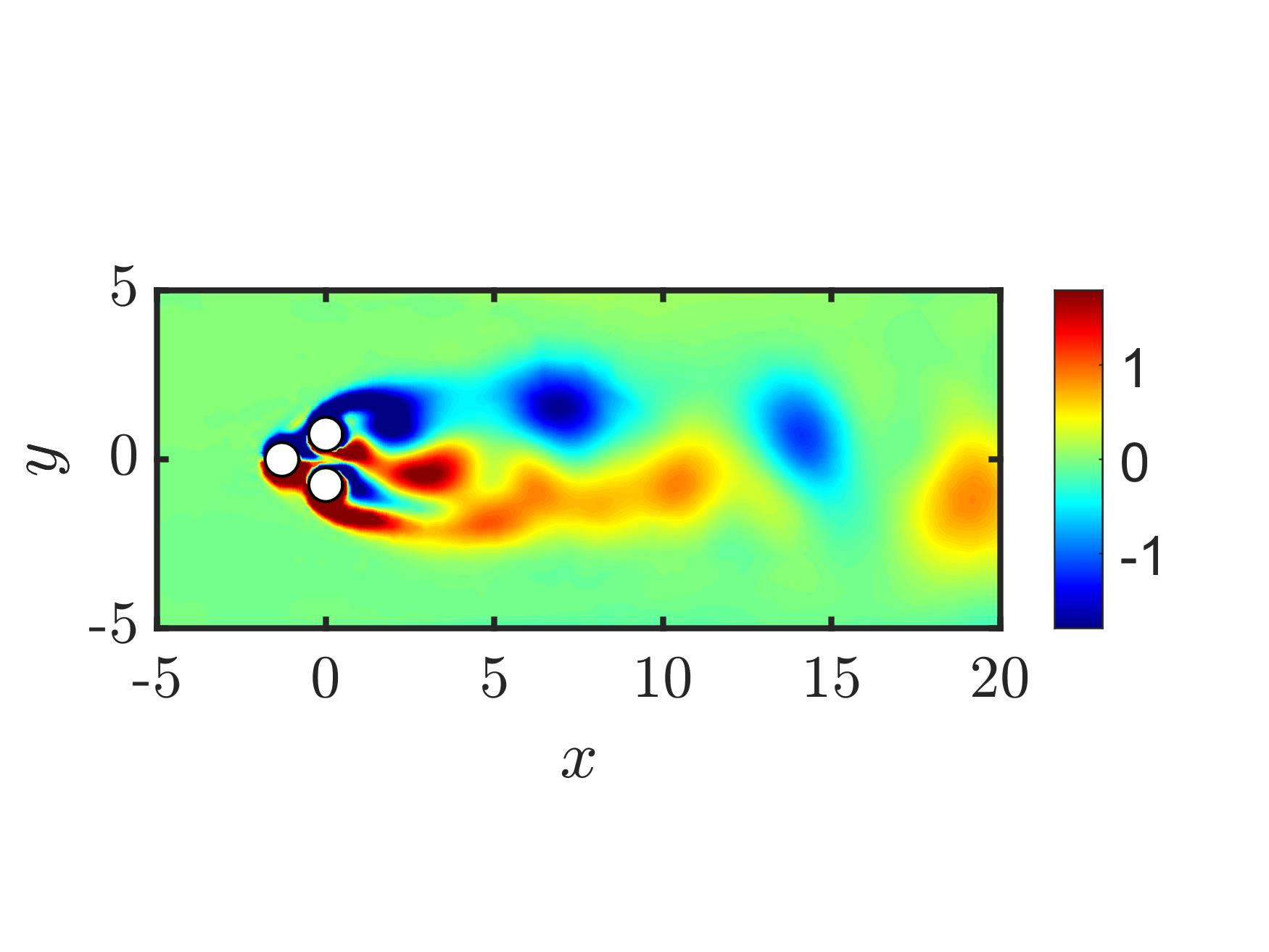}
    \subcaption{DNN}
    \end{subfigure}
    
    \caption{A comparison between the estimated instantaneous vorticity fields and the DNS data in the base bleeding case.
    \textcolor{black}{The DNN estimate is averaged from 20 trained neural networks with different initializations.}}
    \label{fig:ID75_recon}
\end{figure}

In the last example we examine a test case which relates to the base bleeding flow control mechanism.
For base bleeding, the counter-rotation of rear cylinders enables a streamwise fluidic jet in-between, thus attenuate the vortex shedding downstream. 
In this test case, the control command is $\pmb{b}= [-0.2382, -1.2296, 1.1534]^{\text{T}}$, from which a fully chaotic phase pattern is established in \textcolor{black}{\cref{fig:ID75_phase}}.
For the estimation of the first two POD modal coefficients, although both non-linear methods still consistently outperform LSE,
the estimation errors from $k$NN and DNN have been notably increased when comparing to the previous examples.
Comparatively, DNN provides a slightly lower estimation error than $k$NN in this case.
\Cref{fig:ID75_pred} presents the estimation of the POD modal coefficients over time in comparison to the ground truth.
For POD mode 1, results from all estimation methods can basically follow the trend of the ground truth.
However, for higher order modes the estimation curves can only roughly reflect the variation of the ground truth. 
Among these methods, the DNN estimation is more consistent to the DNS data than other two methods.
\textcolor{black}{In addition, the uncertainty of the DNN estimates in \cref{fig:ID75_phase,fig:ID75_pred} becomes more significant in comparison to the previous two cases, indicating the DNN estimate becomes more sensitive to the initialization in comparison to the previous two cases.}
The exemplary flow field estimation is presented in \textcolor{black}{\cref{fig:ID75_recon}}.
In this snapshot, the dominant flow feature of the DNS flow snapshot is the streamwise shedding of the smaller-scaled vortical structures generated by the base bleeding mechanism.
Yet, the LSE estimation can only provide a rough estimation of the flow field before $x=5$, and any flow structure appearing downstream can no longer be featured from the estimated flow field.
$k$NN provides a reasonable flow prediction before $x=8$.
Beyond this point, significant discrepancy from the DNS snapshot can be clearly observed.
Among these estimation methods, DNN can provide a more accurate depiction of the flow.
All dominant vortices in the DNS snapshot can be clearly identified from the DNN estimate.
However, one can still observe disparities from ground truth regarding to the shapes and intensities of downstream eddies in the flow.

\subsection{Overall evaluation of the estimation methods}\label{ssec:evaluation}
\begin{figure}
    \centering
    \includegraphics[width=.8\linewidth]{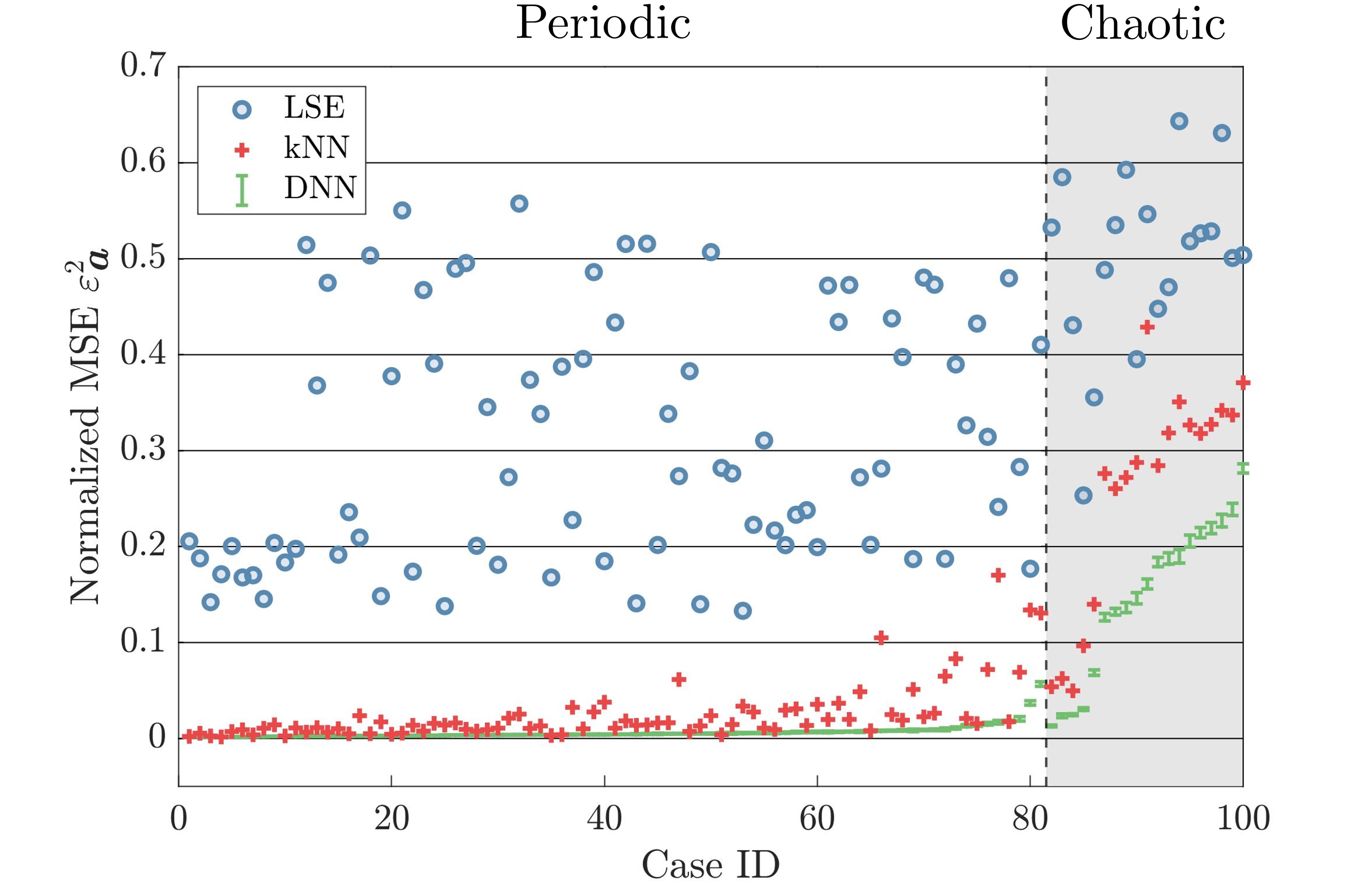}
    \caption{Distribution of normalized mean-squared-error ($\varepsilon^2_{\pmb{a}}$) for the prediction of POD expansion coefficients in all 100 test cases. The cases were classified into periodic (white background) and chaotic (gray background) categories. \textcolor{black}{In each category, case IDs were sorted based on the averaged DNN estimation error. The standard deviation over the 20 trained networks are denoted by the green error bars.}}
    \label{fig:pred_error}
\end{figure}

\begin{figure}
    \centering
    \includegraphics[width=.5\linewidth,trim={15cm 8cm 5cm 8cm},clip]{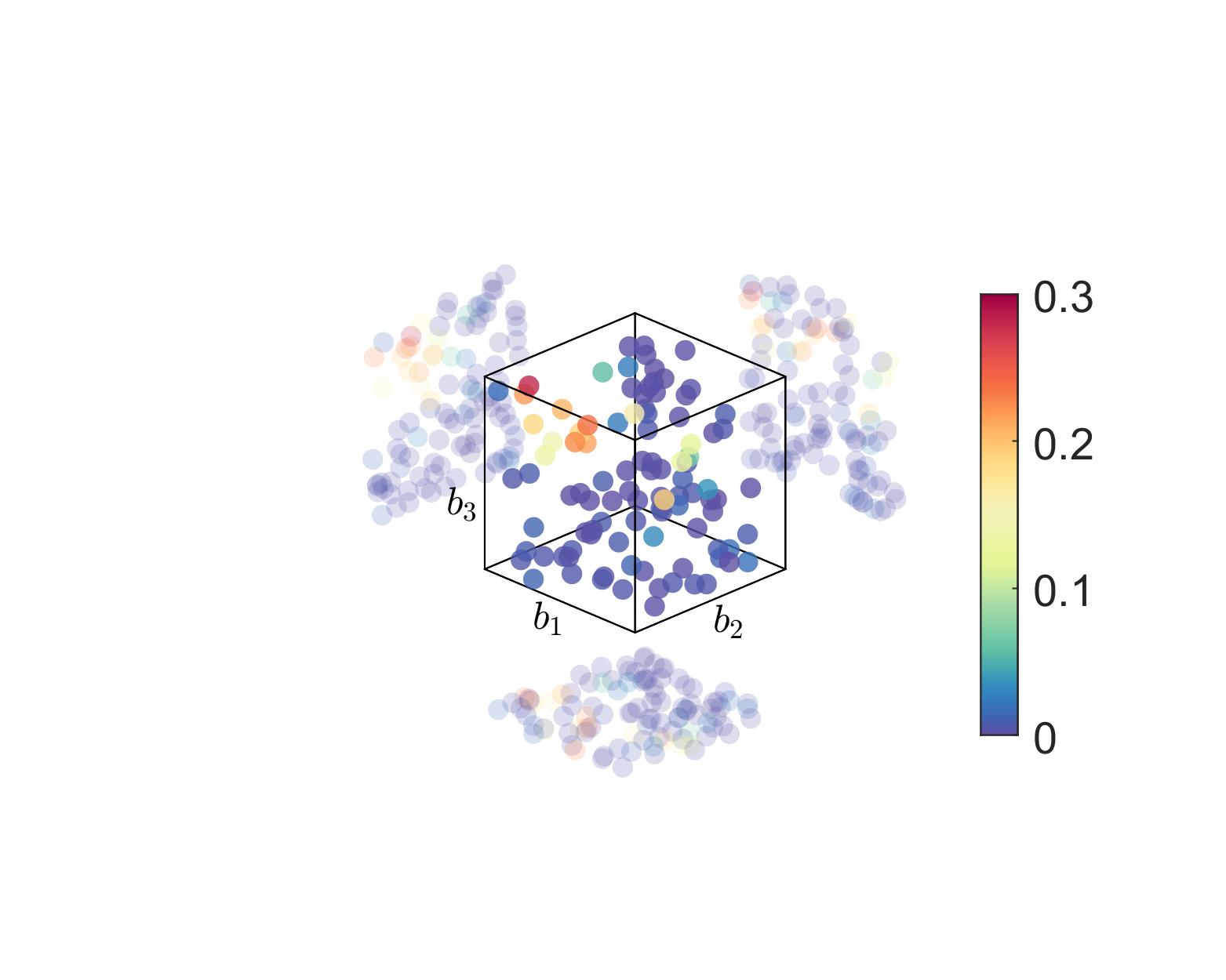}
    \caption{\textcolor{black}{Distribution of squared estimation error $E^2$ from DNN estimator in the space of control command $\bm{b}$. Error distribution is also projected onto the three orthogonal 2D planes for better visualization.}}
    \label{fig:error_b}
\end{figure}

Based on the analysis of the above-mentioned exemplary cases which correspond to specific flow control strategies, we evaluate the performance of the three estimators from all 100 randomly-generated control commands in the test set.
As introduced in \cref{sec:plant}, each test case is composed of 100 consecutive time instances.
For each estimator, the squared estimation error of POD modal coefficients, $\varepsilon_{\pmb{a}}^2$, is calculated from the flow estimates and the results are summarized and presented in \textcolor{black}{\cref{fig:pred_error}}.
For better comparison, the test cases are classified into periodic and chaotic categories according to the classification criteria described in \cref{ssec:s+b}, and all cases in each category are sorted by the error magnitude of DNN in an ascending order.
As an analytically simple approach, LSE estimates always lead to relatively larger error than the other two non-linear approaches. 
This observation confirms the appearance of strong non-linearity in the control-oriented flow estimation problem.
In most periodic cases, both $k$NN and DNN can achieve accurate estimation of the POD modal coefficients with $\varepsilon_{\pmb{a}}^2<0.03$.
Nevertheless, a few outliers from $k$NN can be identified from the figure. 
As the $k$NN estimation relies heavily on the neighborhood data, the estimation result can be easily biased when the first $k$th neighbors are distant from the input data or when a steep variation of the mapping function $\mathcal{F}$ occurs in the neighborhood.
On the other hand, by optimizing a generalized model from the database, DNN is more resistant to these challenges and is capable to provide a more robust estimation result. 
\textcolor{black}{Also, the influence of random initialization to the estimation performance is nearly negligible in all periodic cases.}
\textcolor{black}{When the flow becomes chaotic, the DNN estimator become more sensitive to different initialization conditions with a increased level of uncertainty.
However, the limited level of uncertainty still confirms the robustness of DNN estimation in chaotic cases.}
In the chaotic regime, the estimation errors from both DNN and $k$NN are seen to increase in comparison to the periodic cases.
The increased level of estimation error indicates the appearance of more complex vortical structures generated from the cylinder rotations. 
These structures makes the estimators more difficult to identify the correct flow features from the point-wise sensor measurements.
\textcolor{black}{Nevertheless}, DNN shows its advantage on the consistently better estimation of the chaotic forced flow.
A comparison of the squared overall estimation error $E^2$ from all estimators is included in \textcolor{black}{\cref{tab:conclusion}}. 
This comparison includes the errors from all test cases, as well as the periodic and chaotic components.
\textcolor{black}{In agreement} with the previous analysis, the quantitative assessment of $E^2$ confirms (1) the increased difficulty to accurately estimate the flow state under chaotic cases, (2) the improvement of estimation accuracy using non-linear estimators, and (3) the consistently better performance of DNN.
\textcolor{black}{The standard deviation of $E^2$ from all 20 trained neural networks are 0.00036, 0.00015, and 0.0016 $s$ for all test cases, periodic cases, and chaotic cases, respectively.
Hence the influence of weights initialization to the overall estimation error is negligible in DNN estimation.}
\textcolor{black}{To help understanding the error distribution regarding to different control commands, \cref{fig:error_b} displays the relationship between the control commands and their associated estimation error $E^2$ in the test set. 
Here we only present the error associated with the DNN estimator as it provides a consistently better result.
Consistent to the previous discussion, almost all control commands having relatively high estimation errors locate inside the chaotic region displayed in \cref{fig:LHS}. 
An increased sampling density of the training dataset in the chaotic region is able to further improve the overall estimation accuracy.}

\begin{figure}
\centering
\begin{subfigure}{0.3\linewidth}
    \includegraphics[height=4cm,trim={6cm 0cm 9cm 0cm },clip]{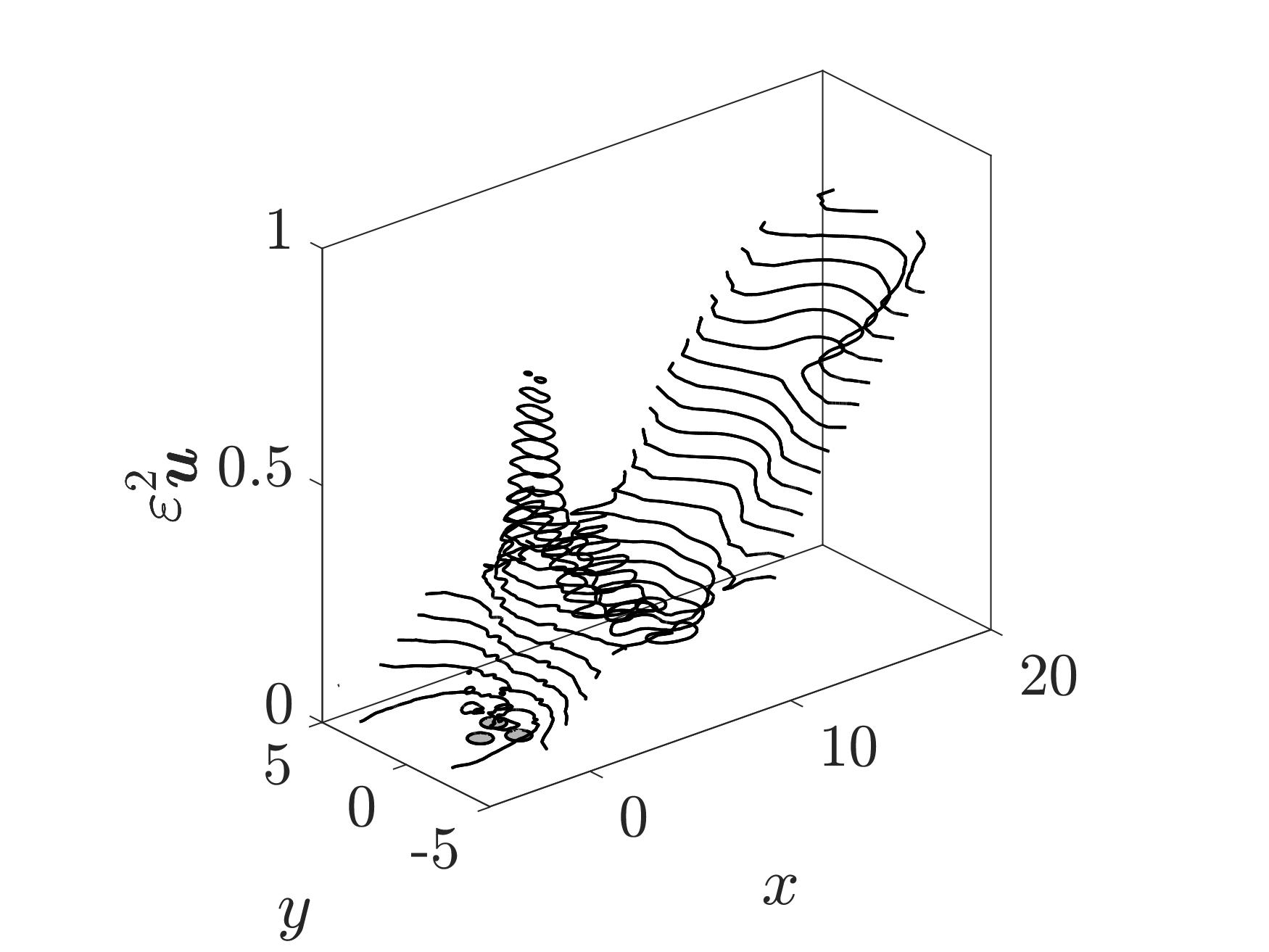}
    \caption{LSE}
\end{subfigure}
\begin{subfigure}{0.3\linewidth}
    \includegraphics[height=4cm,trim={6cm 0cm 9cm 0cm },clip]{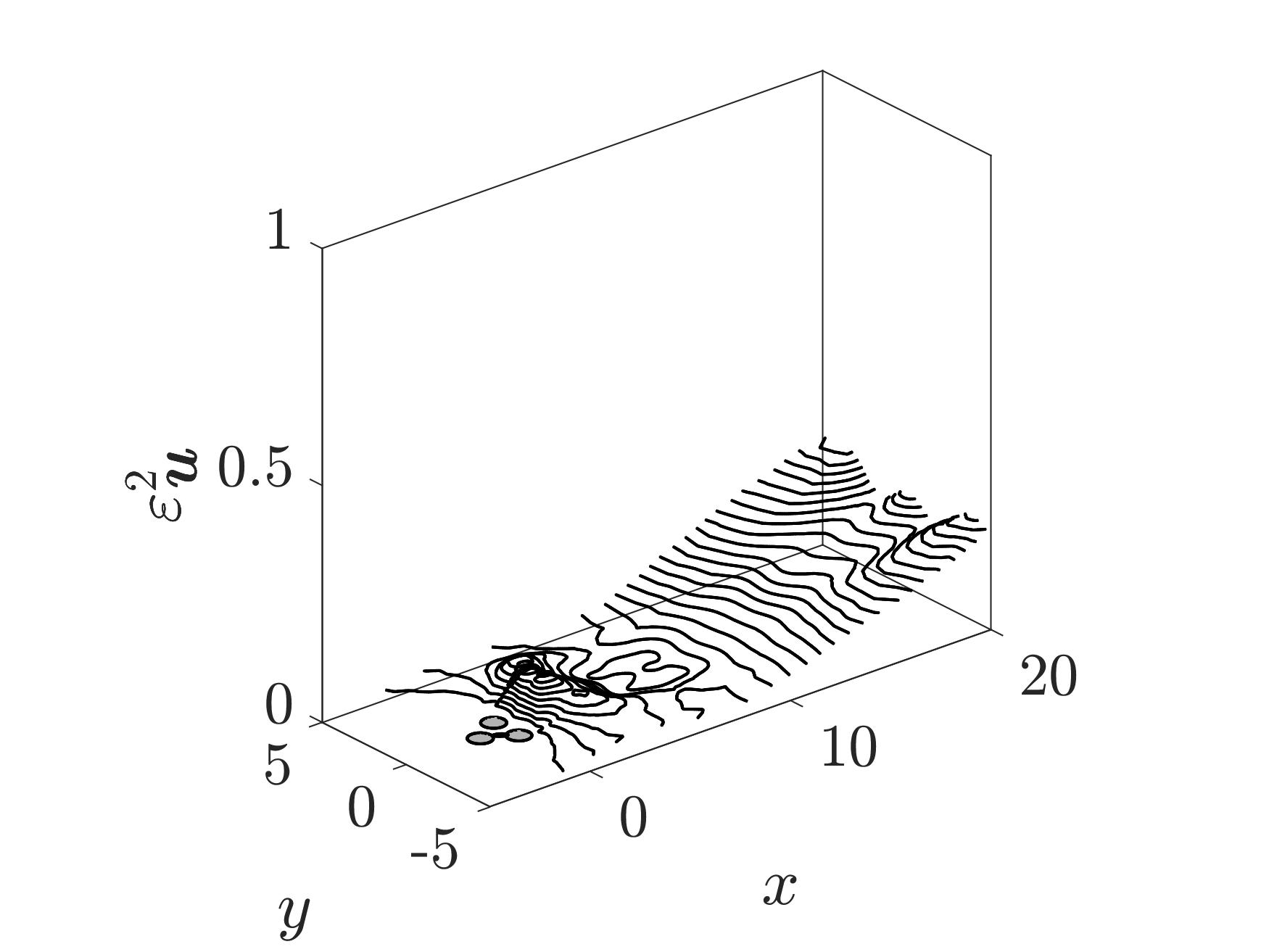}
    \caption{$k$NN}
\end{subfigure}
\begin{subfigure}{0.3\linewidth}
    \includegraphics[height=4cm,trim={6cm 0cm 9cm 0cm },clip]{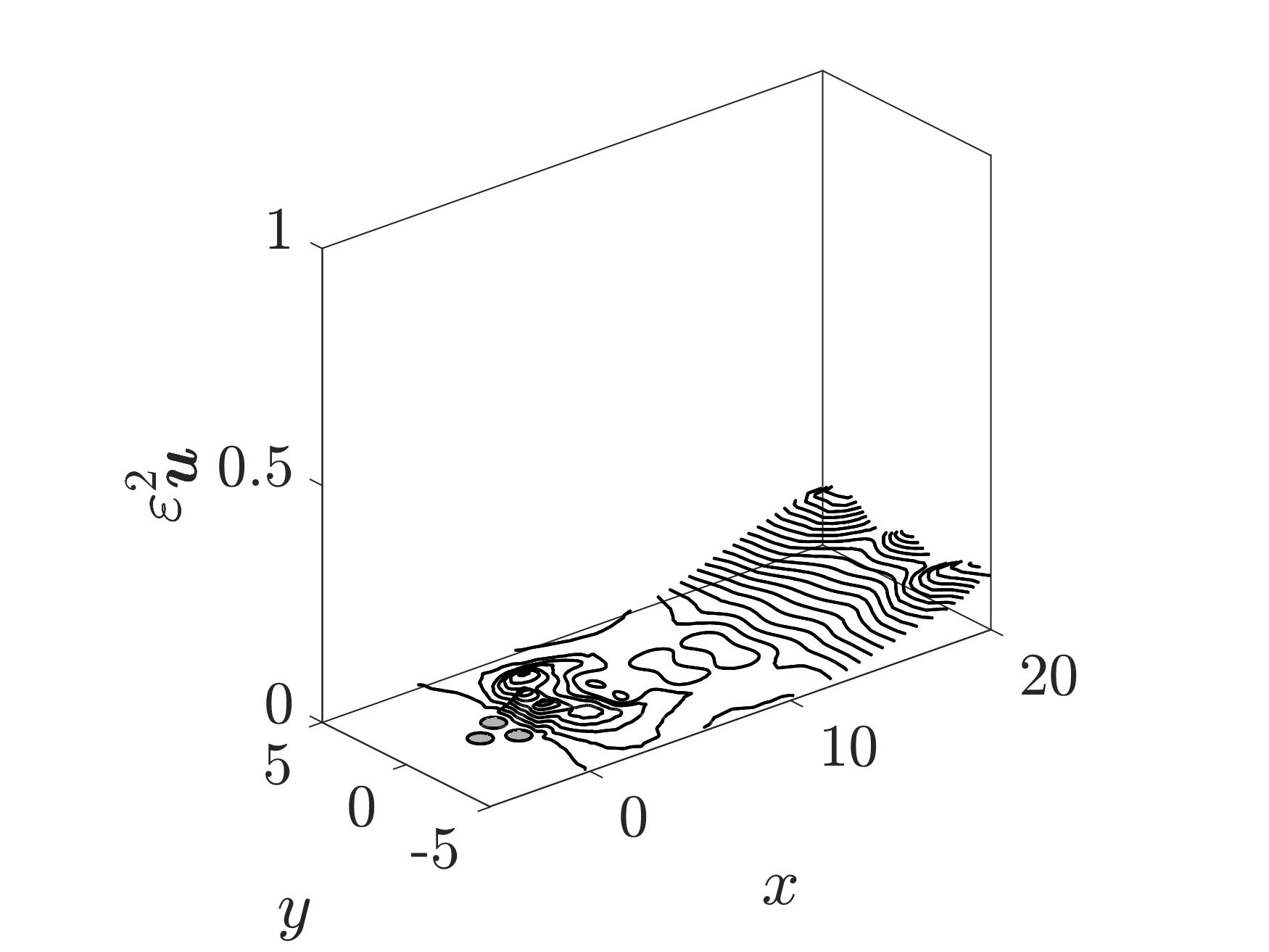}
    \caption{DNN}
\end{subfigure}\\
\caption{The squared normalized velocity estimation error ($\varepsilon^2_{\pmb{u}}$) from each method.}
\label{fig:error_vel}    
\end{figure}

\Cref{fig:error_vel} visualizes the distribution of the squared normalized velocity estimation error $\varepsilon^2_{\pmb{u}}$ in the spatial domain.
The error distribution in space not only evaluates the relative performance of each method, but can also justify the ideal locations of additional sensors once the estimation performance are required be further improved.
Among all estimation methods, LSE generates the the most significant estimation error in the flow domain, which demonstrates its naiveness to handle the complex non-linear problem.
The most significant estimation error \textcolor{black}{from} LSE resides in two regions, one just behind the cylinders and the other one further downstream after $x=10$.
The dominant error source from both $k$NN and DNN resides in the downstream region after $x=13$.
Comparatively the estimation error from DNN is slightly lower than that of $k$NN thanks to the better performance in the chaotic regime according to \textcolor{black}{\cref{fig:pred_error}}. 
Since all in-flow sensors are deployed between $5<x<8$, the flow structures can be better identified from the sensor measurements. 
As a result, the estimation error inside this region remains at a relatively low level for all estimation methods.
\textcolor{black}{In a similar manner, the estimation performance for chaotic flows can also be further improved by deploying addition sensors after $x=10$ to obtain additional input information.
Same principle can be used in real-life applications to improve the estimation accuracy in which the flow will be more complex than this 2D benchmark case.}

Besides the error assessment of each method, \textcolor{black}{\cref{tab:conclusion}} also comprehensively concludes the characteristics of three estimators from multiple aspects. 
This parametric study serves to highlight the relative merits of each method.
The comparison includes the complexity (i.e., the number of tunable variables in each method), as well as the time costs.
All comparison horizons are evaluated using the same workstation which has two 2.25 GHz AMD 64-core processors with paralleled computation.
The LSE algorithm is implemented on MATLAB, and for the two machine learning methods, python with scikit-learn \textcolor{black}{\citep{scikit-learn}} and tensorflow-cpu \textcolor{black}{\citep{tensorflow2015}} packages are utilized to achieve the estimation.
Regarding to the number of tunable parameters, LSE results in a total of 936 parameters which correspond to all elements in the linear transfer matrix $T_{ij}$ in \textcolor{black}{\cref{eqn:lse_1}}.
The only tunable parameter in $k$NN is the number of observations $k$ in the neighborhood.
As previously mentioned the value of $k$ is determined via a convergence test in this work (see \textcolor{black}{\cref{fig:kNN_convergence}}).
Given the complexity of the nonlinear neural network architecture, DNN possesses the largest number of parameters (154,958) which corresponds to the weights and biases from all artificial neurons.

In addition, time consumption of each estimation method is listed in \textcolor{black}{\cref{tab:conclusion}}.
Here both offline and online time costs are proposed and counted \textcolor{black}{in terms of CPU time}.
The offline time represents the time cost to optimize the tunable parameters, and the online time cost stands for the computational time to obtain an output estimate from \textcolor{black}{the test dataset}.
For LSE, the offline time is equivalent to the time consumption solving for $T_{ij}$, which is about 0.01 s starting from the calculation of correlation matrices in \textcolor{black}{\cref{eqn:lse_2}}.
\textcolor{black}{$k$NN represents a model-free approach and only relies on the observations from the neighboring observations, no training is necessary for this method.
However, in the practical implementation, some preliminary data-processing steps (such as database sorting) are executed in the scikit-learn package before the estimation stage. 
We count the corresponding pre-processing time as the offline time cost of $k$NN in this table while reminding readers the training-free nature of this method.}
The offline training of DNN involves the iterative optimization of the loss function based on the randomly-partitioned mini-batches, and this process takes around \textcolor{black}{3,000 CPU seconds} with a total of 500 training epochs.
The addition of GPU may alleviate the training time but the expected training time is still much longer than the other two methods.
However, once the DNN model is well trained, the online estimation can be finished within a short time.
In this study the online estimation of DNN can be realized \textcolor{black}{within 0.4 s CPU time}.
\textcolor{black}{With model compression techniques, such as the optimization of DNN architecture \citep{Ramchoun2016} or symbolic regression \citep{Cranmer2020}, the DNN estimation can be further accelerated and can be potentially applied to feedback flow control as described in \citet{Samimy2007}}.  
\textcolor{black}{$k$NN represents the most data-intensive mapping, in a sense that the training dataset need to be kept during the estimation.}
As $k$NN requires intensive computation to search for the neighborhood observations from the huge database, the corresponding online estimation time becomes significantly longer than the other methods.
In this study the online cost of $k$NN is about 40 s, and 
we can conclude that this data-intensive method can be hardly adopted by any closed-loop control application which requires instantaneous feedback.
\textcolor{black}{For the purpose of reference, the break-even point in number of calls to $k$NN and DNN estimators in terms of CPU cost is 52.}
Although the online time cost for LSE is only 0.008 s,
the linear assumption makes this method still difficult to be adopted by any real applications which require promising estimation accuracy.

\begin{table}
\def~{\hphantom{0}}
\begin{center}
\begin{tabular}{m{1.8cm}P{3cm}P{3cm}P{3cm}P{0.5cm}}
Method&LSE&$k$NN&DNN&\\ \hline
{Schematic}&
\multicolumn{4}{m{11cm}}{
\begin{center}
\begin{minipage}{11cm}
\vspace{0.5em}

\raisebox{-1.1\height}{\includegraphics[height=2.2cm]{figures/lse_new.jpg}}\hspace{0.8cm}
\raisebox{-1.1\height}{\includegraphics[height=2.4cm]{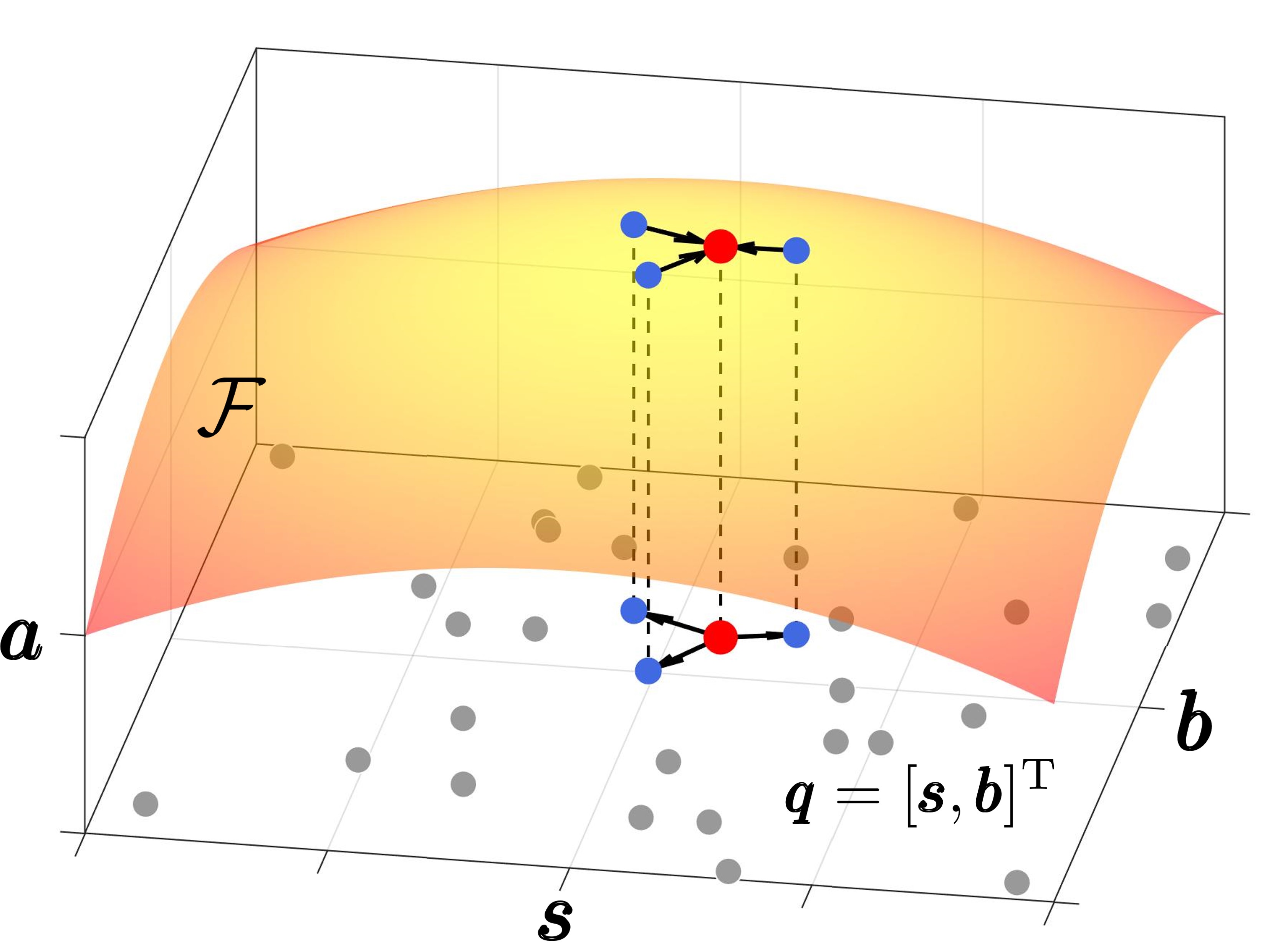}}\hspace{0.5cm}
\raisebox{-1.1\height}{\includegraphics[height=2cm]{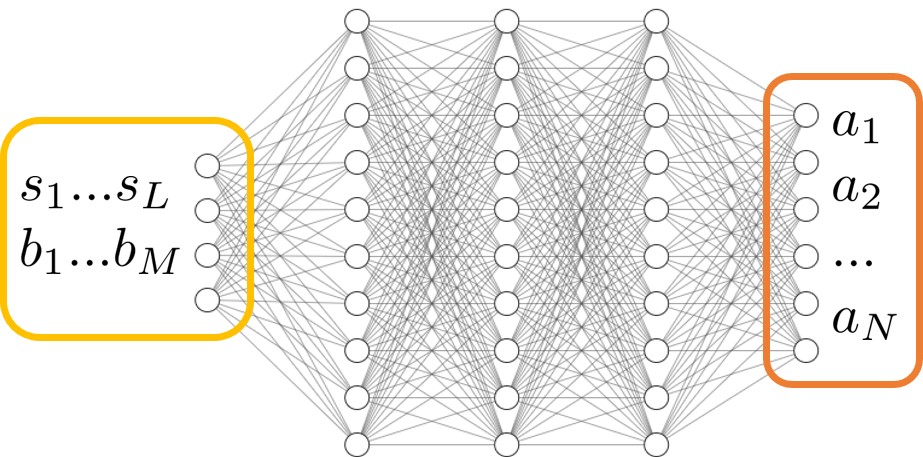}}

\vspace{0.5em}
\end{minipage}
\end{center}
}
\\\hline

Number of parameters&
\multicolumn{4}{c}{
\begin{minipage}{11cm}
\begin{center}
\begin{tikzpicture}
\begin{axis}[
    xlabel={},
    symbolic x coords={LSE,kNN,DNN},
    xtick=data,
    axis x line*=bottom,
    ymode=log,log origin=infty,
    width=11cm, 
    height=3cm,
    axis y line*=right,
    ytick pos=right,
    ylabel={},
    y tick label style={anchor=west,color=black,xshift=0pt},
    ymax = 200000,
    ymin = 1,
    enlarge x limits=0.25,
    ybar=0 pt,
    bar width=30pt,
    hide x axis,
    nodes near coords={\pgfmathprintnumber[fixed, precision=3]{\pgfplotspointmeta}},
    point meta=rawy,
    legend style={at={(0.5,-1em)}, draw=none, anchor=north, legend columns=0,legend  style={/tikz/every even column/.append style={column sep=0.5cm}}}
]

\addplot[orange,fill=orange!30!white,mark=none] coordinates {(LSE, 936) (kNN, 1)  (DNN, 154958)};


\legend{}
\end{axis}
\end{tikzpicture}
\end{center}
\vspace{1em}
\end{minipage}
}\\\hline

\textcolor{black}{CPU time cost (s)}&
\multicolumn{4}{c}{
\begin{minipage}{11cm}
\begin{center}
\begin{tikzpicture}
\begin{axis}[
    xlabel={},
    symbolic x coords={LSE,kNN,DNN},
    xtick=data,
    hide x axis,
    ymode=log,log origin=infty,
    ymin = 0.005,
    ymax = 5000,
    width=11cm, 
    height=3cm,
    axis y line*=right,
    ytick pos=right,
    ylabel={},
    y tick label style={anchor=west,color=black,xshift=0pt},
    ytick={1e-2,1e0,1e2,1e4},
    enlarge x limits=0.25,
    ybar=0 pt,
    bar width=20pt,
    nodes near coords={\pgfmathprintnumber[fixed, precision=3]{\pgfplotspointmeta}},
    point meta=rawy,
    legend style={at={(0.5,-1em)}, draw=none, anchor=north, legend columns=0,legend  style={/tikz/every even column/.append style={column sep=0.5cm}}}
]

\addplot[darkgray, fill=darkgray!30!white, mark=none] coordinates {(LSE, 0.01) (kNN, 29)  (DNN, 3000)};
\addplot[magenta, fill=magenta!30!white, mark=none] coordinates {(LSE, 0.008) (kNN, 40) (DNN, 0.40)};


\legend{Offline,Online}
\end{axis}
\end{tikzpicture}
\end{center}
\end{minipage}
}\\\hline

Squared overall estimation error $E^2$&
\multicolumn{4}{c}{
\begin{minipage}{11cm}
\begin{center}
\begin{tikzpicture}
\begin{axis}[
    xlabel={},
    symbolic x coords={LSE,kNN,DNN},
    xtick=data,
    hide x axis,
    ymin = 0,
    ymax = 0.6,
    width=11cm, 
    height=3cm,
    axis y line*=right,
    ytick pos=right,
    ylabel={},
    y tick label style={anchor=west,color=black,xshift=0pt},
    enlarge x limits=0.25,
    ybar=0 pt,
    bar width=25pt,
    nodes near coords={\pgfmathprintnumber[fixed, precision=3]{\pgfplotspointmeta}},
    legend style={at={(0.5,-1em)}, draw=none, anchor=north, legend columns=0,legend  style={/tikz/every even column/.append style={column sep=0.5cm}}}
]

\addplot [teal, fill=teal!30!white, mark=none,error bars/.cd, y dir=both,y explicit] 
coordinates {(LSE, 0.36884)
(kNN, 0.07287)
(DNN, 0.0360) 
};

\addplot [blue, fill=blue!30!white, mark=none,error bars/.cd, y dir=both,y explicit] 
coordinates {(LSE, 0.32197) 
(kNN, 0.02178) 
(DNN, 0.0061) 
};

\addplot [red, fill=red!30!white, mark=none,error bars/.cd, y dir=both,y explicit] 
coordinates {(LSE, 0.56865) 
(kNN, 0.29068) 
(DNN, 0.1632) 
};


\legend{All,Periodic,Chaotic}
\end{axis}
\end{tikzpicture}
\end{center}
\end{minipage}
}\\\hline

Squared error distribution $\varepsilon_{\pmb{u}}^2$&
\multicolumn{4}{c}{
\begin{minipage}{11cm}
\raisebox{-1.1\height}{\includegraphics[width=\linewidth,trim = {1cm 0cm 0cm 1cm},clip]{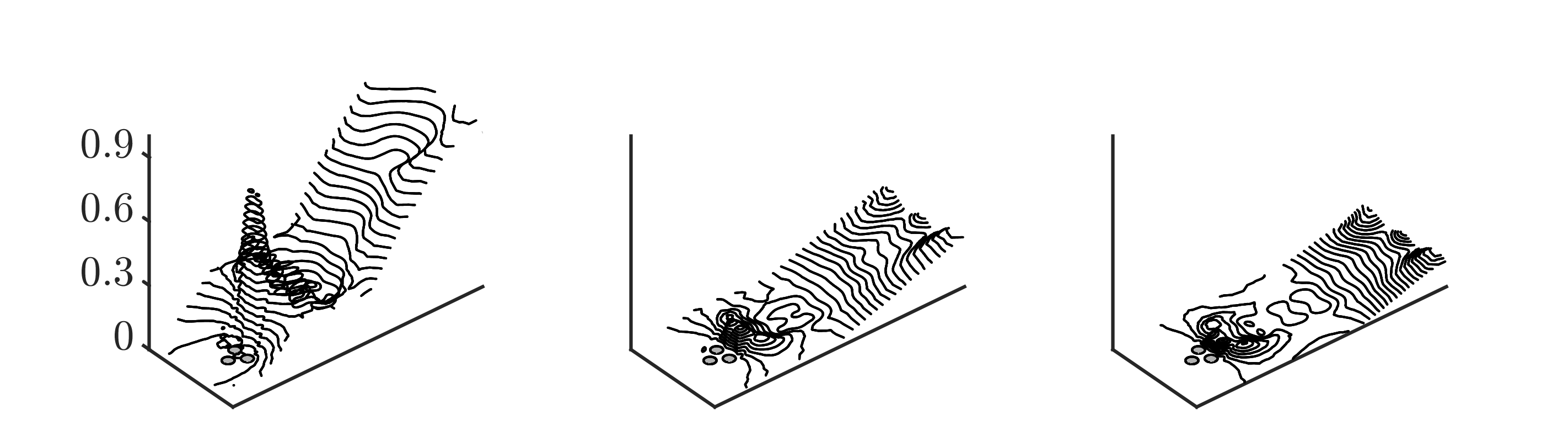}}
\vspace{1em}
\end{minipage}}

\\\hline
Advantage & \multicolumn{1}{l}{\hspace{.3cm}Human interpretable} & \multicolumn{1}{l}{\hspace{.6cm}Intuitive, no training} & \multicolumn{1}{l}{\hspace{0.9cm}Globally non-linear} &\\\hline
Disadvantage &\multicolumn{1}{l}{\hspace{.3cm}Large error} & \multicolumn{1}{l}{\hspace{.6cm}Data-intensive} &\multicolumn{1}{l}{\hspace{.9cm}Significant training} &\\
\end{tabular}
\caption{A comprehensive comparison of all estimation methods from different horizons.}
\label{tab:conclusion}
\end{center}
\end{table}

\section{Conclusions and Outlook}
\label{sec:conclusions}

We propose the first machine-learned, control-oriented flow estimation for multiple-input, multiple-output (MIMO) plants.
In this method, the input vector are the MIMO signals,
i.e.\ comprise the sensor signals and actuation commands. and output is the instantaneous flow field.
Starting point is a data base from simulations
with  representative steady actuation commands,
sensor signals and simultaneously recorded flow snapshots.
The estimator mapping  employs 
three representative data-driven methods:
an analytically simple, a data-centric and a general nonlinear approach.
The analytically simple method adopts Linear Stochastic Estimation (LSE) with the addition of the control commands to the input vector.
The data-centric method employs 
$k$ Nearest Neighbors ($k$NN) regression 
which interpolates the flow state based on closest MIMO signals in the data base.
The general non-linear mapping function is established from the Deep Neural Network (DNN), and the \textcolor{black}{weights and biases} are optimized from the training data.

As benchmark the  fluidic pinball is chosen \citep{Ishar2019,Maceda2021,Li2022}.
This two-dimensional configuration 
consists of three parallel cylinders of same diameter
on an equilateral triangle pointing upstream to the uniform flow.
The flow may be manipulated by the cylinder rotations,
i.e.\ three actuation commands.
The flow is monitored by 9 downstream velocity sensors.
The reference Reynolds number is 100 based on cylinder diameter.
The unforced flow exhibits 6 different Navier-Stokes solutions,
including two stable statistically asymmetric periodic vortex sheddings 
\cite{Deng2020jfm,Deng2021}.
The steady rotations of cylinders lead to rich flow dynamics,
e.g. steady, periodic, quasiperiodic and chaotic behaviour
with eddies at various scales.
The flow estimation problem has a large dynamic complexity
despite geometric simplicity.
With the application of three estimation methods to the fluidic pinball plant, we conclude the key results in the following.

\textcolor{black}{
\Cref{tab:conclusion} concludes the major characteristics of three machine-learning estimation methods. 
As a widely-adopted linear reference point, LSE, helps to identify the degree of non-linearity contained in the fluidic pinball plant under different open-loop actuation commands.
In comparison to non-linear estimators,
LSE leads to the highest estimation error albeit being computationally cheap.
}
In all exemplary cases, this method fails to faithfully capture the phase relationship regarding the downstream vortex shedding.
From the instantaneous flow fields, LSE estimates can hardly identify the correct locations and intensities of the large scale vortical structures in the flow.
\textcolor{black}{
The data-centric ($k$NN) and globally non-linear (DNN) estimators have distinctly lower estimation errors.
These observations confirm the appearance of strong non-linearity in the estimation mapping function when the flow is under different control commands.
}

Both nonlinear estimators based on $k$NN and DNN 
are capable to provide more accurate flow field estimates 
than the generalized linear stochastic estimation.
The nonlinear estimators exhibit similar accuracy
for forced periodic dynamics.
DNN performs consistently better than $k$NN
for forced chaotic dynamics
albeit at lower accuracy than for periodic flow.
DNN seems to have learned 
a mapping which extrapolates better than $k$NN.
\textcolor{black}{Apart from the improved accuracy, 
each nonlinear estimator has its distinct merits.
$k$NN possesses the simplest architecture and no training phase is required.
In contrast, 
the network model in DNN is relatively complicated, 
and requires off-line training.
Yet, the online estimation of DNN is 100 times faster
than $k$NN.
This makes DNN more favorable than $k$NN in time-critical engineering applications.}

The feasibility of the machine-learned, control-oriented flow estimation suggest exciting future directions.
For instance, the noise filtering capability
of Kalman filters of (N)ARMAX methods 
may be integrated by lifting
the sensor signals to feature vectors with time-delay coordinates \citep{Loiseau2018jfm}.
Similarly, unsteady actuation may  also be incorporated
by including time-delay coordinates of the actuation commands.
\textcolor{black}{From then on, the control-oriented flow estimation approaches can be integrated into closed-loop control algorithms.}
\textcolor{black}{The addition of time-delay coordinates also allows an alleviation of the computational cost of the training dataset.
The method to generate the training dataset under time-varying control laws has been detailed in \cite{Bieker2020}.}

In addition, the low-dimensional representation of the flow field will become more efficient by a wide spectrum of dimensionality reduction techniques. 
\textcolor{black}{A striking example is the 
transient oscillatory cylinder dynamics which requires  50 POD modes to resolve a two-dimensional manifold \citep{Loiseau2018jfm}.
Another example is an ensemble of actuated flows: POD modes of all data is less efficient than first clustering the flow snapshots followed by cluster-based (local) POD bases  --- similar to \cite{Hess2019} for bifurcating flows. Following this path, 
reduced-order models based on centroidal Voronoi tessellations (CVT, \citealt{Burkardt2006a, Burkardt2006b}) constitute an alternative to a monolithic POD model.
The Isomap represents another non-linear dimensionality reduction technique which embeds flow snapshots into a low-dimensional manifold \citep{Farzamnik2022}.
}
\citet{Fukami2020} and \cite{Murata2020} have reported that non-linear modal decomposition methods based on Convolutional Neural Network (CNN) and Autoencoder (AE) are capable to extract flow features in a more efficient manner than the POD-based methods.
The latent space of AE is the compact representation of the full-state flow field.
Different from POD which decomposes the flow as spatially distributed flow structures,
AE explores the low-dimensional feature vectors of the flow. 

The combination between control-oriented flow estimation and non-linear reduced-order modelling will lead to a more compact output vector and accelerate the estimation procedure correspondingly.

For $k$NN and DNN, their implementations in this study follow the most generally adopted architectures.
However, the estimation accuracy can be further improved.
\textcolor{black}{The inflexible assignment of the $K$ value in \cref{eqn:kNN} and the intensive computational cost of searching for neighbors can be cured by the adaptive selection of $K$ \citep{Zhang2018}, as well as the clustering-based searching approach \citep{Deng2016}}.
The accuracy of  DNN may be improved  with first principles, i.e. the governing equations,
following Physical-Informed Neural Networks (PINN, \citealt{Raissi2019}).
This type of network assimilates the flow estimation and the governing equations of the investigated flow and is capable to enhance the estimation accuracy.

Finally,  this method can be easily extended to experimental flow control plants  under steady and periodic actuation.
In general, the control commands can be easily adapted to actuation parameters of the most commonly used actuators in engineering applications \citep{Cattafesta2011}.
For passive control devices such as vortex generators, the control command can be the corresponding design parameters, such as height, length, position and the inclination angle, etc.
For active control devices this parameter includes the activation amplitude, frequency, and duty cycle, among others.
The training database can be constructed from synchronized measurements between Particle Image Velocimetry and sensor signals.
The resulting estimators from the machine learning approaches introduced above will provide understanding of relationships between the flow field and sensor signals under various actuation commands in real world applications, and will guide the design of the model-based, closed-loop flow control \citep{Samimy2007}.

\section*{Acknowledgements}
This work is supported 
by the National Science Foundation of China (NSFC) through grants 12172109 and 12172111 and
by the  Natural Science and Engineering grant 2022A1515011492 of Guangdong province, China.
We acknowledge valuable discussions 
with Guy Cornejo Maceda, Nan Deng,  Qixin (Kiki) Lin, Marek Morzy\'nski and Luc Pastur.
\textcolor{black}{We are indebted to the anonymous referees for ample constructive advice.}

\oneappendix   
\section{\textcolor{black}{Effects of Sensor Noise to the Estimation Performance}}
\label{sec:noise}

\begin{figure}
    \centering
    \includegraphics[width=.8\linewidth]{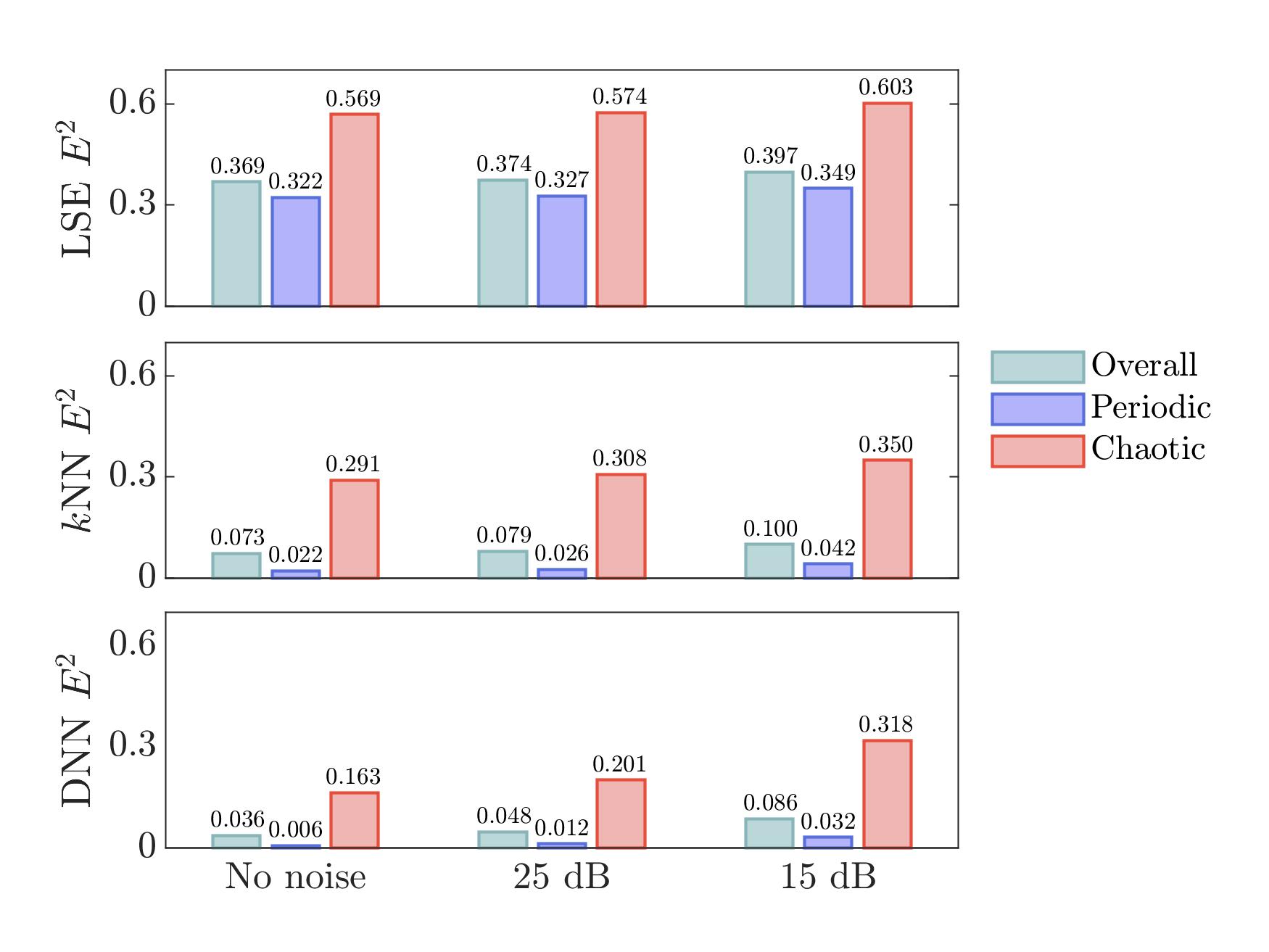}
    \caption{Squared overall estimation error $E^2$ with different noise levels in the sensor signals.}
    \label{fig:noise}
\end{figure}

One major concern to apply control-oriented flow estimation techniques to the real experiment is the robustness against sensor noise.
To evaluate the performance of the estimation methods under noisy conditions, we add two different levels of Gaussian noise to the velocity measurements from the 9 virtual probes.
We adopt two different levels of signal-to-noise ratio (SNR) to all virtual probes, then apply the estimation methods to the noisy dataset following the procedure introduced in \Cref{sec:estimation}.
The two different SNRs we use in this study are 25 dB and 15 dB, respectively.
A 25 dB noise level represents a relatively optimum measurement and a 15 dB SNR will imply really noisy measurements.
\Cref{fig:noise} displays the squared overall estimation error $E^2$ from all three estimators under the proposed SNRs.
For a complete comparison, the squared estimation error without artificial sensor noise is also presented in this figure.
From all methods, the estimation errors only slightly increase with the enhanced noise level.
Comparatively the estimation performance of periodic flows are more robust than that of the chaotic flows as the noise level increases.
Among all three methods, LSE always leads to the highest estimation error under different noise levels, either in periodic cases or chaotic cases.
On the other hand, DNN performs consistently better than $k$NN and LSE.
These results suggest the feasibility of non-linear, control-oriented flow estimation techniques towards their applications in real experiments.

\bibliographystyle{jfm}
\bibliography{references_abbr}


\end{document}